\documentclass[a4paper,11pt,twoside,reqno]{amsart}
\usepackage{preamble}
\usepackage{amsmath,amssymb,amsfonts}
\usepackage{algorithmic}
\usepackage{graphicx}
\usepackage{textcomp}
\usepackage{xcolor}
\usepackage{caption}
\usepackage{subcaption}
\usepackage{xurl}
\newcommand\nnfootnote[1]{%
  \begin{NoHyper}
  \renewcommand\thefootnote{}\footnote{#1}%
  \addtocounter{footnote}{-1}%
  \end{NoHyper}
}
\usepackage{listings}
\def\BibTeX{{\rm B\kern-.05em{\sc i\kern-.025em b}\kern-.08em
    T\kern-.1667em\lower.7ex\hbox{E}\kern-.125emX}}

\begin{document}

\title[Balancing Privacy and Utility in Correlated Data]{Balancing Privacy and Utility in Correlated Data: A Study of Bayesian Differential
Privacy\\{\small \textit{E\lowercase{xtended version}}}}

\author{Martin Lange (*)}
%\orcid{0009-0006-4605-6466}
\address{Karlsruhe Institute of Technology, KASTEL Security Research Labs}
%  \city{Karlsruhe}
%  \country{Germany}
\email{lange@martin-lange.eu}
\author{Patricia Guerra-Balboa (*)}
%\orcid{0009-0003-3035-8230}
\email{patricia.balboa@kit.edu}

\author{Javier Parra-Arnau}
%\orcid{0000-0002-1772-1088}
\address{Universitat Politècnica de Catalunya}
%  \city{Karlsruhe}
%  \country{Germany}
\email{javier.parra@upc.edu}

\author{Thorsten Strufe}
%\orcid{0000-0002-8723-9692}
%  \city{Karlsruhe}
%  \country{Germany}
\email{thorsten.strufe@kit.edu}

\begin{abstract}
Privacy risks in differentially private (DP) systems increase significantly when data is correlated, as standard DP metrics often underestimate the resulting privacy leakage, leaving sensitive information vulnerable. Given the ubiquity of dependencies in real-world databases, this oversight poses a critical challenge for privacy protections. Bayesian differential privacy (BDP) extends DP to account for these correlations, yet current BDP mechanisms indicate notable utility loss, limiting its adoption.

In this work, we address whether BDP can be realistically implemented in common data structures without sacrificing utility---a key factor for its applicability. By analyzing arbitrary and structured correlation models, including Gaussian multivariate distributions and Markov chains, we derive practical utility guarantees for BDP. Our contributions include theoretical links between DP and BDP and a novel methodology for adapting DP mechanisms to meet the BDP requirements.
Through evaluations on real-world databases, we demonstrate that our novel theorems enable the design of BDP mechanisms that maintain competitive utility, paving the way for practical privacy-preserving data practices in correlated settings.

\end{abstract}

\maketitle
\nnfootnote{
(*) These authors contributed equally.}
\section{Introduction}
\noindent
%DP is cool but THERE IS A PROBLEM WITH CORRELATION
\textit{Differential privacy} (DP)~\cite{dwork2014algorithmic} has become the leading framework for preserving privacy in data analysis, providing formal guarantees that protect individuals' sensitive information. 
However, its protection guarantees are limited to statistically independent data records, i.e., DP mechanisms can leak private information when the underlying data is correlated. The limitations of DP for protecting correlated data have been theoretically exposed~\cite{Kifer_2011_FreeLunch,Liu_2016_DDP,WangCorrelated2020,miranda2023sok} and empirically confirmed with attacks on real databases~\cite{Humphries_2023_Membership}.
This is a significant issue, as correlations among data records are common in real-world databases, such as those induced by friendships in social networks~\cite{LibenSocialNetworkNowell_2003} or genetic similarities among  family members~\cite{Almadhoun_2019_Genomic_privacy}.

%BAYESIAN DP AS SOLUTION
As a response to the limitations of DP in the presence of correlation, several instantiations of the Pufferfish framework--a general methodology to define privacy notions--have been proposed to specifically address this challenge~\cite{kifer_pufferfish_2014,Liu_2016_DDP,He_2014_Blowfish,Li_2019_Notion,Chong_2024_Notion}. Among them, \textit{Bayesian Differential Privacy} (BDP)~\cite{Yang_2015_BDP} stands out for its simplicity and generality: it provides a strict strengthening of DP, supports arbitrary correlation structures, and preserves the composability properties of DP--capabilities that are not generally achievable within the Pufferfish framework. BDP also underlies extensions such as prior DP~\cite{Li_2019_Notion} and correlated DP for location data~\cite{Chong_2024_Notion}.

%Short primer on BDP
While DP assumes the adversary knows all records except the target, BDP considers arbitrary priors, including those where unknown records are correlated. It ensures bounded changes in output distributions even when the target record is part of a correlated subset. When data is independent, BDP and DP coincide. Under correlation, however, BDP quantifies worst-case leakage by integrating the mechanism’s output with the data distribution via Bayes' rule, capturing adversarial advantages that DP overlooks. Hence, BDP mitigates correlation-driven reconstruction attacks that breach DP’s guarantees as empirically shown in~\cite{chakrabarti2022optimal}.

%OUR RESEARCH PROBLEM: We need better protection than DP but good utility, can BDP provide that?
While BDP provides a robust framework for assessing privacy leakage under data dependencies, its practical applicability remains uncertain. The few mechanisms that satisfy this notion~\cite{Yang_2015_BDP,chakrabarti2022optimal} are limited to specific correlation models, such as Gaussian Markov random fields--a subclass of multivariate Gaussian distributions forming a Markov random field where missing edges correspond to zeros in the inverse covariance matrix~\cite{rue2005gaussian}--and binary-state Markov chains with a symmetric transition matrix.
Given the scarcity of mechanisms and their applicability restrictions, it remains unclear whether BDP can serve as a usable privacy notion. Moreover, the only solution for Gaussian Markov fields reported highly conservative utility, since noise addition scales linearly with the number of records in the database and their only mitigation is to weaken BDP privacy by incorporating assumptions about the adversary~\cite{Yang_2015_BDP}.

% Research question & our solution
In summary, DP privacy leakage estimation does not provide sufficient protection under data dependencies, and there is a need for improved utility with the robust BDP framework. Motivated by this issue, this paper examines BDP’s utility from both theoretical and practical perspectives, analyzing its limitations and proposing new strategies to reduce utility loss while maintaining  BDP privacy guarantees. Particularly, we present theoretical bounds on the accuracy of BDP mechanisms and derive specific utility guarantees when certain correlation models are assumed. To formally analyze utility, we use the standard utility metric for DP mechanisms, $(\alpha, \beta)$-accuracy~\cite{dwork2014algorithmic,wasserman2010statistical}, due to its mathematical formalism and broad applicability. For the experimental results, we focus on two specific, albeit common, tasks: counting and sum queries~\cite{dwork2014algorithmic}.

Prior impossibility results~\cite{kifer_no_2011,kifer_pufferfish_2014} show that strong utility under BDP without distributional assumptions is fundamentally limited. We extend this insight by proving that, without any assumption on the data correlation model, no BDP mechanism can simultaneously guarantee meaningful $(\alpha,\beta)$-accuracy and valid privacy. Thus, the rest of our work examines whether targeting specific correlation models can improve utility.

Particularly, we analyze the impact of limiting the amount of correlated records, and we investigate the applicability of BDP to both discrete and continuous correlation models. For the discrete case, we analyze data following a Markov chain and, for continuous data, we analyze multivariate Gaussian correlation. We focus on these two particular correlation models following previous work in BDP~\cite{Yang_2015_BDP,Li_2019_Notion} and due to their relevance in many real-world applications such as medical ~\cite{Brainard_1992_Height}, location ~\cite{Gambs_2012_Location_Markov}, or activity data~\cite{Duong_2005_Activity_Markov}. 

For each correlation model studied, we prove novel theorems that bound the BDP leakage of a DP mechanism. Notably, our BDP leakage bound for Gaussian multivariate models is tighter than that provided in~\cite{Yang_2015_BDP}, and our correlation model is broader. 
These privacy bounds provide a systematic way to build BDP mechanisms by adjusting the parameters of existing DP mechanisms. 
Using this approach, we propose novel BDP mechanisms based on Laplace noise. 
Furthermore, we calculate the accuracy of our BDP mechanisms showing the improved accuracy compared to scenarios where protection is required against any correlation.

Finally, we provide insight into how our theoretical results apply in practice to real-world data containing Gaussian and Markov correlations. This allows us to confirm that our results enhance the utility of BDP mechanisms in actual applications. 

In summary, this work makes the following main contributions:
\begin{itemize}
 % \item We establish a new bound on the optimal accuracy of BDP mechanisms under arbitrary correlations, highlighting the need to assume a specific distribution to retain utility.

  \item We prove a bound on the BDP leakage of a DP mechanism with a fixed number of arbitrarily correlated records, showing it is tight. We call this the \textit{general bound}.

  \item We derive a tighter BDP leakage bound for DP mechanisms under multivariate Gaussian correlations, improving on the general bound and prior work. This provides a systematic method for constructing more accurate BDP mechanisms tailored to Gaussian dependencies.

  \item We derive a BDP leakage bound for DP mechanisms under Markovian correlations, improving on the general bound when transition probabilities are similar. This enables the design of more accurate mechanisms than prior approaches in Markov settings.

\end{itemize}

The paper is organized as follows: In \Cref{sec:related_work,sec:background}, we review relevant prior work and provide the necessary preliminaries. We then present our analysis of arbitrary correlation limiting the number of correlated records in \Cref{sec:limited}. In \Cref{sec:gaussian}, we analyze the impact of Gaussian correlation on BDP and provide our improved bound in \Cref{th:gaussian_dp_bdpl_bound}. In \Cref{sec:markov}, we present analogous results for the Markov scenario. Finally, we discuss our empirical study in \Cref{sec:experiments}, demonstrating the practical relevance of our theoretical results, and conclude with a brief summary in \Cref{sec:conclusions}.

This is the extended version of the paper accepted in the Proceedings of the VLDB Endowment (PVLDB), 2025. The code used for our experiments is accessible in \url{https://github.com/lange-martin/privacy-utility-bdp}.

\section{Related Work}\label{sec:related_work}
\noindent
%Pesimist view
The challenge of designing privacy mechanisms that remain robust under arbitrary correlations has been a central concern in the development of privacy frameworks. Foundational work by \citeauthor{Kifer_2011_FreeLunch}~\cite{Kifer_2011_FreeLunch} introduced free-lunch Privacy, the first formalism to consider the impact of correlations on privacy guarantees. Their no-free-lunch theorem shows that, under arbitrary data distributions, achievable utility is fundamentally constrained. 
However, they express utility in terms of discriminants--an abstraction that is neither intuitively interpretable nor translatable into practical utility metrics.
\citeauthor{kifer_pufferfish_2014}~\cite{kifer_pufferfish_2014} further rise this concern defining the general Pufferfish framework for privacy notions proving that any Pufferfish notion protecting against arbitrary correlations will face the same free-lunch utility challenge.

The existing strategy for obtaining Pufferfish privacy~\cite{song_pufferfish_2017} mechanisms requires 
% The main strategy to fulfill pufferfish definitions ~\cite{song_pufferfish_2017} relays on a 
noise calibration based on the Wasserstein distance. 
%However, this approach faces significant limitations. 
It does not, however, provide a closed-form solution, but requires computing the Wasserstein distance between the conditional output distributions corresponding to all pairs of sensitive values.
This is computationally  intractable~\cite{song_pufferfish_2017,nuradha_pufferfish_2023} in the general case. 
While a closed-form mechanism is derived for specific Markov chain models, it relies on a weakened instantiation of Pufferfish that assumes limited adversarial background knowledge, and therefore cannot be meaningfully compared to BDP.

The only concrete evidence of the potential applicability of pure BDP in practice has been provided in the context of Gaussian and Markov correlation models. In their foundational work, \citeauthor{Yang_2015_BDP}~\cite{Yang_2015_BDP} proposed adapting the Laplace mechanism to defend against correlated leakage in Gaussian Markov Random Fields. They also established preliminary theoretical connections between DP and BDP in this setting. Despite these important contributions, the proposed mechanisms face several limitations: (1) the approach is restricted to Gaussian Markov models, which greatly limits its practical scope. (2) Even within this narrow domain, the privacy guarantees degrade linearly with the number of correlated records, resulting in excessive noise that renders the mechanism impractical. Although the authors suggest mitigating this by limiting the adversary’s knowledge, such a compromise weakens the privacy model and undermines the core guarantees of BDP.
(3) The proposed mechanisms remain purely theoretical and have not been evaluated in real-world scenarios, leaving their practical effectiveness uncertain.

A more recent effort by \citeauthor{chakrabarti2022optimal}~\cite{chakrabarti2022optimal} proposes an adaptation of the randomized response to BDP over binary Markov chains. However, this mechanism is extremely constrained: it only applies to lazy, binary, stationary Markov chains and does not provide any general bounds relating DP and BDP leakage. Moreover, the only closed-form expressions for mechanism parameters holds under the restrictive assumption of a symmetric transition matrix limiting its usability even further.

In response to these limitations, several relaxed privacy notions have been proposed to strike a better balance between privacy and utility. Mutual Information Privacy (MI DP)~\cite{cuff_differential_2016} and its extension to Pufferfish~\cite{nuradha_pufferfish_2023}, for example, can be viewed as a relaxation of Pufferfish, offering a framework where traditional mechanisms like Laplace and Gaussian can be recalibrated to account for correlation. These methods yield promising theoretical utility guarantees. However,  MI  guarantees are weaker, in particular, MI characterizes average-case privacy leakage rather than worst-case guarantees, and therefore cannot substitute the BDP framework when worst-case guarantees are desired.

In conclusion, while previous work highlights the limitations of DP protection and the need for BDP as a privacy standard, the challenge of providing utility with BDP protection remains unresolved, and the relationship between DP and BDP is not fully understood.

\section{Background}\label{sec:background}
\noindent
In this section, we present the fundamental definitions and notation (summarized in~\Cref{tab:notation}) necessary to understand this work.
\renewcommand{\arraystretch}{1.2}
\begin{table}
    \small
    \centering
    \begin{tabular}{p{3cm}p{9cm}}
        \textbf{Notation}                                       & \textbf{Description}  \\\hline\hline
        $\mathcal{X}$                                           & Domain of a single record $x \in \mathcal{X}$.  \\\hline
        $\mathcal{M} : \mathcal{X}^n \to \mathcal{Y}$   & Randomized mechanism with input from domain $\mathcal{X}^n$ and output in codomain $\mathcal{Y}$. \\\hline
         $\mathbf{X} = (X_1, \dots, X_n) $ & Random vector representing the input of $\M$.        \\\hline 
        $Y$                                                     & Random variable representing output of $\M$.\\\hline
                                  $[n]$         & Set $\{1, \dots, n\}$ for $n \in \mathbb{N}$. \\\hline
        $\mathbf{X}_K = (X_{i_1}, \dots, X_{i_k})$         & Random vector of a subset $K = \{i_1, \dots, i_k\} \subseteq [n]$ of the random variables $X_1, \dots, X_n$. \\\hline
          $\mathbf{x}_{K} =  (x_{i_1}, \dots, x_{i_k})$          & Database with $k$ records belonging to $\mathcal{X}^{k}$ \\\hline
        
    \end{tabular}
    \caption{Notation summary}
    \label{tab:notation}
    \normalsize
\end{table}
\subsection{Differential Privacy and Metric Privacy}\label{sec:background_dp}
\noindent
In the bounded formulation of DP~\cite{dwork2014algorithmic}, the database is assumed to consist of a finite number \( n \) of rows, 
$
D = (x_1, \dots, x_n) \in \mathcal{X}^n,
$
drawn from the joint distribution of the random vector 
$
\mathbf{X} = (X_1, \dots, X_n),
$
where each row represents data associated with an individual, sampled from a universe of records \( \mathcal{X} \). 
We use 
$
[n] := \{1, \dots, n\}
$
to denote the set of indices. For a subset \( K = \{i_1, \dots, i_k\} \subseteq [n] \), we define the subvector 
$
\mathbf{X}_K \in \mathcal{X}^k$ as
$\mathbf{X}_K := (X_{i_1}, \dots, X_{i_k}).
$
In particular, 
$
\mathbf{X}_{-i}$ denotes $ \mathbf{X}_K $ with $ K = [n] \setminus \{i\}$. The attacker is assumed to know all records except for a target index \( i \in [n] \), for which all possible values \( x_i \) and \( x_i' \) must be indistinguishable. Formally,

\begin{definition}[Differential Privacy~\cite{dwork2014algorithmic}]\label{def:dp}
    A randomized mechanism $\mathcal{M}: \mathcal{X}^n \to \mathcal{Y}$ is called \textit{$\varepsilon$-differentially private}, if for all measurable sets $S \subseteq \mathcal{Y}$ any target index \( i \in [n] \), any target values \( x_i, x_i' \in \mathcal{X} \), and any remaining values \( \mathbf{x} \in \mathcal{X}^{n-1} \), we have
\begin{align*}
        \Pr[Y \in S \mid \mathbf{X}_{-i} = \mathbf{x}, X_i = x_i]\leq
        \e^{\varepsilon}\Pr[Y \in S \mid \mathbf{X}_{-i} =\mathbf{x}, X_i = x_i'].
\end{align*}
\end{definition}
The output of $\M$ is represented by the random variable $Y$, which  depends on the input data. The DP leakage $\varepsilon$ governs the privacy-utility trade-off: a smaller $\varepsilon$ means that the output distributions for neighboring inputs are ``closer together'', resulting in higher privacy  with an opposing effect on utility (See~\Cref{thm:laplace_mechanism_accuracy}).

%All $D, D' \in \mathcal{X}^n$ such that $d_H(D, D') = 1 $ are called \textit{bounded-neighboring databases}~\cite{Kifer_2011_FreeLunch}. This neighborhood definition specifies which information can change, while ensuring that the output probabilities remain similar up to \( e^\varepsilon \). 

We focus on a bounded DP due to its broad applicability and  its close relation to BDP. However, other neighboring definitions, i.e., specifications of which information can change, while ensuring that the output probabilities remain similar up to \( e^\varepsilon \), exist~\cite{Desfontaines_2020_DP_SoK}. For instance, in streaming data applications it is common to use \textit{event-level DP}~\cite{dwork2014algorithmic}: While each stream belongs to an individual, two streams are neighbors if they differ in one single time step value. We will see an example of application of this neighborhood in~\Cref{sec:experiments}. The change of neighborhood allows to encode protection against different privacy threats~\cite{Desfontaines_2020_DP_SoK,Chatzikokolakis_2013_d_privacy}. To obtain a general framework suitable to model a large variety of privacy problems,
\citeauthor{Chatzikokolakis_2013_d_privacy}~\cite{Chatzikokolakis_2013_d_privacy} introduce \emph{metric privacy} as a generalization of DP that encapsulates the neighborhood notion and privacy leakage $\varepsilon$ into a single parameter $d$, which determines the level of indistinguishability  between databases:

\begin{definition}[Metric Privacy~\cite{Chatzikokolakis_2013_d_privacy}]\label{def:d_privacy}
    Given $d\colon\mathcal{X}^{2n} \to \mathbb{R}$ a pseudometric,
    a randomized mechanism $\mathcal{M} : \mathcal{X}^n \to \mathcal{Y}$ is called \textit{$d$-private} if for all databases $D, D' \in \mathcal{X}^n$ and all measurable sets $S \subseteq \mathcal{Y}$ we have
    \begin{equation*}
        \Pr_{\M}[Y \in S \mid \mathbf{X} = D] \leq e^{d(D, D')} \Pr_{\M}[Y \in S \mid \mathbf{X} = D'].
    \end{equation*}
\end{definition}

This definition makes it challenging for an adversary to distinguish between databases $D$ and $D'$ that are ``close'' according to the metric $d$. However, if the two databases are significantly different, the output distributions can differ more, making it easier for the adversary to distinguish them. Note that $d$-privacy is equivalent to DP when considering the Hamming distance scaled by $\varepsilon$.

% SensitivityT
One of the earliest and most common methods proven to satisfy $\varepsilon$-DP is the Laplace mechanism~\cite{dwork2014algorithmic}:

\begin{definition}[Laplace Mechanism~\cite{dwork2014algorithmic}]\label{def:laplace_mechanism}
    Let $f: \mathcal{X}^n \to \mathbb{R}^k$ be a function and its \textit{sensitivity}  defined as
    \begin{equation*}
        \Delta f := \sup_{d_{H}(D,D')=1} || f(D) - f(D') ||_1.
    \end{equation*} 
    Given that sensitivity $\Delta f<\infty$ and $\varepsilon > 0$, the Laplace mechanism is defined for all $D\in\X^n$ as $\M_{\varepsilon,f}(D) = f (D) + (Z_1,\dots, Z_k)$
   where $Z_i$ are i.i.d. random variables that follow the Laplace distribution centered at $0$ and with scale $\frac{\Delta f}{\varepsilon}$.
\end{definition}

While $\M_{\varepsilon,f}$ provides $\varepsilon$-DP, adding noise to the output of a function $f$ undoubtedly has an impact on utility. A well-established metric for quantifying the utility of a private mechanism is the $(\alpha, \beta)$-accuracy~\cite{Liu_2016_DDP,Blum_2013_Accuracy}. It provides a measure of how well the mechanism approximates a true statistic or function while considering the inherent randomness introduced by the mechanism:

\begin{definition}[$(\alpha, \beta)$-Accuracy~\cite{Blum_2013_Accuracy}]\label{def:alpha_beta_accuracy}
    A randomized mechanism $\mathcal{M}$ is \textit{$(\alpha, \beta)$-accurate with respect to function $f$} if for all databases $D \in \mathcal{X}^n$ we have
    \begin{equation*}
        \Pr[| \mathcal{M}(D) - f(D) | \geq \alpha] \leq \beta.
    \end{equation*}
\end{definition}

A randomized mechanism $\mathcal{M}$ is $(\alpha, \beta)$-accurate if an error of magnitude $\alpha$ has a probability of at most $\beta$. Thus, the smaller $\alpha$ and/or $\beta$, the better the accuracy of mechanism $\mathcal{M}$.  Here, $\alpha$ quantifies the error tolerance, and $\beta$ the failure probability. More precisely, it refers to the utility guarantee that with probability at least $1-\beta$, the mechanism's output is within an interval of radius $\alpha$ centered on the true value. For example, the Laplace mechanism  accuracy follows:

\begin{proposition}[\cite{dwork2014algorithmic}]\label{thm:laplace_mechanism_accuracy}
    Let $\mathcal{M}_{\varepsilon, f}$ be the Laplace mechanism. Let $\beta \in (0,1]$ be a probability. Then $\mathcal{M}_{\varepsilon, f}$ is $(\alpha, \beta)$-accurate with respect to $f$ with $\alpha = \ln\left(\beta^{-1}\right) \frac{\Delta f}{\varepsilon}$.
\end{proposition}

This accuracy result for the Laplace mechanism  is tight~\cite{dwork2014algorithmic}.

\subsection{Bayesian Differential Privacy}\label{sec:bck_bdp}
\noindent
BDP~\cite{Yang_2015_BDP} is an instantiation of the general Pufferfish framework that extends  DP privacy guarantees to settings with correlated data. It assumes the adversary is uncertain between two possible records $x_i,x'_i$, analogously as DP. However, it eliminates the notion of neighboring databases in order to consider different possible adversaries with different background knowledge. Formally, the adversary   $(K, i)$ is targeting the record at position $i$ and already knows the values of the sub vector $\mathbf{x}_K$ on the database. Then, for each adversary, Bayesian leakage is defined as follows:

\begin{definition}[Adversary-specific $\mathrm{BDPL}$~\cite{Yang_2015_BDP}]\label{def:bdpl}
    Given $\mathcal{M}: \mathcal{X}^n \to \mathcal{Y}$ a randomized mechanism, $\mathbf{X}$ the input random vector following the distribution $\pi$, the targeted record index $i\in[n]$, and the known record indices $K\subseteq[n]\backslash\{i\}$, the \textit{adversary-specific Bayesian differential privacy leakage} is
    \begin{equation*}\label{eq:bdp_bdpl}
        \mathrm{BDPL}_{(K, i)} = \sup_{x_i, x_i', \mathbf{x}_K, S} \ln \frac{\Pr[Y \in S \mid \mathbf{X}_K = \mathbf{x}_K, X_i = x_i]}{\Pr[Y \in S \mid \mathbf{X}_K = \mathbf{x}_K, X_i = x_i']},
    \end{equation*}
    where the supremum is taken over all the possible target values $x_i,x'_i\in\X$, all the possible known vector values $\mathbf{x}_K\in\X^K$ and all the measurable sets $S\subseteq\mathcal{Y}$.
\end{definition}

When computing the adversary-specific BDPL, the correlation between the unknown and known records modifies the final leakage since given the unknown remaining indices $U$, we have 
{\small
\begin{gather*}
  \Pr[Y \in S \mid \mathbf{x}_K, x_i]=\sum_{\mathbf{x}_U\in\X^u}\Pr[Y \in S \mid  \mathbf{x}_K,x_i, \mathbf{x}_U ]\Pr[\mathbf{x}_U \mid \mathbf{x}_K, x_i],
\end{gather*}
}
where $u=|U|=n-k-1$. The sum must be substituted by an integral in the continuous case. 

While the adversary-specific BDPL only accounts for a particular case, we aim to protect against any possible adversary. Therefore, to compute the worst-case leakage we take the supremum:
    
 \begin{definition}[Bayesian DP~\cite{Yang_2015_BDP}]
    A mechanism $\mathcal{M}$ satisfies \textit{$\varepsilon$-Bayesian differentially privacy}  if 
    \begin{equation*}
        \mathrm{BDPL}(\mathcal{M}) = \sup_{K,i} \mathrm{BDPL}_{(K,i)}(\mathcal{M})\leq\varepsilon,
    \end{equation*}
    where the supremum is taken over all the possible set of indices $i\in[n]$ and $K\subseteq [n] \setminus \{i\}$. $\mathrm{BDPL}(\mathcal{M})$ is called \textit{Bayesian differential privacy leakage}.
\end{definition}

The BDPL has a similar role to the privacy leakage $\varepsilon$ in DP: It measures the extent of a possible privacy violation by comparing the difference in the output probabilities of mechanism $\mathcal{M}$. A lower BDPL corresponds to higher privacy because any adversary will be less likely to differentiate between any two target values $x_i,x'_i\in\X$. Particularly, if $X_i,X_j$ are mutually independent for all $i\neq j\in[n]$ then $\varepsilon$-DP and $\varepsilon$-BDP are equivalent~\cite{Yang_2015_BDP}. 

While we have results on the accuracy loss associated with using DP mechanisms~\cite{vadhan2017complexity}, the impact of BDP protection on utility remains unclear. The following sections  aim to address this question by analyzing various correlation scenarios.

\section{Limited Number of Correlated Variables}\label{sec:limited}
\noindent
 To protect against potential correlations without making distributional assumptions--which are often unclear or hard to estimate~\cite{Sunn_ker_2015_Model_Uncertainty}--a mechanism must satisfy BDP with respect to all possible correlation distributions $\pi$, a condition we call protection under \textit{arbitrary correlation}. However, \citeauthor{kifer_pufferfish_2014} showed that under this assumption, any Pufferfish notion--including BDP--collapses to free-lunch privacy~\cite{Yang_2015_BDP,kifer_no_2011}. This corresponds to a metric privacy model where all dataset pairs are at distance $\varepsilon$, forcing all query outputs $f(D)$ and $f(D')$ to be $\varepsilon$-indistinguishable~\cite{dwork2006calibrating}--intuitively implying a complete loss of utility. To our knowledge, we are the first to formalize this limitation using the standard $(\alpha,\beta)$-accuracy metric, offering a concrete, interpretable, and widely used measure of utility loss that enables clearer reasoning and meaningful comparison across mechanisms.
\begin{proposition}\label{cor:bdp_accuracy_arbitrary_correlation}
Let $\mathcal{M}: \mathcal{X}^n \to \mathbb{R}$ be an $\varepsilon$-BDP mechanism protecting against arbitrary correlation. Let $0 \leq \beta < \frac{1}{e^\varepsilon + 1}$ be a real number and let $f: \mathcal{X}^n \to \mathbb{R}$ be a deterministic function. 
    If $\mathcal{M}$ is $(\alpha, \beta)$-accurate w.r.t.\ $f$, then
    \begin{equation*}
        \alpha > \frac{1}{2} \max_{D, D'} |f(D) - f(D')|.
    \end{equation*}
\end{proposition}
\begin{proof}
     First, note that any BDP mechanism protecting against arbitrary correlation is free-lunch~\cite{kifer_pufferfish_2014}. Therefore for every $S\subseteq \mathbb{R}$,  and for every pair $D,D\in\X^n$,
     \begin{equation}\label{eq:free-lunch}
         \Pr(\M(D)\in S)\leq \e^{\varepsilon}\Pr(\M(D')\in S).
     \end{equation}
     We use this property and proceed by \textit{reductio ad absurdum}. We assume that $\M$ fulfills an  $(\alpha, \beta)$-accuracy respect to $f$ with  $\alpha \leq \frac{1}{2} |f(D) - f(D')|$ and $\beta <  \frac{1}{e^\varepsilon + 1}$ and derive a contradiction for $D'$:
     {\small
    \begin{gather*}
        \Pr[|f(D') - \M(D')| \geq \alpha]
        = \Pr[\M(D') \in \mathbb{R} \setminus (f(D') - \alpha, f(D') + \alpha)]\\
        \geq \Pr[\M(D') \in (f(D) - \alpha, f(D) + \alpha)]  \\
\overset{\text{Eq.\ref{eq:free-lunch}}}{\geq} e^{-\varepsilon} \, \Pr[\M(D) \in (f(D) - \alpha, f(D) + \alpha)] \\
        = e^{-\varepsilon} \, (1 - \Pr[\M(D) \in \mathbb{R} \setminus (f(D) - \alpha, f(D) + \alpha)])  \\
        = e^{-\varepsilon} \, (1 - \Pr[|f(D) - \M(D)| \geq \alpha]) 
        \overset{(*)}{\geq} e^{-\varepsilon} \, (1 - \beta) 
        = \frac{1}{e^\varepsilon + 1} > \beta,
    \end{gather*}
    }
    where $(*)$ follows from the $(\alpha,\beta)$-accuracy assumption.
\end{proof}

Specifically, the result from \Cref{cor:bdp_accuracy_arbitrary_correlation} indicates that  for theoretically relevant privacy levels $\varepsilon\in(0,4)$~\cite{Lee_2011_Epsilon},  the only confidence interval where we can reliably estimate the actual value of our query function $f$, with standard confidence levels (e.g., between 90\% and 99\%), includes almost all possible query values. 
For instance, consider a free-lunch algorithm used to compute $f(D)$, where $f$ counts the number of infections in a database of $n$ individuals. If the algorithm outputs $\frac{n}{2}$, it suggests that half of the population is infected. However, with a 90$\%$ confidence interval, we cannot tell whether there is no infection at all, or whether the entire population is infected.

While designing accurate BDP mechanisms is infeasible under arbitrary correlation--potentially involving all records--it is often reasonable in practice to assume that only subsets of records are correlated. For instance, in the context of genomic data, an individual's genome is strongly correlated with that of their relatives, but not with the entire population~\cite{Almadhoun_2019_Genomic_privacy}. Hence,  we assume that only $m$ of $n$ records in the database are correlated with each other, formally:
\begin{definition}\label{def:gaussian_distribution}
    We say the random vector $\mathbf{X} = (X_1, \dots, X_n)$ has \textit{at most $m \leq n$ correlated random variables} if there exist disjoint sets of indices $C_1, \dots, C_r$ that make up $[n] = \bigcup_{l=1}^r C_l$ so that each set $C_l$ has maximum cardinality $m \geq |C_l|$ for any $l \in [r]$, and for any $l \in [r]$, the random variables $\{ X_j \mid j \in C_l\}$ are independent of the remaining random variables $\{ X_j \mid j \in [n] \setminus C_l\}$.
\end{definition}
This definition considers  multiple groups of up to $m$ correlated records as long as they do not ``overlap'', i.e., the records in one group are independent of the records in the other groups. 
Otherwise, we do not make any further assumptions about the distribution of the data. This allows us to find acceptable utility guarantees in \Cref{cor:arbitrary_accuracy_laplace} as long as $m$ is sufficiently small.

\subsection{Relationship between DP and BDP}
\noindent
We begin by introducing and proving a general bound on the BDP leakage of an $\varepsilon$-DP mechanism. Specifically, we show that if an $\varepsilon$-DP mechanism operates on data drawn from a distribution involving at most $m$ correlated random variables, then it satisfies $m\varepsilon$-BDP. The practice of scaling the DP leakage by the number of correlated records to estimate worst-case leakage under correlation has been used in prior work~\cite{Chen_2013_k_epsilon,Liu_2016_DDP}, but to our knowledge, this approach had not been formally shown to satisfy the BDP definition. We further prove that this bound is tight.
\begin{theorem}[The General Bound]\label{thm:arbitrary_general_bound}
     Let $\mathbf{X} = (X_1, \dots, X_n)$ be a random vector with at most $m \leq n$ correlated random variables that follows a distribution $\pi$.
    Then, any $\varepsilon$-DP mechanism with input data drawn from distribution $\pi$ is $ m\varepsilon$-BDP.
\end{theorem}

\begin{proof}
    Consider any adversary $(K, i)$ with $i \in [n]$, $K \subseteq [n] \setminus \{i\}$ and $k=|K|$. Since $\{C_j\}_{j\in r}$ is a partition of $[n]$, there exists an $l \in [r]$ so that we have target index $i \in C_l$. Thus, $C_l$ contains the index $i$ and all indices of random variables potentially correlated with $X_i$. Let set $\Tilde{C} := C_l \setminus K$ be the indices of random variables correlated with $X_i$ that are not already included in $K$. Then, we first show that the adversary-specific BDPL can be upper bounded  as follows:
    \begin{gather}\label{eq:intermidiate_bound}
        \mathrm{BDPL}_{(K, i)} = \sup_{S, \mathbf{x}_K, x_i, x_i'} \ln \frac{\Pr[Y \in S \mid \mathbf{x}_K,  x_i]}{\Pr[Y \in S \mid \mathbf{x}_K,x_i']} 
        \leq \sup_{S, \mathbf{x}_K, \mathbf{x}_{\Tilde{C}}, \mathbf{x}_{\Tilde{C}}'} \ln \frac{\Pr[Y \in S \mid  \mathbf{x}_K,\mathbf{x}_{\Tilde{C}}]}{\Pr[Y \in S \mid\mathbf{x}_K, \mathbf{x}_{\Tilde{C}}']} 
    \end{gather} 
    %% mini proof of the statement
    Assume that for all $\mathbf{x}_{\Tilde{C}} \in \mathcal{X}^{|\Tilde{C}|}$, we have
    \begin{equation}\label{eq:arbitrary_inequality_to_contradict}
        \Pr[Y \in S \mid \mathbf{X}_K = \mathbf{x}_K, X_i = x_i] > \Pr[Y \in S \mid \mathbf{X}_K = \mathbf{x}_K, \mathbf{X}_{\Tilde{C}} = \mathbf{x}_{\Tilde{C}}].
    \end{equation}
    Now, we bring this to a contradiction, thereby proving the opposite of \cref{eq:arbitrary_inequality_to_contradict}. Let index set $\Tilde{C}_{-i} := \Tilde{C} \setminus \{i\}$ include all indices of $\Tilde{C}$ except for $i$. Then, we have
    \begin{align}
        &\Pr[Y \in S \mid  \mathbf{x}_K, x_i] \notag \\ &= \int_{\mathcal{X}^{|\Tilde{C}_{-i}|}} \!\Pr[Y \!\in \!S |  \mathbf{x}_K, x_i, \mathbf{x}_{\Tilde{C}_{-i}}] \; p_{\mathbf{X}_{\Tilde{C}_{-i}}}\!(\mathbf{x}_{\Tilde{C}_{-i}} |  \mathbf{x}_K,  x_i) \diff \mathbf{x}_{\Tilde{C}_{-i}} \label{eq:arbitrary_total_prob_C} \\
        &= \int_{\mathcal{X}^{|\Tilde{C}_{-i}|}} \Pr[Y \in S \mid \mathbf{x}_K,   (x_i, \mathbf{x}_{\Tilde{C}_{-i}})] \; p_{\mathbf{X}_{\Tilde{C}_{-i}}}\!(\mathbf{x}_{\Tilde{C}_{-i}} \mid \mathbf{x}_K, x_i) \diff \mathbf{x}_{\Tilde{C}_{-i}} \label{eq:arbitrary_combine_to_XC}\\
        &< \int_{\mathcal{X}^{|\Tilde{C}_{-i}|}} \Pr[Y \in S \mid \mathbf{x}_K,  x_i] \; p_{\mathbf{X}_{\Tilde{C}_{-i}}}\!(\mathbf{x}_{\Tilde{C}_{-i}} \mid  \mathbf{x}_K, x_i) \diff \mathbf{x}_{\Tilde{C}_{-i}} \label{eq:arbitrary_apply_wrong_assumption} \\
        &= \Pr[Y \in S \mid  \mathbf{x}_K, x_i] \int_{\mathcal{X}^{|\Tilde{C}_{-i}|}} \; p_{\mathbf{X}_{\Tilde{C}_{-i}}}\!(\mathbf{x}_{\Tilde{C}_{-i}} \mid \mathbf{x}_K,x_i) \diff \mathbf{x}_{\Tilde{C}_{-i}} \label{eq:arbitrary_pull_out_of_integral} \\
        &= \Pr[Y \in S \mid  \mathbf{x}_K, x_i]. \label{eq:arbitrary_int_to_1}
    \end{align}
     where the random variable $X_i$ is already included in the condition, so only indices $\Tilde{C}_{-i}$ need to be added. In \cref{eq:arbitrary_combine_to_XC}, we combine the two conditions $X_i = x_i$ and $\mathbf{X}_{\Tilde{C}_{-i}} = \mathbf{x}_{\Tilde{C}_{-i}}$ into one condition $\mathbf{X}_{\Tilde{C}} = (x_i, \mathbf{x}_{\Tilde{C}_{-i}})$. Notice that this is the same condition, just stated differently. Then, we use \cref{eq:arbitrary_inequality_to_contradict}, which applies to all $\mathbf{x}_{\Tilde{C}} \in \mathcal{X}^{|\Tilde{C}|}$, in \cref{eq:arbitrary_apply_wrong_assumption}. Now, the first probability can be pulled out of the integral in \cref{eq:arbitrary_pull_out_of_integral} as it no longer depends on the value $\mathbf{x}_{\Tilde{C}_{-i}}$. The final \cref{eq:arbitrary_int_to_1} follows because a probability density integrated over its entire domain is always $1$. Note that if the variables are discrete the integral must be changed by a sum and the result follows analogously.
     
     We have shown that the initial probability is strictly smaller than itself--a contradiction. Thus, the opposite of our assumption in \cref{eq:arbitrary_inequality_to_contradict} must be true and there must exist a vector $\mathbf{x}_{\Tilde{C}}$ so that 
    \begin{equation}\label{eq:arbitrary_C_less_eq}
        \Pr[Y \in S \mid \mathbf{X}_K = \mathbf{x}_K, X_i = x_i] \leq \Pr[Y \in S \mid \mathbf{X}_K = \mathbf{x}_K, \mathbf{X}_{\Tilde{C}} = \mathbf{x}_{\Tilde{C}}].
    \end{equation}

    Analogously, we can show that there exists a vector $\mathbf{x}_{\Tilde{C}}' \in \mathcal{X}^{|\Tilde{C}|}$ with the opposite property
    \begin{equation}\label{eq:arbitrary_C_greater_eq}
        \Pr[Y \in S \mid \mathbf{X}_K = \mathbf{x}_K, X_i = x_i'] \geq \Pr[Y \in S \mid \mathbf{X}_K = \mathbf{x}_K, \mathbf{X}_{\Tilde{C}} = \mathbf{x}_{\Tilde{C}}].
    \end{equation}

    Thus, with \cref{eq:arbitrary_C_less_eq} and \cref{eq:arbitrary_C_greater_eq} we have~\Cref{eq:intermidiate_bound}.
    For any values of $\mathbf{x}_K$ and $x_i, x_i'$ included in the left supremum, we can always find values $\mathbf{x}_{\Tilde{C}}$ and $\mathbf{x}_{\Tilde{C}}'$ so that the ratio becomes greater or equal, and these values $\mathbf{x}_{\Tilde{C}}, \mathbf{x}_{\Tilde{C}}'$ are included in the supremum on the right-hand side.
  %%% end of mini proof
    
    Now, we apply~\Cref{eq:intermidiate_bound} to prove the following statement. Let the set $U = [n] \setminus (K \cup \Tilde{C})$, with $u=|U|$, include all remaining indices. Since by hypotheses $|\Tilde{C}|\leq|C_l|\leq m$, for any  known values $\mathbf{x}_K \in \mathcal{X}^{k}$, the correlated values $\mathbf{x}_{\Tilde{C}} \in \mathcal{X}^{|\Tilde{C}|}$ and $\mathbf{x}'_{\Tilde{C}} \in \mathcal{X}^{|\Tilde{C}|}$ we have
    \begin{gather*}
        \Pr_{\M}[Y \in S \mid \mathbf{x}_K, \mathbf{x}_{\Tilde{C}}]\\ = \int_{\mathcal{X}^{u}} \Pr_{\M}[Y\in S \mid \mathbf{x}_K,  \mathbf{x}_{\Tilde{C}}, \mathbf{x}_U] \; p_{\mathbf{X}_U}\!(\mathbf{x}_U \mid \mathbf{x}_K, \mathbf{x}_{\Tilde{C}}) \diff \mathbf{x}_U \\
        \leq \int_{\mathcal{X}^{u}} \e^{m \varepsilon} \Pr_{\M}[Y \in S | \mathbf{x}_K,\mathbf{x}'_{\Tilde{C}},\mathbf{x}_U] \; p_{\mathbf{X}_U}\!(\mathbf{x}_U | \mathbf{x}_K,\mathbf{x}_{\Tilde{C}}) \diff \mathbf{x}_U 
        \\= e^{m \varepsilon} \int_{\mathcal{X}^{u}} \Pr_{\M}[Y \in S \mid  \mathbf{x}_K, \mathbf{x}'_{\Tilde{C}}, \mathbf{x}_U] \; p_{\mathbf{X}_U}\!(\mathbf{x}_U \mid \mathbf{x}_K) \diff \mathbf{x}_U  \\
        =e^{m \varepsilon} \int_{\mathcal{X}^{u}} \Pr[Y \in S |  \mathbf{x}_K,  \mathbf{x}_{\Tilde{C}}', \mathbf{x}_U] \; p_{\mathbf{X}_U}\!(\mathbf{x}_U | \mathbf{x}_K,  \mathbf{x}_{\Tilde{C}}') \diff \mathbf{x}_U  \\
        ={} e^{m \varepsilon} \Pr[Y \in S \mid \mathbf{x}_K, \mathbf{x}_{\Tilde{C}}'] .
    \end{gather*}

Combining both inequalities we obtain the result.
\end{proof}
This bound may seem overly pessimistic, seemly assuming perfect positive correlation--the records are fully dependent, changing in lockstep: when one variable changes, the other changes in the same direction by exactly the same amount. This corresponds to the extreme case of linear dependence, where the Pearson correlation coefficient is $\rho = 1$, an edge case among all possible (including non-linear) correlation models. However, as we show in the following example, the bound remains tight even when this extreme case is excluded. Specifically, we provide a counterexample in which the bound holds even when $\rho$ is arbitrarily small--i.e., the variables do not deterministically determine one another. This confirms both the tightness of our result and that the bound cannot be improved, even in the absence of perfect correlation.
\begin{example}\label{thm:arbitrary_limit_rho}
\begin{table}
    \centering
\begin{tabular}{l|ll|l} 
& $X_1 = 0$ & $X_1 = 1$ & Total \\
\hline
$X_2 = 0$ & $\frac{1}{r^2}$ & $\frac{r - 1}{r^2}$ & $\frac{1}{r}$ 
\\
$X_2 = 1$ & $\frac{1}{r^3}$ & $\frac{r^3 - r^2-1}{r^3}$ & $\frac{r - 1}{r}$ \\ 
\hline
Total & $\frac{1+r}{r^3}$ & $\frac{r^3-r-1}{r^3}$ & $1$ \end{tabular}
\caption[Joint Probability Distribution of Random Variables $X_1$ and $X_2$]{Probability distribution of ~\Cref{thm:arbitrary_limit_rho}}\label{tab:distribution}
\end{table}

Let $r \in \mathbb{N}$. \Cref{tab:distribution} shows a valid probability distribution $\pi$ for $\mathbf{X} = (X_1, X_2)$. Note that
    \[
    \E[X_1X_2]= \frac{r^3 - r^2 - 1}{r^3}, \quad
\mathbb{E}[X_1] = \frac{r^3 - r - 2}{r^3}, \quad
\mathbb{E}[X_2] = \frac{r - 1}{r}. 
    \]
    Hence, the Pearson correlation coefficient satisfies:
\begin{gather*}
\rho_{X_1, X_2} = 
\tfrac{\mathbb{E}[X_1 X_2] - \mathbb{E}[X_1] \mathbb{E}[X_2]}
{\sqrt{\mathbb{E}[X_1^2] - \mathbb{E}[X_1]^2} \cdot \sqrt{\mathbb{E}[X_2^2] - \mathbb{E}[X_2]^2}} 
= \tfrac{\mathbb{E}[X_1 X_2] - \mathbb{E}[X_1] \cdot \mathbb{E}[X_2]}
{\sqrt{ \mathbb{E}[X_1](1 - \mathbb{E}[X_1]) } \cdot \sqrt{ \mathbb{E}[X_2](1 - \mathbb{E}[X_2]) }} \\
=\frac{
\frac{r^{3} - r^{2} - 1}{r^{3}}-\frac{{\left(r^{3} - r - 1\right)} {\left(r - 1\right)}}{r^{4}}
}{
\sqrt{\frac{r^{4} + r^{3} - r^{2} - 2 \, r - 1}{r^{6}}}\sqrt{\frac{r - 1}{r^{2}}}
}
=\tfrac{r^{2} - r - 1}{r^{4} \sqrt{\frac{r - 1}{r^{2}}} \sqrt{\frac{r^{4} + r^{3} - r^{2} - 2 \, r - 1}{r^{6}}}}\\
=\tfrac{r^{2} - r - 1}{ \sqrt{(r - 1)(r^{4} + r^{3} - r^{2} - 2 \, r - 1)}}=\sqrt{
\tfrac{r^{4} - 2 \, r^{3} - r^{2} + 2 \, r + 1}{r^{5} - 2 \, r^{3} - r^{2} + r + 1}
} \xrightarrow{r \rightarrow \infty} 0
\end{gather*}
Moreover, if $\M$ is $\varepsilon$-DP, then there are two neighboring databases $D,D'\in\{0,1\}^2$ for which the privacy loss reaches $\varepsilon$; otherwise, $\varepsilon$ is not tight for $\M$, and a smaller value could be used with the same reasoning. Without loss of generality,  we assume they differ in the first coordinate, otherwise by inverting~\Cref{tab:distribution} we get the same result and we assume
\[
\Pr[A(0, 0) \in S] = \e^{\varepsilon} \Pr[A(0, 1) \in S] = e^{\varepsilon} \Pr[A(1, 0) \in S] = \e^{\varepsilon} \Pr[A(1, 1) \in S],
\]
as is the case, for instance, with the Generalized Randomized Response mechanism~\cite{wang2017locally}.
Then, computing the BDPL we obtain for all $S\subseteq\{0,1\}^2$:
\begin{gather*}
        \e^{\mathrm{BDPL}} \geq \frac{\Pr[Y \in S \mid X_1 = 0]}{\Pr[Y \in S \mid X_1 = 1]} 
        =  \frac{\sum_{x_2 \in \{0,1\}} \Pr[Y \in S \mid X_1=0, X_2=x_2] \, \Pr[X_2=x_2 \mid X_1=0]}{\sum_{x_2 \in \{0,1\}} \Pr[Y \in S \mid X_1=1, X_2=x_2] \, \Pr[X_2=x_2 \mid X_1=1]}  \\
        =  \frac{e^{2\varepsilon} \, \Pr[\M(1, 1) \in S] \, \frac{r}{r+1} + e^\varepsilon \, \Pr[\M(1, 1) \in S] \,\frac{1}{r+1}}{e^\varepsilon \, \Pr[\M(1, 1) \in S] \, \frac{r^2-r}{r^3-r-1} + \Pr[\M(1, 1) \in S] \, \frac{r^3-r^2-1}{r^3-r-1}}  \\
        =\frac{e^{2\varepsilon}  \, \frac{r}{r+1} + e^\varepsilon \,\frac{1}{r+1}}{e^\varepsilon \, \, \frac{r^2-r}{r^3-r-1} +  \, \frac{r^3-r^2-1}{r^3-r-1}},
        \end{gather*}
     for all $r\in\N$. Since
     \[
\lim_{r\to\infty}\frac{e^{2\varepsilon}  \, \frac{r}{r+1} + e^\varepsilon \,\frac{1}{r+1}}{e^\varepsilon \, \, \frac{r^2-r}{r^3-r-1} +  \, \frac{r^3-r^2-1}{r^3-r-1}} = \e^{2\varepsilon},
     \]
     taking the limit when $r$ tends to infinity we have $\mathrm{BDPL}\geq 2\varepsilon$. Since the general upper bound of the BDPL of an $\varepsilon$-DP mechanism is $2 \varepsilon$ we have $\mathrm{BDPL}=2\varepsilon$. Therefore, taking arbitrary large $r$, we have $\rho$ arbitrary close to zero, making impossible perfect correlation, and BDPL arbitrary close to $2\varepsilon$.
\end{example}
\Cref{thm:arbitrary_limit_rho} proves that, without additional hypotheses, the general bound from \Cref{thm:arbitrary_general_bound} is tight, even if we limit the Pearson correlation coefficient $\rho$ to be arbitrarily small. 
\subsection{Accuracy}
\noindent
\Cref{thm:arbitrary_general_bound} enables to use $(\frac{\varepsilon}{m})$-DP mechanisms as $\varepsilon$-BDP mechanisms. However, reducing $\varepsilon$ in a DP mechanism often has a negative impact on utility. In particular, we investigate the impact on the accuracy of the Laplace mechanism. As a consequence of our result \Cref{thm:arbitrary_general_bound} and \Cref{thm:laplace_mechanism_accuracy} we obtain the following result:

\begin{corollary}\label{cor:arbitrary_accuracy_laplace}
    Let  $\mathcal{M}_{\varepsilon, f}$ be the the Laplace $\varepsilon$-DP mechanism that approximates the query $f\colon\X^n\to\mathbb{R}$ with input described by the random vector $\mathbf{X} = (X_1, \dots, X_n)$
    with at most $m \leq n$ correlated random variables that follows distribution $\pi$. 
    Then, if $\mathcal{M}_{\varepsilon, f}$ is $(\alpha, \beta)$-accurate w.r.t. $f$, there exists an $\varepsilon$-BDP mechanism $\mathcal{B}$ whose input is drawn from $\pi$ and that is $(m \alpha, \beta)$-accurate w.r.t. $f$.
\end{corollary}
\begin{proof}
    From \Cref{thm:laplace_mechanism_accuracy}, we know that the $(\alpha, \beta)$-accuracy of $\mathcal{M}_{\varepsilon, f}$ with respect to $f$ is 
    \begin{equation*}
        \alpha = \ln\left(\frac{1}{\beta}\right) \cdot \frac{\Delta f}{\varepsilon}
    \end{equation*}
    for any $\beta \in (0, 1]$ because $\mathcal{M}_{\varepsilon, f}$ uses the Laplace mechanism.

    The idea is to also use the Laplace mechanism for $\mathcal{B}$, but to use an adjusted DP privacy parameter $\varepsilon'$ so that $\mathcal{B}$ is $\varepsilon$-BDP. We will see that this results in $(m \alpha, \beta)$-accuracy.
    With the general bound from \Cref{thm:arbitrary_general_bound}, we know that the mechanism $\mathcal{B}$ is $m \varepsilon'$-BDP. Thus, we have
    \begin{equation*}
        m \varepsilon' = \varepsilon \ \Leftrightarrow \ \varepsilon' = \frac{1}{m} \varepsilon.
    \end{equation*}
    Now, we can calculate the accuracy of mechanism $\mathcal{B}$ because it also uses the Laplace mechanism and we now know the used DP privacy leakage $\varepsilon' = \varepsilon / m$. Mechanism $\mathcal{B}$ is $(\alpha', \beta)$-accurate with
    \begin{align*}
        \alpha' &= \ln\left(\frac{1}{\beta}\right) \cdot \frac{\Delta f}{\varepsilon'} = \ln\left(\frac{1}{\beta}\right) \cdot \frac{m \Delta f}{\varepsilon} = m \alpha.
    \end{align*}
    Thus, mechanism $\mathcal{B}$ is $(m \alpha, \beta)$-accurate.
\end{proof}
This result shows that the error $\alpha$ of the Laplace mechanism increases proportionally with the number of correlated records when moving from $\varepsilon$-DP to $\varepsilon$-BDP, and while making no assumption about the distribution of the records. This may be acceptable when the number of correlated records $m$ is small. For example, if $m=2$, the error $\alpha$ doubles when transitioning from DP to BDP. If the DP mechanism’s error is small, this increase may be acceptable. However, utility sharply decreases as $m$ grows.

Since we have shown that our bound on BDPL is tight under the assumption of arbitrary correlation, the utility bound cannot be improved, even when the Pearson correlation coefficient is close to zero. This motivates the next two sections, where we investigate whether additional assumptions on the correlation model can lead to tighter bounds, enabling reduced noise and improved utility while still protecting against correlation attacks.
\section{Multivariate Gaussian Correlation}\label{sec:gaussian}
\noindent
A wide variety of phenomena are effectively modeled using a Gaussian distribution~\cite{Panaretos_2016_Statistics}. For example, physiological measures such as height and weight are correlated among family members, and the joint distribution of height and weight in a large population is well fit by a bivariate Gaussian distribution~\cite{Brainard_1992_Height}. Consequently, we explore the applicability of BDP to multivariate Gaussian data. 

When we are dealing with a database of $n$ records, and each record is drawn from a Gaussian distribution, we can model the joint distribution of all records as a multivariate Gaussian distribution. This model also captures linear correlation between records~\cite{Shao_2003_Statistics}.  
\begin{definition}[Multivariate Gaussian Distribution~\cite{Shao_2003_Statistics}]\label{def:gaussian_data_set} %p. 19
    Let $\mathbf{X} = (X_1, \dots, X_n)$ be a random vector, let vector $\mu \in \mathbb{R}^n$ be real and let matrix $\Sigma \in \mathbb{R}^{n \times n}$ be symmetric and positive definite. 
    We say $\mathbf{X}$ follows the \textit{multivariate Gaussian distribution with mean $\mu$ and covariance $\Sigma$} if the probability density of $\mathbf{X}$ for any point $\mathbf{x} \in \mathbb{R}^n$ is
    \begin{equation*}
        p_\mathbf{X}(\mathbf{x}) = \frac{1}{\sqrt{(2\pi)^n |\Sigma|}} \exp(-\frac{1}{2}(\mathbf{x} - \mu)^\top \Sigma^{-1}(\mathbf{x} - \mu)),
    \end{equation*}
    where $|\Sigma|$ is the determinant of $\Sigma$.
    We %denote that $\mathbf{X}$ follows the multivariate Gaussian distribution by 
    write $\mathbf{X} \sim \mathcal{N}( \mu, \Sigma )$.
\end{definition}

 We establish a relationship between DP and BDP for data drawn from a multivariate Gaussian distribution, based on the maximum Pearson correlation coefficient, which is calculated directly from the covariance matrix~\cite{Shao_2003_Statistics}. This provides a new, tighter upper bound for the BDPL that improves upon the specific Gaussian bound given in~\cite{Yang_2015_BDP} and upon the general bound $n\varepsilon$ for any correlation model. 

However, our bound applies only to a specific class of mechanisms: those that satisfy both DP and metric privacy under the $\ell_1$ metric. We show in~\Cref{sec:gaus_relation} that the clipped Laplace mechanism meets these criteria and develop a practical application in~\Cref{sec:utility_experiment_data_and_queries}. To establish this result, we first connect metric privacy with an analogous form of BDP, termed Bayesian metric privacy, which we define below.

\subsection{Relationship between Metric Privacy and Bayesian Metric Privacy}
\noindent
%\texorpdfstring{$d$}{d} 
Unbounded continuous data domains, such us $\mathbb{R}^n$, usually produce challenges on DP application due to infinite sensitivities~\cite{andres2013geo}. In the context of BDP, \citeauthor*{Yang_2015_BDP}~\cite{Yang_2015_BDP} defined a relaxation to work in those domains: If the data domain is equivalent to the real numbers (i.e., $\mathcal{X}^n = \mathbb{R}^n$), they defined a modified leakage, $\mathrm{BDPL}(\M; M )$, where they only take into account the leakage between points with a distance smaller than $M \in \mathbb{R}$, i.e.,
    \begin{equation*}\label{eq:bdp_bdpl_real}
         \sup_{|x_i - x_i'| \leq M, \mathbf{x}_K, S} \ln \frac{\Pr[Y \in S \mid \mathbf{X}_K = \mathbf{x}_K, X_i = x_i]}{\Pr[Y \in S \mid \mathbf{X}_K = \mathbf{x}_K, X_i = x_i']}.
    \end{equation*}
Applying this BDP relaxation leaves indistinguishability between records at distances greater than $M$ entirely uncontrolled. While this may increase applicability, it reduces privacy and limits insights into the impact of correlation.

However, metric privacy provides a solution to quantify privacy leakage as the distance $d(D,D')$ for each pair of databases $D,D'$ when the maximum privacy leakage cannot be bounded~\cite{Chatzikokolakis_2013_d_privacy}. 
Therefore, we define Bayesian metric privacy as equivalent to metric privacy where the indistinguishability between two records $x,x'$ depends on the distance $d(x,x')$ between them. Note that the change from databases to records is necessary because BDP does not apply to neighboring databases, but to target records, as we describe in~\Cref{sec:bck_bdp}. In this way, we can work with $\mathbb{R}^n$ as the data domain without losing information about the privacy leakage.
\begin{definition}[Target Dependent Leakage]
    Given a randomized mechanism $\mathcal{M}: \mathcal{X}^n \to \mathcal{Y}$, $\mathbf{X}$ the input random vector following the distribution $\pi$, the targeted record index $i\in[n]$, and the known record indices $K\subseteq[n]\backslash\{i\}$, the \textit{adversary-specific target dependent} BDPL of $\mathcal{M}$ w.r.t. adversary $(K, i)$ for any target values $x, x' \in \mathcal{X}$ is
    \begin{equation*}\label{eq:gaussian_target_dependent_bdpl}
        \mathrm{BDPL}_{(K,i)}(x, x') = \sup_{\mathbf{x}_K, s} \ln \frac{p_{Y}(s \mid \mathbf{X}_K = \mathbf{x}_K, X_i = x)}{p_{Y}(s \mid \mathbf{X}_K = \mathbf{x}_K, X_i = x')}.
    \end{equation*}
\end{definition}
Given that we understand the leakage for each pair of data records we can simply define Bayesian metric privacy analogously to the original metric privacy notion:
\begin{definition}[Bayesian Metric Privacy]
    Let $d$ be a (pseudo)metric on $\mathcal{X}^{2}$. A mechanism $\mathcal{M}$ is \textit{Bayesian $d$-private} if for all $x,x'\in\X$, 
    \begin{equation*}
        \mathrm{BDPL}(x, x') = \sup_{i, K} \mathrm{BDPL}_{(K,i)}(x, x')\leq d(x,x'),
    \end{equation*}
    where the supremum is taken over all the possible set of indices $i\in[n]$ and $K\subseteq [n] \setminus \{i\}$. $\mathrm{BDPL}(x, x')$ is called  \textit{target dependent} BDPL.
\end{definition}

The only difference between BDP and Bayesian $d$-privacy is that Bayesian $d$-privacy does not take the supremum over $x,x'$. Moreover, both notions are equivalent when the distance metric is defined as $d(x, x') = \varepsilon$ for $x \neq x'$ and $d(x, x') = 0$ otherwise.

Now we can prove the relation between a $d$-private and a Bayesian $d$-private mechanism when the data distribution is a multivariate Gaussian. Particularly, we focus on the $\ell_1$ distance due to its direct application to the Gaussian case. 
We formalize the conditions needed to obtain our bound:

\begin{definition}\label{def:gaussian_limited_covariance_matrix}
    For $\rho \in [0, 1]$ and $n \in \mathbb{N}$, we call the matrix $\Sigma_\rho \in \mathbb{R}^{n \times n}$ a \textit{limited covariance matrix} if
    \begin{itemize}
        \item the matrix $\Sigma_\rho$ is symmetric and positive definite,
        \item the diagonal of $\Sigma_\rho$ is constant, i.e., there is a variance $\sigma^2 > 0$ so that $\Sigma_{\rho,ii} = \sigma^2$ for all $i \in [n]$ and,
        \item any pairwise correlation is limited by $\rho$. I.e., for all $i \neq j$ we have $|\Sigma_{\rho,ij}| \leq \rho \sigma^2$.
    \end{itemize}
\end{definition}
The first condition is required to be a valid covariance matrix for a Gaussian distribution (see \Cref{def:gaussian_distribution}). The second condition ensures that no records have a deviating variance, i.e., all records are drawn from the same one-dimensional distribution. The final condition imposes that the maximum Pearson correlation coefficient between any two random variables $X_i$ and $X_j$ is bounded by $\rho$. If we limit $\rho$ to be small enough, we get a novel bound on the BDPL (See~\Cref{thm:gaussian_d-privacy_bdpl_bound}). However, before we can prove \Cref{thm:gaussian_d-privacy_bdpl_bound}, we require the following lemma that establishes the maximum BDPL of a single adversary.

\begin{lemma}\label{thm:gaussian_correlation_bdpl}
 Let $\mathcal{M}$ with data domain $\mathbb{R}^{n}$ be an $(\varepsilon\ell_1$)-private mechanism where $\varepsilon > 0$  with input data drawn from a multivariate Gaussian distribution $\mathcal{N}(\mu, \Sigma)$ where $m \geq 2$ is the maximum number of correlated variables.

 Let $K = \{m - k, \dots, m - 1\}$ be the set of known indices for the adversary $H$ correlated with the target $X_m$, with $k\leq m-2$, $T=K\cup\{m\}$ and $U$ the set of unknown records correlated with $X_m$. If the principal submatrix $\Sigma_{T}$ spanning $k+1$ rows and columns is invertible, then the adversary-specific target dependent $\mathrm{BDPL}$ of $\M$ for any target values $x_m, x_m' \in \mathbb{R}$ is bounded by 
    \begin{equation*}
        \mathrm{BDPL}_{(H, m)}(x_m, x_m') \leq \varepsilon|x_m-x_m'|\left(\|\Sigma_{U;T} \Sigma^{-1}_{T} \mathbf{e}_{k+1}\|_1 + 1\right)
    \end{equation*}
    where  $\mathbf{e}_{k+1}\equiv(0, \dots, 0, 1)^\top \in \mathbb{R}^{k+1}$ and the notation of the Gaussian distribution $\mathcal{N}(\mu, \Sigma)$ is reordered as
    \begin{gather}
        \mu = \begin{pmatrix}
            \mu_U \\
            \mu_{T}\\
            \mu_S
        \end{pmatrix}, 
        \Sigma = \begin{pmatrix}
            \Sigma_U            & \Sigma_{U;T}& \mathbf{0} \\
            \Sigma_{U;T}^\top & \Sigma_{T}& \mathbf{0}\\
            \mathbf{0} & \mathbf{0}& \Sigma_{S}
        \end{pmatrix}.
    \end{gather}\label{eq:notation_gaus}
\end{lemma}
\begin{proof}
The proof is derived from the multivariate Gaussian distribution properties. We introduce the following notation:
\begin{itemize}
    \item From the set of unknown records $V$, $U$ are correlated with $X_m$ and $W$ are independent.
    \item From the set of known records $H$, $K$ are correlated with $X_m$ and $L$ are independent. $T=K\cup\{m\}$ and $R=H\cup\{m\}$.
\end{itemize}
Note that, by definition of the covariance matrix, $\Sigma_{ij} = 0$ for all pairs of independent variables $X_i \perp X_j$, which leads to the zero submatrices in Eq.~\ref{eq:notation_gaus}.

First, we prove that for any $\mathbf{x}_V\in\mathbb{R}^v$ and $\mathbf{x}_T,\mathbf{x}'_T\in\mathbb{R}^{n-v}$, there exists a $\gamma\in\mathbb{R}^{u}$, such that\begin{equation}\label{eq:gaussian_conditional_equivalence}
       p_{\mathbf{X}_V}\!(\mathbf{x}_V \mid \mathbf{X}_{T}=\mathbf{x}_{T})\equiv p_{\mathbf{X}_V}\!(\mathbf{x}_{U},\mathbf{x}_{W} \mid \mathbf{X}_{T}=\mathbf{x}_{T})=p_{\mathbf{X}_U}\!(\mathbf{x}_{U}+ \gamma,\mathbf{x}_{W}\mid \mathbf{X}_{T} = \mathbf{x}'_{T}).  
    \end{equation}
Then, the combination of this property with the $\varepsilon\ell_1$ condition gives the result.

 We prove \Cref{eq:gaussian_conditional_equivalence} by using that the conditional distribution of a Gaussian distribution also follows the Gaussian distribution~\cite{Petersen_2008_Matrix}, i.e., $\mathbf{X}_U \mid \mathbf{X}_T = \mathbf{x}_T \sim \mathcal{N}(\hat{\mu}_U, \hat{\Sigma}_U)$ with
    \begin{gather*}
        \hat{\mu}_U = \mu_U + \Sigma_{U;T} \Sigma_{T}^{-1}(\mathbf{x}_{T} - \mu_{T}),\quad 
        \hat{\Sigma}_U = \Sigma_U - \Sigma_{U;T} \Sigma_{T}^{-1} \Sigma_{U;T}^\top. 
    \end{gather*}
    
    While the conditional mean $\hat{\mu}_U$ depends on the specific value $\mathbf{x}_{T}$, the conditional covariance $\hat{\Sigma}_U$ remains fixed. Therefore, the two distributions (conditioned on $\mathbf{x}_{T}$ and $\mathbf{x}_{T}' \in \mathbb{R}^{k+1}$ respectively) only differ by a translation, i.e., for any $\mathbf{x}_U\in\mathbb{R}^u$ we have 
    {\small
    \begin{align*}
        \MoveEqLeft[4] p_{\mathbf{X}_U}\!({\mathbf{x}}_U + \Sigma_{U;T} \Sigma_{T}^{-1}(\mathbf{x}_{T} - \mathbf{x}_{T}') \mid \mathbf{X}_{T} = \mathbf{x}_{T}) \notag \\
        &= \frac{1}{\sqrt{(2 \pi)^u |\hat{\Sigma}_U|}} \exp\Bigl(-\frac{1}{2}(\mathbf{x}_U + \Sigma_{U;T} \Sigma_{T}^{-1}(\mathbf{x}_{T} - \mathbf{x}_{T}') - \hat{\mu}_U)^\top \hat{\Sigma}_U^{-1} (\mathbf{x}_U + \Sigma_{U;T} \Sigma_{T}^{-1}(\mathbf{x}_{T} - \mathbf{x}_{T}') - \hat{\mu}_U)\Bigr) \notag\\
        &= \frac{1}{\sqrt{(2 \pi)^u |\hat{\Sigma}_U'|}} \exp\Bigl(-\frac{1}{2}({\mathbf{x}}_U - \hat{\mu}_U')^\top \hat{\Sigma}_U'^{-1} ({\mathbf{x}}_U - \hat{\mu}_U')\Bigr)\notag  \\
        &= p_{\mathbf{X}_U}\!(\mathbf{x}_U \mid \mathbf{X}_{T} = \mathbf{x}_{T}'). 
    \end{align*}
   }
   
    Therefore, we have shown that the condition of the probability density $p_{\mathbf{X}_U}$ can simply be changed by additively shifting the input for the density, where the shift is $\gamma=\Sigma_{U;T} \Sigma_{T}^{-1}(\mathbf{x'}_{T} - \mu_{T})$.

   Second, we can use the fact that mechanism $\M$ is $(\varepsilon\ell_1)$-private, therefore, for all $\mathbf{x}_n,\mathbf{x}'_n\in\mathbb{R}^{n}$,
   \begin{align}\label{eq:d_privacy_inequality}
        p_{Y}(s \mid{} \mathbf{X}_n =\mathbf{x}_n)\leq \exp(\varepsilon\|(\mathbf{x}_n,\mathbf{x}'_n)\|_1 ) \, p_{Y}(s \mid \mathbf{X}_n = \mathbf{x}'_n) 
    \end{align}

    Now, applying \Cref{eq:gaussian_conditional_equivalence} and \Cref{eq:d_privacy_inequality} to the density function computation we can bound the adversary-specific target dependent BDPL. Let $Y$ be the random variable that represents the output of mechanism $\M$. Let random vector $\mathbf{X} = (X_1, \dots, X_n)$ follow the multivariate Gaussian distribution $\mathbf{X} \sim \mathcal{N}(\mu, \Sigma)$. To simplify notation, we combine the random vector $\mathbf{X}_H$ and random variable $X_n$ into one target-inclusive vector $\mathbf{X}_{R}$ with $\mathbf{x}_{R} = (\mathbf{x}_H, x_m)^\top$ and $\mathbf{x}_{H}' = (\mathbf{x}_K, x_m')^\top$.
    Considering the definition of the adversary-specific target dependent BDPL of $(H, m)$ we have 
    \begin{equation*}\label{eq:gaussian_bdpl_density}
        \mathrm{BDPL}_{(H,m)}(x_m, x_m') = \sup_{\mathbf{x}_H, s} \ln \frac{p_{Y}(s \mid \mathbf{X}_H = \mathbf{x}_H, X_m = x_m)}{p_{Y}(s \mid \mathbf{X}_K = \mathbf{x}_K, X_m = x_m')}\equiv \sup_{\mathbf{x}_R,\mathbf{x}'_R, s} \ln \frac{p_{Y}(s \mid \mathbf{X}_R = \mathbf{x}_R)}{p_{Y}(s \mid \mathbf{X}_R = \mathbf{x}'_R) }
    \end{equation*}
    with the supremum taken over $s \in \mathbb{R}$ and known values $\mathbf{x}_H \in \mathbb{R}^h$. 
    
    We calculate the adversary-specific target dependent BDPL by rewriting the density $p_{Y}(s \mid \mathbf{x}_H, x_m)$ in terms of $p_{Y}(s \mid  \mathbf{x}_H, x_m')$:

    \begin{align}
    \MoveEqLeft[7] p_{Y}(s \mid \mathbf{X}_H = x_H, X_n = x_n) \equiv p_{Y}(s \mid \mathbf{x}_{R}) \\
    ={}& \int_{\mathbb{R}^v} p_{Y}(s \mid  \mathbf{x}_V,  \mathbf{x}_{R}) \, p_{\mathbf{X}_V}(\mathbf{x}_V \mid  \mathbf{x}_{R}) \, \mathrm{d}\mathbf{x}_V \\
    ={}& \int_{\mathbb{R}^v} p_{Y}(s \mid  \mathbf{x}_{U},  \mathbf{x}_{W},  \mathbf{x}_{R}) \,
    p_{\mathbf{X}_{W}}(\mathbf{x}_{W} \mid \mathbf{x}_{L}) \,
    p_{\mathbf{X}_{U}}(\mathbf{x}_{U} \mid  \mathbf{x}_{T}) \, \mathrm{d}\mathbf{x}_V\label{eq:gaussian_substitution}\\
    ={}& \int_{\mathbb{R}^v} p_{Y}(s \mid  \Tilde{\mathbf{x}}_{U}+\gamma,  \mathbf{x}_{W},  \mathbf{x}_{R}) \,
    p_{\mathbf{X}_{W}}(\mathbf{x}_{W} \mid \mathbf{x}_{L}) \,
    p_{\mathbf{X}_{U}}(\Tilde{\mathbf{x}_{U}} \mid  \mathbf{x'}_{T}) \, \mathrm{d}\mathbf{x}_V\label{eq:gaussian_apply_conditional_equivalence} \\
    \leq &\exp((\|\gamma\|_1 + |x_m - x_m'|)\varepsilon) \,\int_{\mathbb{R}^v} p_{Y}(s \mid  \Tilde{\mathbf{x}}_{U},\mathbf{x}_{W}\mathbf{x}'_{R}) \, p_{\mathbf{X}_{W}}\!(\mathbf{x}_{W} \mid  \mathbf{x}_{L})\,p_{\mathbf{X}_{U}}(\Tilde{\mathbf{x}}_{U} \mid  \mathbf{x'}_{T}) \diff \Tilde{\mathbf{x}}_V \label{eq:gaussian_apply_sum_inequality} \\
    = &\exp((\|\gamma\|_1 + |x_m - x_m'|)\varepsilon) \, p_{Y}(s \mid \mathbf{X}_{H} = x_{H}, X_m = x'_m) \notag
    \end{align}
    
    Eq.~\ref{eq:gaussian_substitution} is obtained applying that $\mathbf{X_{W}}\perp \mathbf{X}_T$ and $\mathbf{X_{U}}\perp \mathbf{X}_L$. Then we substitute $\mathbf{x}_U$ for $\Tilde{\mathbf{x}}_U + \gamma$ in \cref{eq:gaussian_substitution} using the change of variable theorem for multiple integrals~\cite[p. 310]{Shurman_2016_calculus_euclidean}. The substitution is linear so the domain $\mathbb{R}^v$ over which we integrate does not change, and the determinant of the Jacobian of the substitution is simply $1$. 
    Then, we apply \cref{eq:gaussian_conditional_equivalence} to obtain \cref{eq:gaussian_apply_conditional_equivalence}. Finally, we use the inequality from \cref{eq:d_privacy_inequality} to derive \cref{eq:gaussian_apply_sum_inequality}, since $\|(\mathbf{x}_U,\mathbf{x}_T) - (\mathbf{x}_U+\gamma,\mathbf{x}'_T)\|_1=\|\gamma\|_1 + |x_m - x_m'|$.

    Now, we can formulate the upper bound of the adversary-specific target dependent BDPL for $(H, m)$ for all $x_m, x_m' \in \mathbb{R}$:
    \begin{align}
        \mathrm{BDPL}_{(H,m)}(x_m, x_m') &\leq (\|\gamma\|_1 + |x_m - x_m'|)\varepsilon \notag\\
        &= (\|\Sigma_{U;T}\Sigma_{T}^{-1}(\mathbf{x}_{T} - \mathbf{x}_{T}')\|_1 + |x_m - x_m'|)\varepsilon \notag\\
        &= (\|\Sigma_{U;T}\Sigma_{T}^{-1}((\mathbf{x}_K, x_m)^\top - (\mathbf{x}_K, x_m')^\top)\|_1 + |x_m - x_m'|)\varepsilon \notag\\
        &= (\|\Sigma_{U;T}\Sigma_{T}^{-1}(\mathbf{0}, x_m - x_m')^\top\|_1 + |x_m - x'_m|)\varepsilon \notag\\
        &= (\|\Sigma_{U;T}\Sigma_{T}^{-1}\mathbf{e}_{k+1}\|_1 + 1)|x_m - x_m'|\varepsilon. \qedhere \notag
    \end{align}
\end{proof}
Applying previous lemma, when we limit $\rho$ to be small enough, specifically, smaller $\rho(m-2)<1$ we get the following bound:

\begin{theorem}\label{thm:gaussian_d-privacy_bdpl_bound}
    Let $\mathcal{M}$ with data domain $\mathbb{R}^{n}$ be an $(\varepsilon\ell_1$)-private mechanism where $\varepsilon > 0$ with input data drawn from a multivariate Gaussian distribution $\mathcal{N}(\mu, \Sigma_\rho)$ with mean $\mu \in \mathbb{R}^n$ and limited covariance matrix $\Sigma_\rho \in \mathbb{R}^{n \times n}$ (Def.~\ref{def:gaussian_limited_covariance_matrix}) and limited number of correlated records $m\leq n$, such that $\rho ({m - 2})< 1$ is the maximum correlation coefficient.
    Then, for any $x, x' \in \mathbb{R}$ we have
    \begin{equation*}\label{eq:bdpl_bound_to_prove}
        \mathrm{BDPL}(x, x') \leq \left(\frac{m^2}{4(\frac{1}{\rho} - m + 2)} + 1\right)|x'-x|.
    \end{equation*}
\end{theorem}
\begin{proof}
    To prove an upper bound for the target dependent BDPL, we must bound the adversary-specific target dependent BDPL of every possible adversary $(H, i)$ with $i \in [n]$ and $H \subseteq [n] \setminus \{i\}$. The remaining indices besides $H$ and $i$ make up the unknown indices $V = [n] \setminus \{H,i\}$. We differentiate between two cases.

    \textbf{Case 1:} There are no unknown indices correlated with the target $i\in[n]$, i.e., we have $U = \varnothing\subseteq V$. Therefore, 
    we can calculate the target dependent BDPL of this adversary by using the fact that $\mathcal{M}$ is $(\varepsilon \ell_1)$-private.
    We following the same notation as in previous lemma: Set of unknown records $V$, $U$ are correlated with $X_i$ and $W$ are independent. Set of known records $H$, $K$ are correlated with $X_i$ and $L$ are independent, so in particular $i\not\in L$. $T=K\cup\{i\}$ and $R=H\cup\{i\}$. Hence, we obtain:
    \begin{align}
    \MoveEqLeft[7] p_{Y}(s \mid \mathbf{X}_H = x_H, X_i = x_i) \equiv p_{Y}(s \mid \mathbf{x}_{R}) \\
    ={}& \int_{\mathbb{R}^v} p_{Y}(s \mid  \mathbf{x}_V,  \mathbf{x}_{R}) \, p_{\mathbf{X}_V}(\mathbf{x}_{V} \mid  \mathbf{x}_{R}) \, \mathrm{d}\mathbf{x}_V \\
    ={}& \int_{\mathbb{R}^v} p_{Y}(s \mid  \mathbf{x}_V,  \mathbf{x}_{R}) \, p_{\mathbf{X}_V}(\mathbf{x}_V \mid  \mathbf{x}_{L}) \, \mathrm{d}\mathbf{x}_V \\
    \leq{}& \int_{\mathbb{R}^u} \e^{\varepsilon|x_i-x'_i|}p_{Y}(s \mid  \mathbf{x}_V,  \mathbf{x'}_{R}) \, p_{\mathbf{X}_V}(\mathbf{x}_V \mid  \mathbf{x}_{L}) \, \mathrm{d}\mathbf{x}_V \\
    ={}& \e^{\varepsilon|x_i-x'_i|} p_{Y}(s \mid \mathbf{X}_H = x_H, X_i = x'_i).
    \end{align}
    Consequently,
    \begin{align*}
        \mathrm{BDPL}_{(H,i)}(x_i, x_i') &= \sup_{\mathbf{x}_H, s} \ln \frac{p_Y(s \mid \mathbf{X}_H = \mathbf{x}_H, X_i = x_i)}{p_Y(s \mid \mathbf{X}_H = \mathbf{x}_H, X_i = x_i')}\notag \\
        &= \ln e^{\varepsilon |x_i - x_i'|} = \varepsilon |x_i - x_i'|\notag  \\
        &\leq \left(\frac{m^2}{4(\frac{1}{\rho} - m + 3)} + 1\right)\varepsilon|x_i-x_i'|.\notag
    \end{align*}

    \textbf{Case 2:} There is at least one unknown record correlated with the target, i.e., $U \neq \varnothing$. 
    
    Let $k = |K|$ be the number of known records correlated with the target and $u = |U|$ the number of unknown ones. With out loss of generality we have $K = \{m - k, \dots, m - 1\}$ and $i = m$. Otherwise, we simply reorder the components of the random vector $\mathbf{X}$ so that the statements about $K_i$ and $i$ apply.

    Choose $\Sigma_U \in \mathbb{R}^{u \times u}$, $\Sigma_{U;T} \in \mathbb{R}^{u \times (k+1)}$ and $\Sigma_{T} \in \mathbb{R}^{(k+1) \times (k+1)}$ so that the following holds:
    \begin{equation*}
        \Sigma_\rho = \begin{pmatrix}
            \Sigma_U            & \Sigma_{U;T}& \mathbf{0} \\
            \Sigma_{U;T}^\top & \Sigma_{T}& \mathbf{0}\\
            \mathbf{0} &\mathbf{0}& \Sigma_{S}
            \end{pmatrix}
    \end{equation*}
    In order to use \Cref{thm:gaussian_correlation_bdpl}, we require the principal submatrix $\Sigma_{T}$ spanning the last $k+1$ rows and columns to be invertible. We separate in two subcases:

    \textbf{Case 2.1} ($m=2$).
    In such case, given than $U\neq \varnothing$ we have that $k=0$ and
    \begin{equation*}
        \Sigma_\rho = \begin{pmatrix}
            \sigma_1^2           & \rho\sigma_1\sigma_2  &\mathbf{0}\\
            \rho\sigma_1\sigma_2 & \sigma_2^2  &\mathbf{0}\\
            \mathbf{0}&\mathbf{0}&\sigma_{S}
        \end{pmatrix}
    \end{equation*}
    Therefore, for all $\sigma_2\neq 0$ we have that $\gamma=\frac{\rho\sigma_1\sigma_2}{\sigma_2^2}=\frac{\rho\sigma_1}{\sigma_2}\leq \rho$. Applying~\Cref{thm:gaussian_correlation_bdpl} we obtain
    \begin{equation}
        \mathrm{BDPL}_{(H,i)}(x_i, x_i') \leq (\|\gamma\|_1 + 1)\varepsilon|x'_i-x_i|=(\rho+1)\varepsilon|x'_i-x_i|.
    \end{equation}
    
    \textbf{Case 2.2} ($m> 2$).
    We proceed finding a strictly positive lower bound for the eigenvalues of $\Sigma_{T}$. Conveniently, we can later use this fact to bound the entries of $\Sigma_{T}^{-1}$.
    We denote the individual cells of $\Sigma_{T}$ as $a_{jl}$ for all $j,l \in [k+1]$.
    
    According to the Gershgorin circle theorem~\cite{Gershgorin_1931_Eigenwerte}, every eigenvalue of a real matrix such as $\Sigma_{T}$ is contained in a closed disc on the complex number plane with center $a_{jj}$ and radius $\sum_{l \neq j} |a_{jl}|$ for $j \in [k+1]$. Since the eigenvalues of a symmetric matrix must be real~\cite[p. 1]{Hawkins_1975_Spectral}, we are only concerned with the real part of this disc, i.e., the  interval $[a_{jj} - \sum_{l \neq j} |a_{jl}|, a_{jj} + \sum_{l \neq j} |a_{jl}|]$. We can construct a lower bound of the smallest eigenvalue $\lambda_-$ of $\Sigma_{T}$ by finding the lowest border of these intervals.
    \begin{align}
        \lambda_- &\geq \min_j a_{jj} - \sum_{l \neq j} |a_{jl}| \\
        &\geq \min_j \sigma^2 - \sum_{l \neq j} |\rho \sigma^2| \label{eq:gaussian_apply_limit_on_rho} \\
        &= \sigma^2 - k \rho \sigma^2 \\
        &= (1 - k\rho)\sigma^2 \label{eq:gaussian_all_eigenvalues_positive}\\
        &> (1 - (m - 2) \, \frac{1}{m - 2}) \sigma^2 \label{eq:gaussian_bound_positive} = 0 
    \end{align}
    In \cref{eq:gaussian_apply_limit_on_rho}, we use the fact that every random variable has the same variance $\sigma^2$ and the correlation between any two random variables in $\mathbf{X}$ is in the interval $[-\rho,\rho]$. %See \Cref{def:gaussian_limited_covariance_matrix} on the limited covariance matrix.
    Then we show in \cref{eq:gaussian_bound_positive} that the bound is positive since  $k$ must be $m - 2$ or smaller because there is one targeted record and $U \neq \varnothing$ and the maximum correlation $\rho$ is bounded with $\rho < \frac{1}{m - 2}$. Thus, we have shown that each eigenvalue of $\Sigma_{T}$ is strictly positive and the matrix is therefore invertible. 
    
    Now, by direct application of~\Cref{thm:gaussian_correlation_bdpl}  we have an upper bound of the adversary-specific target dependent BDPL for any $x_i, x_i' \in \mathbb{R}$ with 
    \begin{equation}\label{eq:gaussian_specific_bdpl_bound}
        \mathrm{BDPL}_{(H,i)}(x_i, x_i') \leq (\|\Sigma_{U;T} \Sigma_{T}^{-1} \mathbf{e}_{k+1}\|_1 + 1)\varepsilon|x'_i-x_i|. 
    \end{equation}
    This bound depends on the adversary-specific matrices $\Sigma_{U;T}$ and $\Sigma_{T}$. Our goal is to find a bound for the total target dependent BDPL, irrespective of the specific adversary.  We denote the individual cells of $\Sigma_{U;T}$ as $b_{jl}$ for all $j\in[u]$, $l \in [k+1]$ and the cells of $\Sigma_{T}^{-1}$ as $\alpha_{jl}$ for all $j,l\in [k+1]$.

    Since $\Sigma_{U;T}$ only contains covariances between random variables from $U$ and  $K$ or $n$ and the covariance is bounded by $\rho \sigma^2$, for all indices $j,l \in [u]$ we obtain that
    \begin{equation}
        |b_{jl}| \leq \rho \sigma^2.
    \end{equation}

    Now, we bound the entries of the inverse matrix $\Sigma_{T}^{-1}$. The 2-norm of any symmetric matrix $A \in \mathbb{R}^{m \times m}$ is defined as
    \begin{equation*}
        \|A\|_2 := \sup_{\mathbf{x}\in\mathbb{R}^m} \{ \|A\mathbf{x}\|_2 \colon \|\mathbf{x}\|_2 = 1\}.
    \end{equation*}
    The 2-norm of $A$ is equal to its maximum singular value (i.e., the largest absolute eigenvalue for a symmetric matrix)~\cite[p. 47]{Trop_2004_Norm}. Thus, we can use the 2-norm to bound the entries of a symmetric matrix by its largest absolute eigenvalue $|\lambda_+|$. Let $\mathbf{e}_j \in \mathbb{R}^m$ be the vector with 1 at position $j$ and 0 elsewhere. For every entry $a_{ij}$ in $A$, we have 
    \begin{align}
        |a_{ij}| &= \sqrt{a_{ij}^2} \leq \sqrt{\sum_{k \in [m]} a_{kj}^2} = \|A\mathbf{e}_j\|_2 \leq \|A\|_2 = |\lambda_+|. \label{eq:gaussian_bound_entry_by_ev}
    \end{align}
    Additionally, the eigenvalues of an inverse $A^{-1}$ can be determined knowing the eigenvalues of $A$. Let $\lambda \in \mathbb{R}$ be any eigenvalue of $A$; $\lambda$ cannot be zero because $A$ is invertible, consequently,
    \begin{alignat}{2}
        \quad&& Ax &= \lambda x \notag\\
        \Leftrightarrow \quad&& A^{-1}Ax &= \lambda A^{-1} x \notag\\
        \Leftrightarrow \quad&& x &= \lambda A^{-1}x \notag \\
        \Leftrightarrow \quad&& \frac{1}{\lambda} x &= A^{-1}x \notag
    \end{alignat}
    Thus, the eigenvalues of $A^{-1}$ are the inverses of the eigenvalues of $A$.
    Putting it all together, the entries of $\Sigma_{T}^{-1}$ are smaller or equal to the inverse of the smallest absolute eigenvalue of $\Sigma_{T}$. In \cref{eq:gaussian_all_eigenvalues_positive}, we have shown that all eigenvalues of $\Sigma_{T}$ are positive and larger than $(1 - k\rho)\sigma^2$. Therefore, the smallest \textit{absolute} eigenvalue of $\Sigma_{T}$, denoted $\lambda_{-}$ is also larger than this bound. Now, these two facts (the entries of a matrix can be bounded by the largest absolute eigenvalue, and the eigenvalues of $A^{-1}$ are the inverses of the eigenvalues of $A$) are brought together to bound the entries of $\Sigma_{T}^{-1}$:
    \begin{align}
        \alpha_{jl} &\leq \frac{1}{\lambda_-} \leq \frac{1}{(1 - k\rho)\sigma^2} \label{eq:gaussian_bound_ev_by_gregoshin}
    \end{align}
    
    With the bounds for the entries of $\Sigma_{U;T}$ and $\Sigma_{T}^{-1}$ in hand, we can find a general upper bound for the adversary-specific target dependent BDPL for any $x_i, x_i' \in \mathbb{R}$.

    \begin{align}
        \mathrm{BDPL}_{(H,i)}(x_i, x_i') &\leq (\|\Sigma_{U;T} \Sigma_{T}^{-1} \mathbf{e}_{k+1}\|_1 + 1)|x'_i-x_i|\varepsilon\label{eq:gaussian_apply_theorem} \\
        &= (\|\Sigma_{U;T} (\alpha_{1, k+1}, \dots, \alpha_{k+1, k+1})^\top\|_1 + 1)|x'_i-x_i|\varepsilon\label{eq:gaussian_multiply_matrix_vector} \\
        &= \left(\|(\sum_{l=1}^{k+1} \alpha_{l, k+1} \, b_{1, l}, \dots, \sum_{l=1}^{k+1} \alpha_{l, k+1} \, b_{u, l})^\top\|_1 + 1\right)|x'_i-x_i|\varepsilon\label{eq:gaussian_multiply_matrix_result} \\
        &= \left(\sum_{j = 1}^u | \sum_{l=1}^{k+1} \alpha_{l, k+1} \, b_{j, l} | + 1\right)|x'_i-x_i|\varepsilon\label{eq:gaussian_rewrite_L1} \\
        &\leq \left(\sum_{j = 1}^u | \sum_{l=1}^{k+1} \frac{\rho \sigma^2}{(1 - k\rho)\sigma^2} | + 1\right)|x'_i-x_i|\varepsilon\label{eq:gaussian_apply_entry_bounds} \\
        &= \left(\frac{u(k+1)\rho}{1 - k\rho} + 1\right)|x'_i-x_i|\varepsilon\label{eq:gaussian_replace_sum_with_cardinality} \\
        &= \left(\frac{u(k+1)}{\frac{1}{\rho} - k} + 1\right)|x'_i-x_i|\varepsilon\\
        &\leq \left(\frac{(\frac{m}{2})^2}{\frac{1}{\rho} - m + 2} + 1\right)|x'_i-x_i|\varepsilon\label{eq:gaussian_final_bound} \\
        &= \left(\frac{m^2}{4(\frac{1}{\rho} - m + 2)} + 1\right)|x'_i-x_i|\varepsilon
    \end{align}

    We first use the inequality from \Cref{thm:gaussian_correlation_bdpl} in \cref{eq:gaussian_apply_theorem}. Then, we multiply $\Sigma_{T}^{-1}$ and $\mathbf{e}_{k+1}$; only the last column of $\Sigma_{T}^{-1}$ remains in \cref{eq:gaussian_multiply_matrix_vector}. The result is multiplied with $\Sigma_{U;T}$ in \cref{eq:gaussian_multiply_matrix_result}. Afterwards, in \cref{eq:gaussian_rewrite_L1} we apply the $\ell_1$-distance to the remaining vector. The bounds for the entries $\alpha$ and $b$ are used in \cref{eq:gaussian_apply_entry_bounds}. Keep in mind that the denominator is always positive because $k \leq m - 2$ and $\rho < \frac{1}{m-2}$. Then the two sums can be simplified by multiplying by their cardinality in \cref{eq:gaussian_replace_sum_with_cardinality} since the entries of the sum no longer depend on $j$ or $l$. Finally, we bound the numerator and denominator in \cref{eq:gaussian_final_bound}: The numerator $u(k + 1)$ is smaller or equal to $(m/2)^2$ because $u$ and $k + 1$ are positive and together form $m = u + k + 1$. Thus, their product is maximal if they meet at the exact midpoint to $m$. As mentioned previously, the denominator is always positive. It therefore becomes minimal (and the entire expression maximal) if $k$ becomes maximal. This is the case for $k = m - 2$.

    In both cases we were able to bound adversary-specific target dependent BDPL as required. Thus, the proof is complete.
\end{proof}
\Cref{thm:gaussian_d-privacy_bdpl_bound} provides a concrete formula for the increase in privacy leakage due to linear correlation relative to independent data. Higher Pearson coefficients lead to greater leakage. Additionally, we can extend this result to derive a relation between DP and BDP.
\subsection{Relationship between DP and BDP}\label{sec:gaus_relation}
\noindent
 Observe that any $d$-private mechanism is an $\varepsilon$-DP mechanism with $\varepsilon=\sup_{D\sim D'}d(D,D')$. Moreover, any Bayesian $d$-private mechanism is an $\varepsilon$-BDP mechanism with $\varepsilon= \sup_{x,x'}d(x,x')$. By leveraging these relationships between privacy notions we can establish a connection between DP and BDP. However, since this supremum may be unbounded, it can lead to undesirable privacy guarantees. To manage this relationship effectively we apply clipping techniques, resulting in~\Cref{th:gaussian_dp_bdpl_bound}, which enables the construction of BDP mechanisms from DP mechanisms. Formally, clipping is defined as:

\begin{definition}\label{def:gaussian_clipping_and_dp_alg}
    For any interval $I = [a, b] \subset \mathbb{R}$, we define the \textit{clipping function} $c_{I} : \mathbb{R}^n \to \mathbb{R}^n$, which, for all $D \in \mathbb{R}^n$ and all $i \in [n]$, outputs  \[
c_{I}(D)_i = \max(a, \min(b, D_i)).
\]
Let $\mathcal{M} : \mathbb{R}^n \to \mathbb{R}$ be a mechanism. We define its \textit{clipped version} $\mathcal{M}_{I} : \mathbb{R}^n \to \mathbb{R}$ as  $
\mathcal{M}_{I} = \mathcal{M} \circ c_{I}.
$

\end{definition}
Due to the data domain reduction, we can bound the DP leakage of $\varepsilon\ell_1$-private mechanisms.

\begin{lemma}\label{lem:clipped_algo_DP}
    If $\mathcal{M} : \mathbb{R}^n \to \mathbb{R}$ is $\varepsilon \ell_1$-private, then its clipped version $\mathcal{M}_{I}$ is $\varepsilon \ell_1$-private and $(M\varepsilon)$-DP with $M=|b-a|$.
\end{lemma}
\begin{proof}
    We begin by showing that $\mathcal{M}_{I}$ is $\varepsilon \ell_1$-private. Let $D_1, D_2 \in \mathbb{R}^n$ be arbitrary and $S \subseteq \mathbb{R}$ be any measurable set. We have
    \begin{align}
        \Pr[\mathcal{M}_{I}(D_1) \in S] &= \Pr[\mathcal{M}(c_{I}(D_1)) \in S] \\
        &\leq e^{\varepsilon \|c_{I}(D_1) - c_{I}(D_2)\|_1} \Pr[\mathcal{M}(c_{I}(D_2)) \in S] \label{eq:gaussian_use_l1_privacy} \\
        &\leq e^{\varepsilon \|D_1 - D_2\|_1} \Pr[\mathcal{M}(c_{I}(D_2)) \in S] \label{eq:gaussian_l1_clipped} \\
        &= e^{\varepsilon \|D_1 - D_2\|_1} \Pr[\mathcal{M}_{I}(D_2) \in S] \label{eq:gaussian_l1_private_end}.
    \end{align}
    In \cref{eq:gaussian_use_l1_privacy} we use that $\mathcal{M}$ is $\varepsilon \ell_1$-private. Then, we apply the fact that the $\ell_1$-distance of two clipped data sets is smaller or equal to the $\ell_1$-distance of the original data sets in \cref{eq:gaussian_l1_clipped}. Finally, \cref{eq:gaussian_l1_private_end} shows that $\mathcal{M}_{I}$ is $\varepsilon \ell_1$-private.

    Now, we will show that $\mathcal{M}_{I}$ is also DP. Let $D, D' \in \mathbb{R}^n$ be arbitrary neighboring data sets, i.e., there only exists a single index $i \in [n]$ with $D_i \neq D_i'$. For any measurable set $S \subseteq \mathcal{Y}$ we have
    \begin{align}
        \Pr[\mathcal{M}_{I}(D) \in S] &= \Pr[\mathcal{M}(c_{I}(D)) \in S] \\
        &\leq e^{\varepsilon \|c_{I}(D) - c_{I}(D')\|_1} \Pr[\mathcal{M}(c_{I}(D')) \in S] \label{eq:gaussian_use_l1_privacy_again} \\
        &= e^{\varepsilon \sum_{j \in [n]}|\max(a, \min(b, D_j)) - \max(a, \min(b, D_j'))|} \Pr[\mathcal{M}(c_{I}(D')) \in S] \label{eq:gaussian_l1_dist_and_clip} \\
        &= e^{\varepsilon |\max(a, \min(b, D_i)) - \max(a, \min(b, D_i'))|} \Pr[\mathcal{M}(c_{I}(D')) \in S] \label{eq:gaussian_only_one_diff} \\
        &\leq e^{\varepsilon (b - a)} \Pr[\mathcal{M}(c_{I}(D')) \in S] = e^{\varepsilon (b - a)} \Pr[\mathcal{M}_{I}(D') \in S].
    \end{align}
    Once again, we use that $\mathcal{M}$ is $\varepsilon \ell_1$-private in \cref{eq:gaussian_use_l1_privacy_again}. We expand the definition of the $\ell_1$-distance and of the clipping function $c_{I}$ in \cref{eq:gaussian_l1_dist_and_clip}. Then, we use for \cref{eq:gaussian_only_one_diff} that $D$ and $D'$ only differ for index $i$. Finally, we can bound the difference between the two entries because they are clipped to the interval $[a, b]$. We have shown that $\mathcal{M}_{I}$ is $(b - a)\varepsilon$-DP.
\end{proof}

With \Cref{lem:clipped_algo_DP} and \Cref{thm:gaussian_d-privacy_bdpl_bound}, we can directly show that this class of DP-mechanisms has a limited BDPL. %Note that we are now directly establishing a relationship to BDP, not to $d$-BDP. \Cref{rmk:gaussian_dp_bdpl_bound} is the Gaussian bound.

\begin{theorem}[The Gaussian Bound]\label{th:gaussian_dp_bdpl_bound}
    Let $\mathcal{M}_{I}$ with data domain $\mathbb{R}^{n}$ be the clipped version of an $(\varepsilon\ell_1)$-private mechanism $\mathcal{M} $ where $\varepsilon > 0$ and with input data drawn from a multivariate Gaussian distribution $\mathcal{N}(\mu, \Sigma_\rho)$ with mean $\mu \in \mathbb{R}^n$, maximum of  $m\leq n$ correlated variables and limited covariance matrix $\Sigma_\rho \in \mathbb{R}^{n \times n}$ (Def.~\ref{def:gaussian_limited_covariance_matrix}) such that $\rho(m-2) < 1$ is the maximum correlation coefficient.
    Then, the clipped mechanism $\mathcal{M}_{I}$ is
    \begin{equation*}
        \left(\frac{m^2}{4(\frac{1}{\rho} - m + 2)} + 1\right) M \varepsilon \textrm{-BDP.}
    \end{equation*}
    where $M$ is the diameter of the interval $I$.
\end{theorem}
\begin{proof}
    We know that $\mathcal{M}_{I}$ is a $\varepsilon \ell_1$-private query because of \Cref{lem:clipped_algo_DP}. Thus, we can apply \Cref{thm:gaussian_d-privacy_bdpl_bound} to find a universal bound of the target dependent BDPL in this situation. Therefore, the idea is to show how the BDPL is bounded by the target dependent BDPL, and to then apply \Cref{thm:gaussian_d-privacy_bdpl_bound}.
    \begin{align}
        \mathrm{BDPL} &\coloneqq \sup_{K, i} \mathrm{BDPL}_{(K, i)} \label{eq:gaussian_expand_BDPL_1} \\
        &= \sup_{K, i} \left( \sup_{\mathbf{x}_K, s, |x_i - x_i'| \leq M} \ln \frac{p_Y(s \mid \mathbf{X}_K = \mathbf{x}_K, X_i = x_i)}{p_Y(s \mid \mathbf{X}_K = \mathbf{x}_K, X_i = x_i')} \right) \label{eq:gaussian_expand_BDPL_2} \\
        &= \sup_{K, i} \left( \sup_{|x_i - x_i'| \leq M} \mathrm{BDPL}_{(K, i)}(x_i, x_i') \right) \label{eq:gaussian_bdpl_to_target_dependent} \\
        &= \sup_{|x_i - x_i'| \leq M} \left( \sup_{K, i}\mathrm{BDPL}_{(K, i)}(x_i, x_i') \right) \label{eq:gaussian_switch_supremums} \\
        &= \sup_{|x_i - x_i'| \leq M}  \mathrm{BDPL}(x_i, x_i') \\
        &\leq \sup_{|x_i - x_i'| \leq M} \left(\frac{m^2}{4(\frac{1}{\rho} - m + 2)} + 1\right) |x'_i-x_i|\varepsilon\label{eq:gaussian_use_bound_from_cor} \\
        &= \sup_{|x_i - x_i'| \leq M} \left(\frac{m^2}{4(\frac{1}{\rho} - m + 2)} + 1\right) |x_i - x_i'| \varepsilon \\
        &= \left(\frac{m^2}{4(\frac{1}{\rho} - m + 2)} + 1\right) M \varepsilon.
    \end{align}
    In \cref{eq:gaussian_expand_BDPL_1,eq:gaussian_expand_BDPL_2} we expand the definition of BDPL. Then, in \cref{eq:gaussian_bdpl_to_target_dependent} we use that the adversary-specific \textit{target dependent} BDPL is defined equivalently, except it does not take the supremum over $x_i, x_i'$. We switch the order of the suprema in \cref{eq:gaussian_switch_supremums} to subsequently plug in the definition of the general target dependent BDPL. Then, we use \Cref{thm:gaussian_d-privacy_bdpl_bound} to derive \cref{eq:gaussian_use_bound_from_cor}. Finally, the last supremum can be resolved by writing out $d(x_i, x_i')$ and bounding it with $M \varepsilon$. It follows that clipped mechanism $\mathcal{M}$ is BDP since the BDPL is limited.
\end{proof}

\Cref{th:gaussian_dp_bdpl_bound} allows us to systematically build a BDP mechanism by recalibrating the noise of a DP mechanism when $\rho(m-2)<1$. For instance, given the clipped Laplace mechanism $\mathcal{M}_{I}$ that adds noise to a data point $x\in\mathbb{R}$ following $\mathrm{Lap}(\frac{M}{\tau})$, where
\begin{equation}\label{eq:gaussian_tao}
\tau=\varepsilon\frac{4(\frac{1}{\rho} - m + 2)}{m^2+4(\frac{1}{\rho} - m + 2)}
\end{equation}
 we obtain an $\varepsilon$-BDP mechanism. Moreover,
\begin{equation}\label{rmk:gaus_impr_general}
        \begin{aligned}
        && \frac{m^2}{4(\frac{1}{\rho} - m + 2)} + 1 &\leq m %\\
        \text{ if and only if } \rho \leq \frac{m - 1}{\frac{5}{4}m^2 - 3m + 2}. \\
        \end{aligned}
\end{equation}

\begin{figure}
    \centering
    \includegraphics[width=0.6\textwidth]{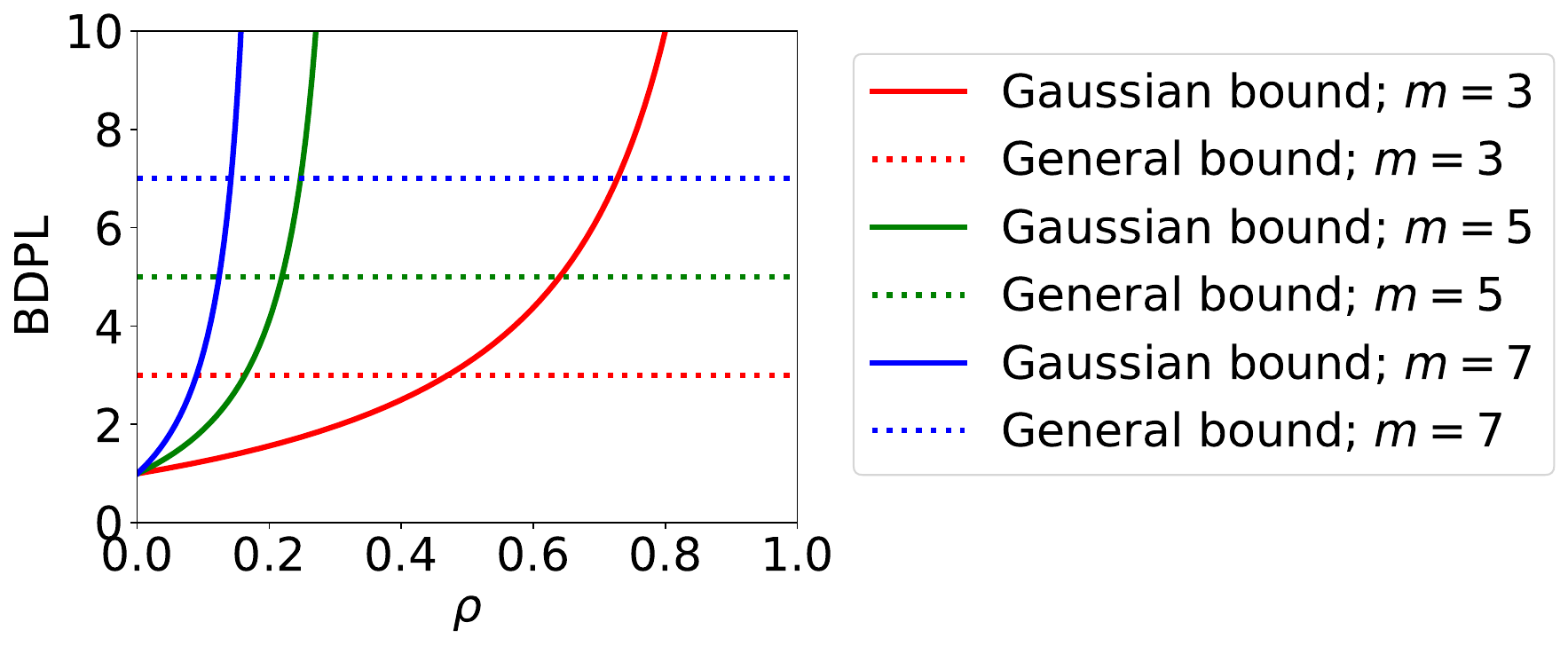}
    \caption[Bound comparison]{Gaussian-specific bound compared to the general bound, which coincides with the s-o-t-a one~\cite{Yang_2015_BDP}.}
    \label{fig:gaus_bound}
\end{figure}

Hence, the Gaussian bound improves on the general bound if $\rho$ is on the order of $\rho \approx \frac{1}{m}$ (See~\Cref{fig:gaus_bound}). 
The higher the number of correlated records $m$, the better the relative improvement of the Gaussian specific bound compared to the general bound for small correlation.
Importantly,~\citeauthor{Yang_2015_BDP}~\cite{Yang_2015_BDP} establish a bound for Gaussian Markov random fields. They establish that a clipped $\M_{\varepsilon,f}$ satisfies $(nM\varepsilon)$-BDP, which coincides with the general bound when all records are correlated. \Cref{th:gaussian_dp_bdpl_bound} applies to this particular case since a Gaussian Markov random field is an example of Gaussian Multivariate distribution. Moreover, our bound improves over theirs in the same cases it improves over the general bound.

\subsection{Accuracy}
\noindent
%The Gaussian bound on the BDPL of a DP mechanism allows us to use $q(\varepsilon)$-DP mechanisms as $\varepsilon$-BDP mechanisms when we know that the data follows a multivariate Gaussian distribution. Notably, 
When the Pearson correlation is bounded as specified in~\Cref{rmk:gaus_impr_general}, it is guaranteed that a larger $\varepsilon'$ than $\frac{\varepsilon}{m}$ is sufficient to guarantee $\varepsilon$-BDP via an $\varepsilon'$-DP mechanism. 
Since a larger privacy budget generally correlates with improved utility, we can therefore anticipate enhanced utility results. In particular, we express the accuracy improvement of the Laplace mechanism when it is calibrated to protect data drawn from a multivariate Gaussian distribution. As a consequence of our \Cref{th:gaussian_dp_bdpl_bound} and \Cref{thm:laplace_mechanism_accuracy} from~\cite{dwork2014algorithmic} we obtain the following result:

\begin{corollary}\label{cor:gaussian_accuracy_laplace}
    Let $\mathcal{M}_{\varepsilon, f_{I}}$ be the clipped Laplace $\varepsilon$-DP mechanism that approximates the query $f_{I}$ as defined in~\ref{def:gaussian_clipping_and_dp_alg} with input data drawn from a multivariate Gaussian distribution $\mathcal{N}(\mu, \Sigma_\rho)$ with mean $\mu \in \mathbb{R}^n$ and limited covariance $\Sigma_\rho \in \mathbb{R}^{n \times n}$ with a maximum number of correlated variables $m\leq n$ such that $\rho(m-2)<1$.
    Then, if the Laplace mechanism $\mathcal{M}_{\varepsilon, f_{I}}$ is $(\alpha, \beta)$-accurate w.r.t. $f_{I}$, there exists an $\varepsilon$-BDP mechanism $\mathcal{B}$ whose input is drawn from $\pi$ and that is $(h \alpha, \beta)$-accurate w.r.t. $f_{I}$ with 
    \[
    h=\frac{m^2}{4(\frac{1}{\rho}-m+2)}+1.
    \]
\end{corollary}
\begin{proof}
    The idea of this proof is to construct mechanism $\mathcal{B}$ with the Laplace mechanism as well, but to choose a carefully selected privacy leakage $\varepsilon' < \varepsilon$ so that mechanism $\mathcal{B}$ is (1) $\varepsilon$-BDP and (2) $(h \alpha, \beta)$-accurate. 

    First, we determine the accuracy of mechanism $\mathcal{M}_{\varepsilon, f_{I}}$. According to \Cref{thm:laplace_mechanism_accuracy}, the $(\alpha, \beta)$-accuracy of the Laplace mechanism for a given probability $\beta \in (0, 1]$ and privacy parameter $\varepsilon$ is 
    \begin{equation}
        \alpha = \ln \left(\frac{1}{\beta}\right) \, \frac{\Delta f_{I}}{\varepsilon} .
    \end{equation} 
    So this is the $(\alpha, \beta)$-accuracy of $\mathcal{M}_{\varepsilon, f_{I}}$.
    We have to show that there exists an $\varepsilon$-BDP mechanism $\mathcal{B}$ which is $(h \alpha, \beta)$-accurate.
    We choose $\mathcal{B}$ as the Laplace mechanism applied to $f_{I}$ with an adjusted privacy parameter $\varepsilon' > 0$. Thus, $\mathcal{B}$ will be $\varepsilon'$-DP. Observe that $\mathcal{B}$ is $d$-private with $d(D, D') = \frac{\varepsilon'}{M}\|D - D'\|_1$. Therefore, we can use \Cref{th:gaussian_dp_bdpl_bound} to show that $\mathcal{B}$ is BDP. We must choose $\varepsilon'$ in a way that ensures that the BDPL is limited to $\varepsilon$, so that we have $\varepsilon$-BDP.
    With \Cref{th:gaussian_dp_bdpl_bound}, $\mathcal{B}$ is
    \begin{equation}\label{eq:gaussian_bdpl_goal}
        \left(\frac{m^2}{4(\frac{1}{\rho} - m + 2)} + 1 \right) \varepsilon'\textrm{-BDP.}
    \end{equation}
    Therefore, to achieve $\varepsilon$-BDP, we must have
    \begin{equation}
        \left(\frac{m^2}{4(\frac{1}{\rho} - m + 2)} + 1\right) \varepsilon' = \varepsilon \quad \Leftrightarrow \quad \varepsilon' = \varepsilon \, \left(\frac{m^2}{4(\frac{1}{\rho} - m + 2)} + 1\right)^{-1}.
    \end{equation}
    Now, we can calculate the accuracy of $\mathcal{B}$ because it also uses the Laplace mechanism. Then, we find an upper bound for this accuracy.
    Mechanism $\mathcal{B}$ is $(\alpha', \beta)$-accurate, with
    \begin{align}
        \alpha' &= \ln (\frac{1}{\beta}) \, \frac{\Delta f_{I}}{\varepsilon'} = \ln (\frac{1}{\beta}) \, \frac{\Delta f_{I}}{\varepsilon} \, (\frac{m^2}{4(\frac{1}{\rho} - m + 2)} + 1) \\
        &= \alpha \left(\frac{m^2}{4(\frac{1}{\rho} - m + 2)} + 1\right) \\
        &= \alpha h
    \end{align}
    Finally, we find that $\mathcal{B}$ is $(h \alpha, \beta)$-accurate.
\end{proof}

\begin{figure}
    \centering
    \includegraphics[width=0.5\textwidth]{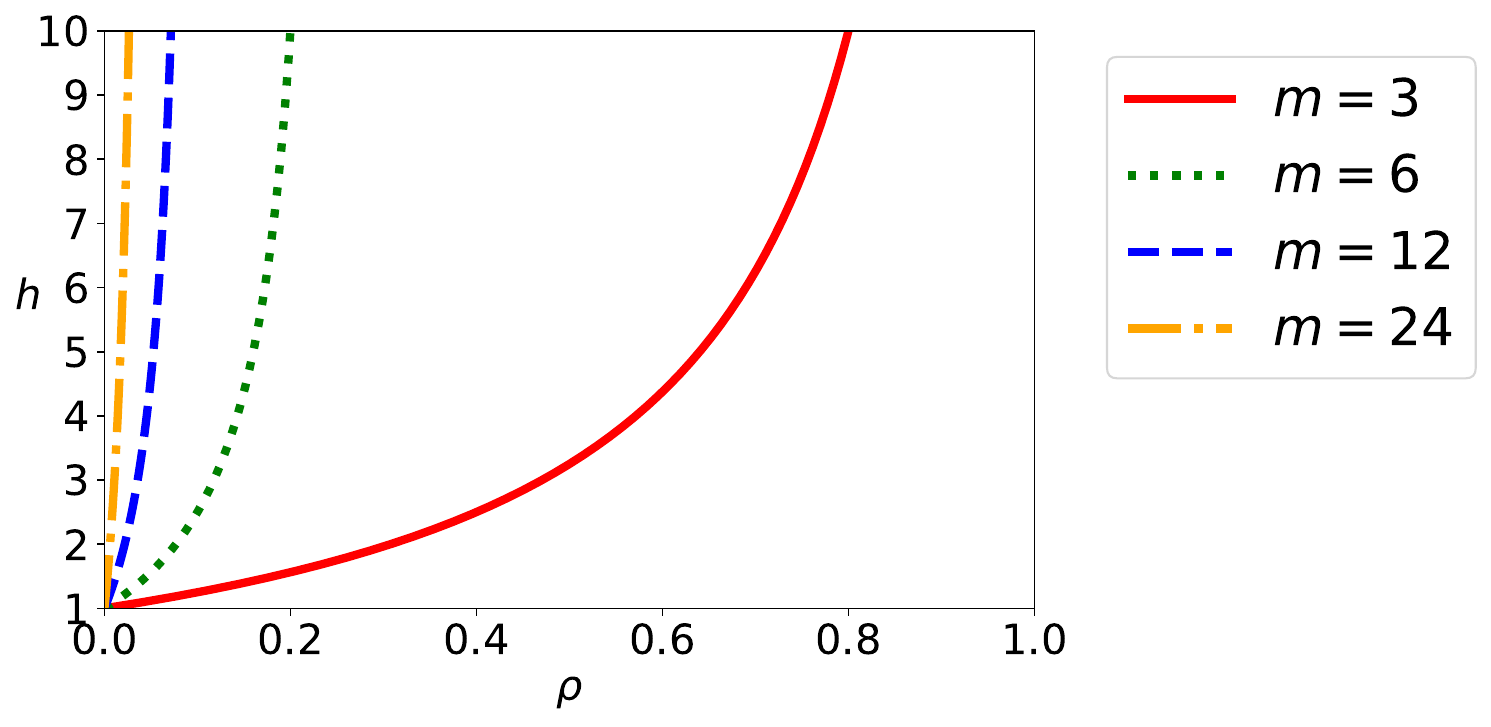}
    \caption[Relative Accuracy $h$ of an $\varepsilon$-BDP mechanism to an $\varepsilon$-DP mechanism for Gaussian Data]{Relative accuracy of an $\varepsilon$-BDP mechanism to an $\varepsilon$-DP mechanism for a Multivariate Gaussian distribution.}
    \label{fig:lap_gaus_accu}
\end{figure}

The statement of \Cref{cor:gaussian_accuracy_laplace} is visualized in \Cref{fig:lap_gaus_accu}. This figure shows that in order to provide similar utility to DP, $\rho$ must be small. The larger the number of correlated records $m$, the smaller $\rho$ has to be to provide similar utility.
The results in this section enable the protection of weakly correlated data drawn from a multivariate Gaussian distribution. Furthermore, a comparison of the accuracy achieved by our method versus the state-of-the-art bound from~\cite{Yang_2015_BDP} and the general BDP bound is presented in~\Cref{fig:results_gaussian}, demonstrating a consistent improvement enabled by our approach.

\section{Markov Chain Correlation Model}\label{sec:markov}
\noindent
In streaming processes or time series data, states at successive time steps are often  correlated, meaning that the state at a given time step depends on the state at the previous one. For example, a user's location at time step $t$ is correlated with their location at $t-1$. This dependency pattern is commonly modeled using Markov chains~\cite{Behrends_2000_Markov}.

Consequently, in this section we investigate the impact of correlations following a Markov model  on the privacy leakage and
utility of BDP mechanisms. Particularly, we prove \Cref{cor:markov_bdpl_bound}, a new bound on the BDPL of any $\varepsilon$-DP mechanism when data is correlated corresponding to a Markov chain.  %We call it the Markov chain bound. 
Additionally, we use our results to elaborate on the utility gain compared to protecting against arbitrary correlation. 

For the remainder of this work, we adopt the definition of a Markov chain from~\cite{Behrends_2000_Markov}, which specifically refers to finite, time-homogeneous Markov chains, i.e., those with finite state spaces and time-invariant transition probabilities. Formally,

\begin{definition}[Markov Chain~\cite{Behrends_2000_Markov}]\label{def:markov_correlation}
    Let $\mathcal{S}$ be a finite set of possible states of size $s\in\N$  and let $\mathbf{X} = (X_1, \dots, X_n)$ be a random vector.
    We say $\mathbf{X}$ is a \textit{Markov chain} with transition matrix $P \in \mathbb{R}^{s \times s}$ and initial distribution $w \in \mathbb{R}^s$ if all of the following holds.
    \begin{enumerate}
        \item For all states $x, y \in\mathcal{S}$ and all indices $i \in [n-1]$ we have $\Pr[X_{i+1} = x | X_i = y] = P_{y,x}$.
        \item For all states $x \in\mathcal{S}$ we have $\Pr[X_1 = x] = w_x$.
        \item The Markov property: 
        For all indices $i \in [n-1]$ and for all states $x_1, \dots, x_i, x_{i+1} \in \mathcal{S}$ we have 
        \begin{align*}
            \Pr&[X_{i+1} = x_{i+1} \mid X_1 = x_1, \dots, X_i = x_i] \\
            &= \Pr[X_{i+1} = x_{i+1} \mid X_i = x_i].
        \end{align*}
    \end{enumerate}
\end{definition}
Note that the Markov property from ~\Cref{def:markov_correlation} holds not only when the full history is known, but also when only the partial
history is known as we show in the following remark.
\begin{remark}\label{rm:gen_mark}
Given an index \( i \in [n-1] \), and a set \( A \subseteq \{1, \dots, i-1\} \) containing only indices smaller than \( i \).  Then, for any states \( x, y \in \mathcal{S} \) and any state tuple \( \mathbf{x}_A \in \mathcal{S}^{|A|} \), we have
    \begin{align*}
        \Pr[& x_{i+1} \mid x_i, \mathbf{x}_A]= \sum_{\mathbf{x}_B \in \S^{b}} \Pr[x_{i+1} |x_i, \mathbf{x}_A,\mathbf{x}_B] \, \Pr[\mathbf{x}_B | x_i, \mathbf{x}_A]  \\
        \overset{(*)}{=}& \sum_{\mathbf{x}_B \in \S^{b}} \Pr[x_{i+1} \mid x_i] \, \Pr[ \mathbf{x}_B \mid x_i,  \mathbf{x}_A]
        = \Pr[x_{i+1} \mid x_i] \sum_{\mathbf{x}_B \in \S^{b}} \Pr[\mathbf{x}_B \mid x_i ,  \mathbf{x}_A] = \Pr[x_{i+1}\mid x_i],
    \end{align*}
   where \( B = [n] \setminus (A \cup \{i\}) \) is the set of remaining indices, and \( b = |B| \). We use the law of total probability to introduce all remaining indices in \( B \). Then, in \((*)\), we apply the Markov property. Finally, \( \Pr[x_{i+1} \mid x_i] \) can be factored out of the sum because it no longer depends on \( \mathbf{x}_B \). The remaining sum adds up to \( 1 \), so we are left with only the condition of the direct predecessor.

\end{remark}

\subsection{Relationship between DP and BDP}
\noindent
In this subsection, we show that it is possible to obtain a bound on the BDPL of any DP mechanism based on the maximum ratio between the largest and smallest transition probabilities in the Markov chain. The intuition is that if all transition probabilities are similar, changing the random variable $X_i$ from state $x_i$ to state $x'_i$ will have minimal impact on the subsequent time steps of the Markov chain. However, if the transition probabilities differ significantly, this change could have a large effect over many time steps. 
To prove our main result~\Cref{cor:markov_bdpl_bound} we need the auxiliary~\Cref{lm:markov_generalized,lm:key_markov}.

\begin{lemma}[Generalized Markov Property]\label{lm:markov_generalized}
   Given $\mathbf{X}$ a Markov chain, for all sets of indices $A\in[n] \setminus \{i\}$:
\begin{equation*}
    \Pr[ x_i \mid \mathbf{x}_{A} ]= \Pr[ x_i \mid  x_{\ell},x_{r}]
\end{equation*}
where $\ell,r$ are the nearest indices to $i$ in $A$ both left and right., i.e., $\ell,r\in A$ with $\ell<i$ and $i <r$ so that for all indices $j\in A$ we have  $j<\ell$ or $r<j$. Notably, if all indices in $A$ are smaller than $i$ we only need to consider $\ell$ and if all are above we only need to consider $r$.
\end{lemma}
\begin{proof}
    We derive this statement directly from probability rules and the Markov property.  Let $i,j \in [n]$ be indices and $A' \subseteq [n] \setminus \{i,j\}$ be a set of indices so that there exists an index $\ell \in A'$, $|A'|=a$ that is ``in between'' $i$ and $j$, i.e., we have $i > \ell > j$ or $i < \ell < j$. 
    If for all states $x_i, x_j \in \S$ and for all state tuples $\mathbf{x}_{A'} \in \S^{a}$ we have
    \begin{equation}\label{eq:markov_remove_condition}
        \Pr[ x_i \mid \mathbf{x}_{A'},\, x_j] = \Pr[ x_i \mid \mathbf{x}_{A'}],
    \end{equation}
    then it follows
    \begin{align}
        \Pr[x_i \mid \mathbf{x}_{A} ] &=\Pr[x_i \mid x_{i_1},\dots x_{\ell},x_{r},\dots,x_{i_a} ]\\
        &=\Pr[x_i \mid x_{i_1},\dots x_{\ell},x_{r} ] \label{eq:gen_markov_prop_right}\\
        &=\Pr[x_i \mid x_{\ell},x_{r} ]. \label{eq:gen_markov_prop_left}
    \end{align}
    Where \cref{eq:gen_markov_prop_right} holds because for all $i_{j}>r$, there exists $r\in A'$ such that $i<r<i_{j}$; therefore we can apply \Cref{eq:markov_remove_condition}. Analogously \cref{eq:gen_markov_prop_left} holds because for all $i_{j}<\ell$, there exists $\ell\in\{\ell,r\}$ such that  $i_{j}<\ell<i$.

    Consequently, for the rest of the proof we focus on proving~\Cref{eq:markov_remove_condition}.
    
    We proceed separately for the case in which the ``irrelevant'' index $j$ is smaller or left, i.e., there exists $\ell\in A$ such that $j<\ell<i$ and for the case in which $j$ is above or right, i.e., exist $r\in A$ such that $i<r<j$ . 
    
    \textbf{Case 1:} First, we show that \Cref{eq:markov_remove_condition} holds if $A_1\subseteq\{1,\dots,i-1\}$ and $\ell\in A_1$ such that $j<\ell$ where $\ell \in A_1$ and $j<\ell<i$ so that no other index in $A_1$ lies between $i$ and $l$.
    This means that the set of indices between $\ell$ and $i$---defined as $B = \{\ell+1, \dots, i-1\}$---is disjoint with $A_1$.
    
    If $B$ is empty, then \cref{eq:markov_remove_condition} follows immediately from~\Cref{rm:gen_mark} because index $\ell$ is the direct predecessor of index $i$, i.e., $\ell=i-1$. 
    
    If $B \neq \varnothing$, using the law of total probability, we have
    \begin{align}
        \Pr[&x_i | \mathbf{x}_{A_1},x_j]
        = \sum_{\mathbf{x}_B \in \S^{b}} \Pr[x_i | \mathbf{x}_{A_1},x_j, \mathbf{x}_B] \, \Pr[\mathbf{x}_B | \mathbf{x}_{A_1}, x_j] \\
        &=\sum_{\mathbf{x}_B \in \S^{b}} \Pr[ x_i |\mathbf{x}_{A_1},\mathbf{x}_B] \, \Pr[ \mathbf{x}_B | \mathbf{x}_{A_1}] =\Pr[ x_i |  \mathbf{x}_{A_1}],\label{eq:markov_use_conditional_independence_and_prev}
    \end{align}
    where \Cref{eq:markov_use_conditional_independence_and_prev} follows by applying~\Cref{rm:gen_mark}, since
    
    \begin{align}
        \Pr[&\mathbf{x}_B |  \mathbf{x}_{A_1},x_j]=\Pr[x_{\ell+1}, \dots, x_{i-1} | \mathbf{x}_{A_1}, x_j] \label{eq:markov_write_out_components_X_M} \\
        = &\Pr[ x_{\ell+1}, \dots,x_{i-2} |  \mathbf{x}_{A_1},x_j] \, \Pr[ x_{i-1} | \mathbf{x}_{A_1}, x_j,  x_{\ell+1},\dots, x_{i-2}] \label{eq:markov_pull_apart_and} \\
        = &\dots \notag\\
        = &\Pr[x_{\ell+1} | \mathbf{x}_{A_1},  x_j] \, \prod_{k=l+2}^{i-1} \Pr[x_k | \mathbf{x}_{A_1},x_j, x_{\ell+1}, \dots, x_{k-1}] \label{eq:markov_repeat_steps} \\
        = &\Pr[x_{\ell+1} | \mathbf{x}_{A_1}] \, \prod_{k=l+2}^{i-1} \Pr[x_k |\mathbf{x}_{A_1}, x_{\ell+1}, \dots,x_{k-1}] \label{eq:markov_l_is_in_I} \\
        = &\Pr[\mathbf{x}_B |  \mathbf{x}_{A_1}] \label{eq:markov_reverse_pull_apart}.
    \end{align}

In \cref{eq:markov_pull_apart_and}, we rewrite the joint probability of $\mathbf{X}_B$ as the two parts $X_{i-1}$ and $\mathbf{X}_{B\setminus\{i-1\}}$. This uses the fact that a joint probability can be rewritten using $\Pr[A \cap B \mid C] = \Pr[A \mid C] \, \Pr[B \mid C, A]$.
Here, event $A$ corresponds to conditions $X_{\ell+1} = x_{\ell+1}, \dots, X_{i-2} = x_{i-2}$, event $B$ to condition $X_{i-1} = x_{i-1}$ and event $C$ to conditions $\mathbf{X}_{A_1} = \mathbf{x}_{A_1}, X_j = x_j$.
This step is repeatedly used to fully split $\mathbf{X}_B$ into its components and derive \cref{eq:markov_repeat_steps}.
Then, we use the~\Cref{rm:gen_mark} for \cref{eq:markov_l_is_in_I}: Random variable $X_{\ell+1}$ is conditionally independent of $X_j$ given the direct predecessor $X_{\ell}$ with index $\ell \in A$. Similarly, random variable $X_k$ is conditionally independent of $X_j$, given the direct predecessor $X_{k-1}$.

Note that, this result can be extended by induction to say that given any set of indices $C\subseteq[\ell+1,\dots,n]$, if $A_1\subseteq\{1,\dots,i-1\}$, $\ell$ the biggest index $A_1$ and $j<\ell$, then
\begin{align}
        \Pr[x_c | \mathbf{x}_{A_1},x_j] = \Pr[ x_c |  \mathbf{x}_{A_1}]\label{eq:extended_to_vector}
\end{align}

For $|C|=1$,  we have just proven \Cref{eq:extended_to_vector}. Now we assume it true for all $|C|\leq n-1$ and we prove it for $|C|=n$:
\begin{align*}
     \Pr[x_c |\mathbf{x}_{A_1}, x_j]=&\Pr[x_{c_1},\dots,x_{c_{n}}\mid \mathbf{x}_{A_1}, x_j]\\
     =&\Pr[x_{c_n}| \mathbf{x}_{A_1},x_j,\mathbf{x}_{C\setminus\{c_{n}\}}]
     \Pr[\mathbf{x}_{C\setminus\{c_{n}\}}\mid\mathbf{x}_{A_1},  x_j]\\
     =&\Pr[x_{c_n}| \mathbf{x}_{A_1},\mathbf{x}_{C\setminus\{c_{n}\}}]
     \Pr[\mathbf{x}_{C\setminus\{c_{n}\}}\mid \mathbf{x}_{A_1}]\\
     =&\Pr[ x_c |  \mathbf{x}_{A_1}]
\end{align*}
where the last equality follows directly from the induction hypothesis since $|C\setminus\{c_{n}\}|=n-1$.

Now, we derive that~\Cref{eq:markov_remove_condition} also holds for an arbitrary set of indices $A$, not necessarily all smaller than $i$, i.e., $A\subseteq[n]\setminus\{i\}$. We can partition $A$ into the indices before $i$ and after $i$, i.e., $A=A_1\cup A_2$  where $A_1=\{i_{j}\in A\colon i_{j}<i\}$ and $A_2=\{i_{j}\in A\colon i_{j}>i\}$. Then, we have
\begin{align}
        \Pr[ x_i | \mathbf{x}_A,x_j]=&\Pr[ x_i | \mathbf{x}_{A_1}, \mathbf{x}_{A_2}, x_j]
        \coloneqq\frac{\Pr[ x_i,\mathbf{x}_{A_1}, \mathbf{x}_{A_2},x_j]}{\Pr[\mathbf{x}_{A_1}, \mathbf{x}_{A_2},x_j]}\\
        =&\frac{\Pr[ x_i,\mathbf{x}_{A_1}, x_j]\Pr[\mathbf{x}_{A_2}\mid x_i,\mathbf{x}_{A_1},x_j]}{
         \Pr[\mathbf{x}_{A_1}, x_j]\Pr[ \mathbf{x}_{A_2}\mid \mathbf{x}_{A_1}, x_j]}\\
        =& \Pr[x_i | \mathbf{x}_{A_1},x_j]\frac{\Pr[\mathbf{x}_{A_2}\mid  x_i, \mathbf{x}_{A_1},x_j]}{\Pr[\mathbf{x}_{A_2}\mid \ \mathbf{x}_{A_1},x_j]}\\
         =& \Pr[ x_i | \mathbf{x}_{A_1}]\frac{\Pr[\mathbf{x}_{A_2}\mid  x_i, \mathbf{x}_{A_1}]}{\Pr[\mathbf{x}_{A_2}\mid \mathbf{x}_{A_1}]}\label{eq:step}\\
         =&\Pr[ x_i | \mathbf{x}_{A_1}, \mathbf{x}_{A_2}]=\Pr[ x_i | \mathbf{x}_A, x_j]
    \end{align}
where~\Cref{eq:step} follows from~\Cref{eq:extended_to_vector}.

\textbf{Case 2:} There exists an index $r \in A$ with $i < l < r$. 
    Here, \Cref{eq:markov_remove_condition} is obtained by applying Bayes' rule to reduce the problem to the already proven first case:
    \begin{align}
        \Pr[x_i \mid\mathbf{x}_A,  x_j] &= \frac{\Pr[x_j \mid  \mathbf{x}_A, x_i] \, \Pr[x_i \mid  \mathbf{x}_A]}{\Pr[x_j \mid  \mathbf{x}_A]} \label{eq:markov_use_bayes_rule} \\
        &= \frac{\Pr[x_j \mid \mathbf{x}_A] \, \Pr[ x_i \mid   \mathbf{x}_A]}{\Pr[x_j \mid   \mathbf{x}_A]} \label{eq:markov_apply_first_statement} \\
        &= \Pr[x_i \mid   \mathbf{x}_A] \label{eq:markov_shorten_ratio}
    \end{align}
    We use Bayes' theorem in \cref{eq:markov_use_bayes_rule}. The first probability of the numerator is now in the situation of Case~1, since it is the probability of random variable $X_j$, conditioned on $\mathbf{X}_A$ and $X_i$ with $i<r<j$ and $r\in A$. \Cref{eq:markov_remove_condition} has already been proven for that case, so we apply it to derive~\cref{eq:markov_apply_first_statement}. Finally, \cref{eq:markov_shorten_ratio} follows directly by simplifying.
\end{proof}
\begin{lemma}\label{lm:key_markov}
      Let random vector $\mathbf{X} = (X_1, \dots, X_n)$ be a Markov chain with transition probabilities $P \in \mathbb{R}^{s \times s}$ and initial distribution $w \in \mathbb{R}^s$ with the following properties:
    \begin{itemize}
        \item[(H1)] Every cell of $P$ is positive, i.e., for all $k,l \in \S$ we have $P_{k,l} > 0$.
        \item[(H2)] Vector $w$ is an eigenvector of $P$ to the eigenvalue $1$, i.e., $w P = w$.
    \end{itemize}
    Let $i \in [n]$ be the target index and let sets $U,K \subseteq [n] \setminus \{i\}$ be disjoint, with $[n] = U \cup K \cup \{ i \}$ and at least one index in $U$. 
    Then, for any unknown states $\mathbf{x}_U \in \S^{u}$, known states $\mathbf{x}_K \in \S^{k}$ and target states $ x_i, x_i' \in \S$ we have
    \begin{equation*}\label{eq:markov_main_theorem_equation}
        \frac{\Pr[\mathbf{X}_U = \mathbf{x}_U \mid  \mathbf{X}_K = \mathbf{x}_K, X_i = x_i]}{\Pr[\mathbf{X}_U = \mathbf{x}_U \mid  \mathbf{X}_K = \mathbf{x}_K, X_i = x_i']} \leq \left(\frac{\max_{kl} P_{kl}}{\min_{kl} P_{kl}}\right)^4\equiv\gamma^4.
    \end{equation*}
\end{lemma}
\begin{proof}
First, combining~\Cref{lm:markov_generalized}  and Bayes' rule we obtain that, for all $A\in[n]\backslash\{i\}$:
\begin{gather}~\label{eq:makov_bayes}
    \Pr[x_i \mid \mathbf{x}_{A} ]= 
    \frac{\Pr[x_r \mid   x_i,x_{\ell}] \, \Pr[ x_i \mid  x_{\ell}]}{\Pr[x_r \mid  x_{\ell}]}
    =\frac{\Pr[ x_r \mid  x_i ]\, \Pr[ x_i \mid x_{\ell}]}{\Pr[ x_r \mid   x_{\ell}]}
\end{gather}
where \ref{eq:makov_bayes} follows by application of \Cref{lm:markov_generalized} since $i$ is closer to $r$ than $\ell$, i.e., $\ell<i<r$.

Second, we use (H1) and (H2) to prove that given $\mathbf{X}$ a Markov chain, for all indices $i,j$, not necessarily consecutive, and for all $x_i,x'_i,y_j,y'_j\in\S$,
 \begin{equation}\label{eq:gamma}
      \frac{\Pr[X_i=x_i]}{\Pr[X_i=x'_i]}\leq \gamma\text{, }\frac{\Pr[X_i=x_i|X_j=y_j]}{\Pr[X_i=x_i|X_j=y'_j]}\leq \gamma \text{ and } \frac{\Pr[X_i=x_i|X_j=y_j]}{\Pr[X_i=x'_i|X_j=y_j]}\leq \gamma 
 \end{equation}

  We begin by proving $ \frac{\Pr[x_i]}{\Pr[{x'_i}]}\leq\gamma $. First, we show that $w$ -- as an eigenvector of $P$ -- not only contains the prior probabilities of $X_1$, but of any random variable $X_i$ for $i \in [n]$. I.e., the equality $\Pr[X_i = x_i] = w_{x_i}$ holds for any state $x_i \in \S$ because $w$ is the equilibrium distribution. We proceed by induction, therefore we assume it true for $i-1$ and prove it for $i$:
    \begin{align}
        \Pr[X_i = x_i] &= \sum_{y \in \S} \Pr[X_i = x_i \mid X_{i-1} = y] \, \Pr[X_{i-1} = y] \label{eq:markov_induction_law_total_prob}\\
        &= \sum_{y \in \S} P_{y,x_i} w_y \label{eq:markov_replace_with_P_w}\\
        &= (w P)_{x_i} \label{eq:markov_matrix_vector_prod}\\
        &= w_{x_i} \label{eq:markov_is_eigenvector}
    \end{align}
    We apply the law of total probability in \cref{eq:markov_induction_law_total_prob}. Then, we use the transition matrix $P$ and the induction hypothesis ($\Pr[X_{i-1} = y] = w_y$) to replace the probabilities with entries of $P$ and $w$ in \cref{eq:markov_replace_with_P_w}. Then, we rewrite the sum as a matrix-vector product in \cref{eq:markov_matrix_vector_prod}. Finally, we take advantage of the fact that $w$ is an eigenvector of $P$ in \cref{eq:markov_is_eigenvector}. With the basis $\Pr[X_1 = x] = w_x$ (which is the definition of $w$, see \Cref{def:markov_correlation}) and this derivation, we prove by induction that the entries of $w$ are equal to the prior probabilities of any random variable $X_i$.

    Now we can bound the entries of $w$, thereby also bounding the probabilities $\Pr[X_i = x]$.
    We prove the upper bound $w_x \leq \max_{k,l\in\S} P_{k,l}$ by contradiction; the lower bound $w_x \geq \min_{k,l\in\S} P_{k,l}$ follows analogously.
    Assume that there exists a state $y \in \S$ so that its prior probability in $w$ is greater than any transition probability, i.e., $w_y > \max_{k,l \in \S} P_{k,l}$. This leads to a contradiction as follows.
    \begin{align}
        w_y &= (w P)_y \label{eq:markov_eigenvector_contradiction} \\
        &= \sum_{k \in \S} P_{k,y} \, w_k \\
        \Leftrightarrow (1 - P_{y,y})w_y &= \sum_{k \neq y} P_{k,y} \, w_k \\
        \Leftrightarrow 1 - P_{y,y} &= \sum_{k \neq y} \frac{P_{k,y}}{w_y} \, w_k  \\
        &< \sum_{k \neq y} w_k \label{eq:markov_use_wrong_assumption}\\
        &= 1 - w_y \label{eq:markov_sum_is_1_minus_wj} \\
        \Leftrightarrow 1 + w_y &< 1 + P_{y,y} \\
        \Leftrightarrow w_y &< P_{y,y} 
    \end{align}
    
    We use that $w$ is an eigenvector in \cref{eq:markov_eigenvector_contradiction} and subsequently rewrite the matrix-vector multiplication. In \cref{eq:markov_use_wrong_assumption}, we apply the assumption that $w_y$ is greater than any transition probability, so $P_{k,y} / w_y$ must be strictly smaller than one. \Cref{eq:markov_sum_is_1_minus_wj} follows because the entries of $w$ must sum to one as $w$ is a probability distribution (see \Cref{def:markov_correlation}). Finally, we arrive at the statement $w_y < P_{y,y}$ which is contradictory to our assumption that $w_y$ is bigger than every $P_{k,y}$. Thus, this assumption was false and probability $w_y$ must be smaller or equal to $\max_{k,l\in \S} P_{kl}$ for any state $y \in \S$.

     Second, we prove that $\frac{\Pr[x_i \mid y_j]}{\Pr[x_i \mid  y'_j]} = \frac{P_{y_j,x_i}}{P_{y'_j,x_i}}\leq\gamma$. Let indices $i,j \in [n]$ such that $j<i$ and let states $x_i, y_j, y'_j \in\S$. If random variables $X_i$ and $X_j$ are direct neighbors, i.e., $i = j + 1$, then the probabilities are transition probabilities from matrix $P$ and the bound follows trivially.
    
    In the other case (i.e., $j+1<i$),  we have
    \begin{align}
        \Pr[&X_i = x_i \mid X_j = y_j] \notag\\ = &\sum_{y_{j+1} \in \S} \Pr[x_i \mid y_j, X_{j+1} = {y_{j+1}}] \, \Pr[X_{j+1} = y_{j+1} \mid y_j] \\
        = &\sum_{y_{j+1} \in \S}  \Pr[x_i \mid y_{j+1}] \, \Pr[y_{j+1} \mid y_j] \label{eq:markov_drop_indepdent_condition}\\
        = &\sum_{y_{j+1} \in \S}  \Pr[x_i \mid  y_{j+1}] \, \Pr[y_{j+1} \mid y'_j] \, \frac{\Pr[y_{j+1} \mid y_j]}{\Pr[y_{j+1} \mid y'_j]} \label{eq:markov_add_independent_condition} \\
        = &\sum_{y_{j+1} \in \S}  \Pr[x_i \mid y'_j, y_{j+1}] \, \Pr[y_{j+1} \mid y'_j] \,  \frac{P_{y_{j},y_{j+1}}}{P_{y'_{j},y_{j+1}}} \label{eq:markov_to_trans_probs} \\
        \leq&\sum_{y_{j+1} \in \S} \Pr[x_i \mid y'_j, y_{j+1}] \, \Pr[y_{j+1}\mid y'_j] \, \gamma \label{eq:markov_bound_ratio} \\
        \leq &\gamma\Pr[X_i = x \mid X_j = y'] \, . \label{eq:markov_reverse_law_of_total_probability}
    \end{align}
    We use \Cref{lm:markov_generalized} to remove $X_j = x_j$ from the condition of the first probability in \cref{eq:markov_to_trans_probs} because $j+1$ is closer to $i$. If $i<j$ then the results follow from applying Bayes' rule and the previous case.

    Finally, we prove the last inequality from \Cref{eq:gamma}, in a similar fashion to the one before. Given $x_i, x'_i, y_j \in \S$. If random variables $X_i$ and $X_j$ are direct neighbors (i.e., $j = i - 1$), the ratio can once again be bounded straightforwardly as it only contains probabilities from $P$.
    
    Otherwise for $j < i - 1$ , we have
    \begin{align}
        \Pr[&X_i = x_i | X_j = y_j] \notag\\ = &\sum_{x_{i-1} \in \S} \Pr[x_i| y_j, x_{i-1} ] \, \Pr[x_{i-1} | y_j] \\
        = &\sum_{x_{i-1} \in \S} \Pr[x_i | x_{i-1}] \, \Pr[x_{i-1} \mid y_j ] \label{eq:markov_use_lemma_1}\\
        = &\sum_{x_{i-1} \in \S} \Pr[x'_i \mid x_{i-1}] \, \frac{\Pr[x_i \mid x_{i-1} ]}{\Pr[x'_i | x_{i-1}]} \, \Pr[x_{i-1} | y_j ] \label{eq:markov_replace_probability_by_multiplying_with_1} \\
        = &\sum_{x_{i-1} \in \S} \Pr[x'_i \mid x_{i-1}]\,\Pr[x_{i-1} | y_j ] \, \frac{P_{x_{i-1},x_i}}{P_{x_{i-1},x'_i}} \\
        \leq&\sum_{x_{i-1} \in \S} \Pr[x'_i \mid x_{i-1},y_j]\,\Pr[x_{i-1} | y_j ] \, \gamma
         \label{eq:markov_bound_ratio_from_P}\\
        = &\gamma \Pr[X_i = x' | X_j = y] \,.\label{eq:markov_reverse_law_prob}
    \end{align}
    \Cref{lm:markov_generalized} is used to drop $X_j = y$ from the condition in \cref{eq:markov_use_lemma_1} and add it again in~\Cref{eq:markov_bound_ratio_from_P}.  If $i<j$ then the results follow form applying the Bayes rule and the previous case.

Combining~\Cref{eq:makov_bayes} and~\Cref{eq:gamma} we obtain that
\begin{equation}\label{eq:gamma2}
     \frac{\Pr[x_i \mid \mathbf{x}_{A} ]}{\Pr[x'_i \mid \mathbf{x}_{A} ]}=
     \frac{\Pr[ x_r \mid  x_i] \, \Pr[x_i \mid   x_{\ell}]}{\Pr[ x_r \mid  x'_i] \, \Pr[x'_i \mid   x_{\ell}]}
     \leq \gamma^2
\end{equation}
Note that if $A$ only contains indices smaller than $i$, then previous equation gets simplified to
\begin{equation*}
       \frac{\Pr[x_i \mid \mathbf{x}_{A} ]}{\Pr[x'_i \mid \mathbf{x}_{A} ]}=
     \frac{\Pr[x_i \mid   x_{\ell}]}{ \Pr[x'_i \mid   x_{\ell}]}
     \leq\gamma\leq \gamma^2
\end{equation*}
and the analogous holds if $A$ only contains indices bigger than $i$.

Finally, combining  Bayes' rule with ~\Cref{eq:gamma2} applied to $A=K$ and $A=[n]\backslash\{i\}$, we obtain the result:
\begin{gather*}
         \frac{\Pr[ \mathbf{x}_U |  \mathbf{x}_K, x_i]}{\Pr[ \mathbf{x}_U | \mathbf{x}_K,x_i']} 
         =\frac{\Pr[ x_i | \mathbf{x}_K,\mathbf{x}_U] \, \Pr[ x_i' | \mathbf{x}_K]}{\Pr[ x_i' |  \mathbf{x}_K,  \mathbf{x}_U] \, \Pr[ x_i |  \mathbf{x}_K]}
         =\frac{\Pr[ x_i |  \mathbf{x}_{-i}] \, \Pr[ x_i' | \mathbf{x}_K]}{\Pr[ x_i' |\mathbf{x}_{-i} ] \, \Pr[x_i | \mathbf{x}_K]}
         \leq \gamma^2\gamma^2=\gamma^4
        \end{gather*}
         Note that if $K=\varnothing$ the previous expression gets simplified to 
        \begin{gather*}
         \frac{\Pr[\mathbf{x}_U |  x_i]}{\Pr[ \mathbf{x}_U |  x_i']} 
         =\frac{\Pr[x_i |  \mathbf{x}_U] \, \Pr[ x_i']}{\Pr[ x_i' |  \mathbf{x}_U] \, \Pr[ x_i ]}
         \leq\gamma^3\leq\gamma^4
        \end{gather*}
\end{proof}
Finally, combining previous lemmas we obtain our novel bound:
\begin{theorem}[The Markov Chain Bound]\label{cor:markov_bdpl_bound}
    Let $s \in \mathbb{N}$ be the number of states. Let $\mathcal{M} : \mathcal{S}^n \to \mathcal{Y}$ be an $\varepsilon$-DP mechanism. Let the databases follow a Markov chain with transition matrix $P \in \mathbb{R}^{s \times s}$ and initial distribution $w \in \mathbb{R}^s$ with the following properties:
    \begin{itemize}
        \item[(H1)] For all $x,y \in \mathcal{S}$ we have $P_{x,y} > 0$ and,
        \item[(H2)]  $w P = w$.
    \end{itemize}
    Then, $\mathcal{M}$ is an $(\varepsilon + 4 \ln\gamma)$-BDP mechanism where \[
    \gamma\coloneqq \frac{\max_{x,y\in\mathcal{S}} P_{xy}}{\min_{x,y\in\mathcal{S}} P_{xy}}.\]
\end{theorem}
\begin{proof}
    If there are no unknown indices $U = \varnothing$ the adversary knows every index $K = [n] \setminus \{i\}$ except the target index and the adversary-specific $\mathrm{BDPL}_{(K,i)}$ becomes the same as the DP privacy leakage~\cite{Yang_2015_BDP}. Thus, since $ \varepsilon \leq \varepsilon + 4 \gamma$ for all $\gamma\geq 1$, the inequality is trivially satisfied.
    
    We denote by $u=|U|$ the size of unknown indices. If there is at least one unknown index for the adversary, i.e., $u\geq1$ then we have
    \begin{gather*}
        \Pr_{\M}[Y \in S | \mathbf{x}_K,  x_i]  = \sum_{\mathbf{x}_U \in \S^{u}} \Pr_{\M}[Y \in S | \mathbf{x}_K,x_i, \mathbf{x}_U] \, \Pr_{\pi}[ \mathbf{x}_U |\mathbf{x}_K, x_i] \\
        = \sum_{\mathbf{x}_U \in \S^{u}} \Pr[Y \in S |  \mathbf{x}_K, x_i, \mathbf{x}_U] \, \Pr[\mathbf{x}_U |\mathbf{x}_K, x_i'] \, 
        \frac{\Pr[\mathbf{x}_U | \mathbf{x}_K, x_i]}{\Pr[ \mathbf{x}_U | \mathbf{x}_K, x_i']}   \\
        \overset{\text{(\Cref{lm:key_markov})}}{\leq} \sum_{\mathbf{x}_U \in \S^{u}} \Pr[Y \in S | \mathbf{x}_K,  x_i, \mathbf{x}_U] \, \Pr[\mathbf{x}_U |  \mathbf{x}_K,  x_i'] \, 
        \left(\frac{\max_{kl} P_{kl}}{\min_{kl} P_{kl}}\right)^4 \\
        = \left(\frac{\max_{kl} P_{kl}}{\min_{kl} P_{kl}}\right)^4 \sum_{\mathbf{x}_U \in \S^{u}} \Pr[Y \in S | \mathbf{x}_K,  x_i,  \mathbf{x}_U] \,
        \Pr[\mathbf{x}_U | \mathbf{x}_K, x_i'] \\
        \leq \left(\frac{\max_{kl} P_{kl}}{\min_{kl} P_{kl}}\right)^4 \, e^\varepsilon \sum_{\mathbf{x}_U \in \S^{u}} \Pr[Y \in S | \mathbf{x}_K, x_i',\mathbf{x}_U] \,  
        \Pr[ \mathbf{x}_U |\mathbf{x}_K, x_i'] \\
        = \left(\frac{\max_{kl} P_{kl}}{\min_{kl} P_{kl}}\right)^4 \, e^\varepsilon \, \Pr[Y \in S |  \mathbf{x}_K,x_i']   
    \end{gather*}
    
    Therefore, for every possible adversary with $u\geq 1$ we have that $\mathrm{BDPL}_{(K,i)}\leq \varepsilon+4\ln\gamma$.
    Since we have bounded the $\mathrm{BDPL}_{(K,i)}$ of every possible adversary $(K, i)$, we also bound the total BDPL.
\end{proof}
%Discussion of the hypotheses
(H1) states that all entries in the transition matrix  are strictly positive, while (H2) requires that the initial distribution is a \textit{stationary distribution}, meaning the distribution over states $w_t$ (without considering the previous one) remains constant at each time--a common modeling assumption in various data mining tasks such as weather forecasting~\cite{wilks2011statistical} or electricity consume~\cite{ardakanian2011markovian}. Notably, condition (H1) implies that the chain is both irreducible and aperiodic, which in turn guarantees the existence of a unique stationary distribution $w$~\cite{levin2017markov}, thereby satisfying (H2). Furthermore, for any initial distribution $w'$, the distribution at time $t$ converges geometrically fast to $w$ as $t$ increases~\cite{levin2017markov}. Consequently, even when the initial distribution is not stationary, it asymptotically approaches the stationary distribution, effectively satisfying (H2) after discarding a sufficient initial portion of the process.

\begin{figure}[t]
    \centering
    \includegraphics[width=0.5\textwidth]{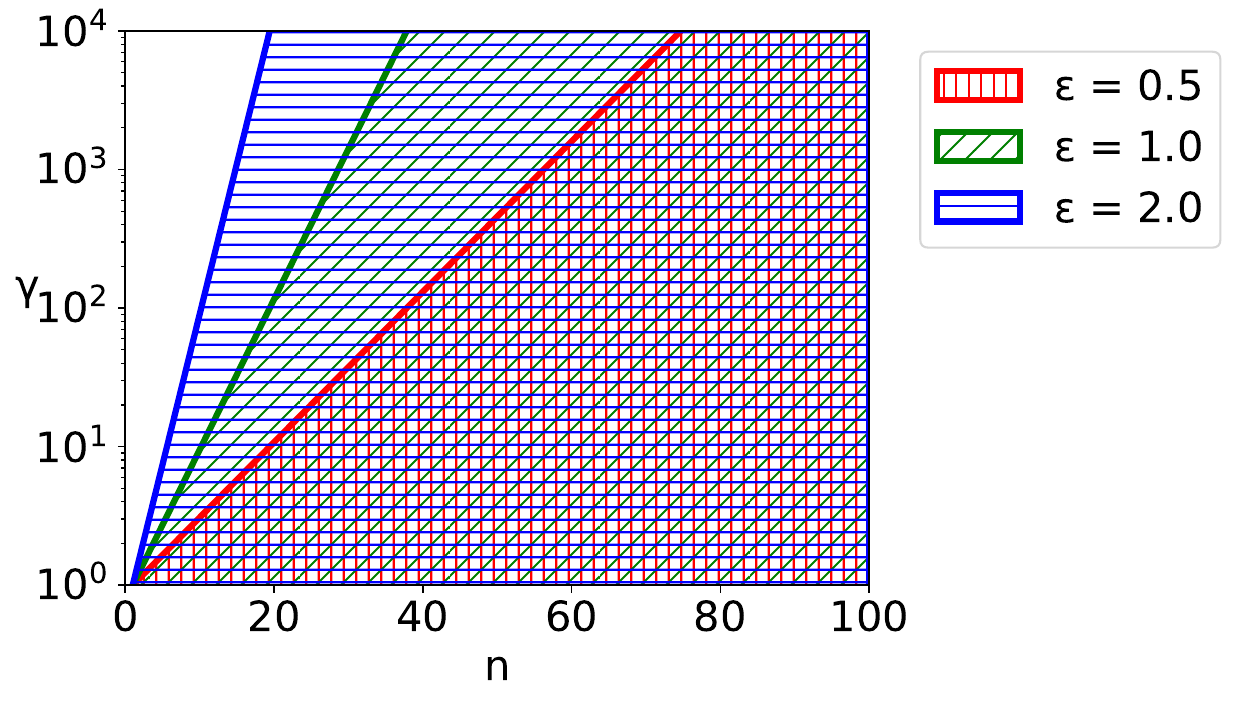}
    \caption[Comparison of Markov Bound to General Bound]{Comparison of Markov-specific bound to general bound. The Markov-specific bound improves upon the general bound for values of $n$ and $\gamma$ in the respective shaded area.}
\label{fig:markov_bound_improvement}
\end{figure}

While prior work provides a mechanism for BDP protection of lazy binary Markov chains with a symmetric transition matrix~\cite{chakrabarti2022optimal}, we present the first direct and general relationship between DP and BDP leakage for arbitrary Markov chains, including non-binary ones. When compared this novel bound with the general one we obtain that for any $\varepsilon > 0$, and maximum transition probability ratio $\gamma \geq 1$ we have
    \begin{gather}\label{eq:markov_vs_general}
       \varepsilon + 4 \ln \gamma < n \varepsilon 
       %\ \Leftrightarrow \  4 \ln \gamma < (n - 1)\varepsilon \\
       %\Leftrightarrow \ \ln \gamma < \frac{n - 1}{4}\varepsilon 
       \ \text{ if and only if} \ \gamma < \exp\left(\frac{n - 1}{4}\varepsilon \right).
     \end{gather}
Therefore, the Markov chain bound outperforms the general bound in most cases. For instance, with an $\varepsilon$-DP mechanism where $\varepsilon = 0.5$ and a database size of $n = 80$, it remains tighter even when the largest transition probability is $10,000$ times the smallest. For the same $\varepsilon = 0.5$, the Markov bound only becomes looser than the general one when the number of correlated records is small, e.g., $n = 20$, and the transition probability ratio $\gamma$ is as high as $100$, which still represents a significant disparity (See~\Cref{fig:markov_bound_improvement}).
Moreover, \Cref{cor:markov_bdpl_bound} enables the systematic design of BDP mechanisms by adjusting the noise of an existing DP mechanism.
Noise calibration depends only on the maximum ratio between the Markov transition probabilities of the model, $\gamma$, and the adjusted mechanism must be calibrated to $\varepsilon'=\varepsilon-4\ln(\gamma)$. 
Note that the best BDPL privacy achievable using~\Cref{cor:markov_bdpl_bound} is $\varepsilon = 4\ln(\gamma)$, since $\varepsilon' \geq 0$. 
Consequently, the minimum achievable $\varepsilon$ is fundamentally constrained by the structure of the underlying Markov model---specifically, by the maximum transition ratio $\gamma$. We illustrate how the transition matrix changes the minimum $\varepsilon$ in theoretical settings in~\Cref{fig:markov_sota}, and in real-world data in~\Cref{sec:experiments}.

\subsection{Accuracy}
\begin{figure}
    \centering
\includegraphics[width=0.5\textwidth]{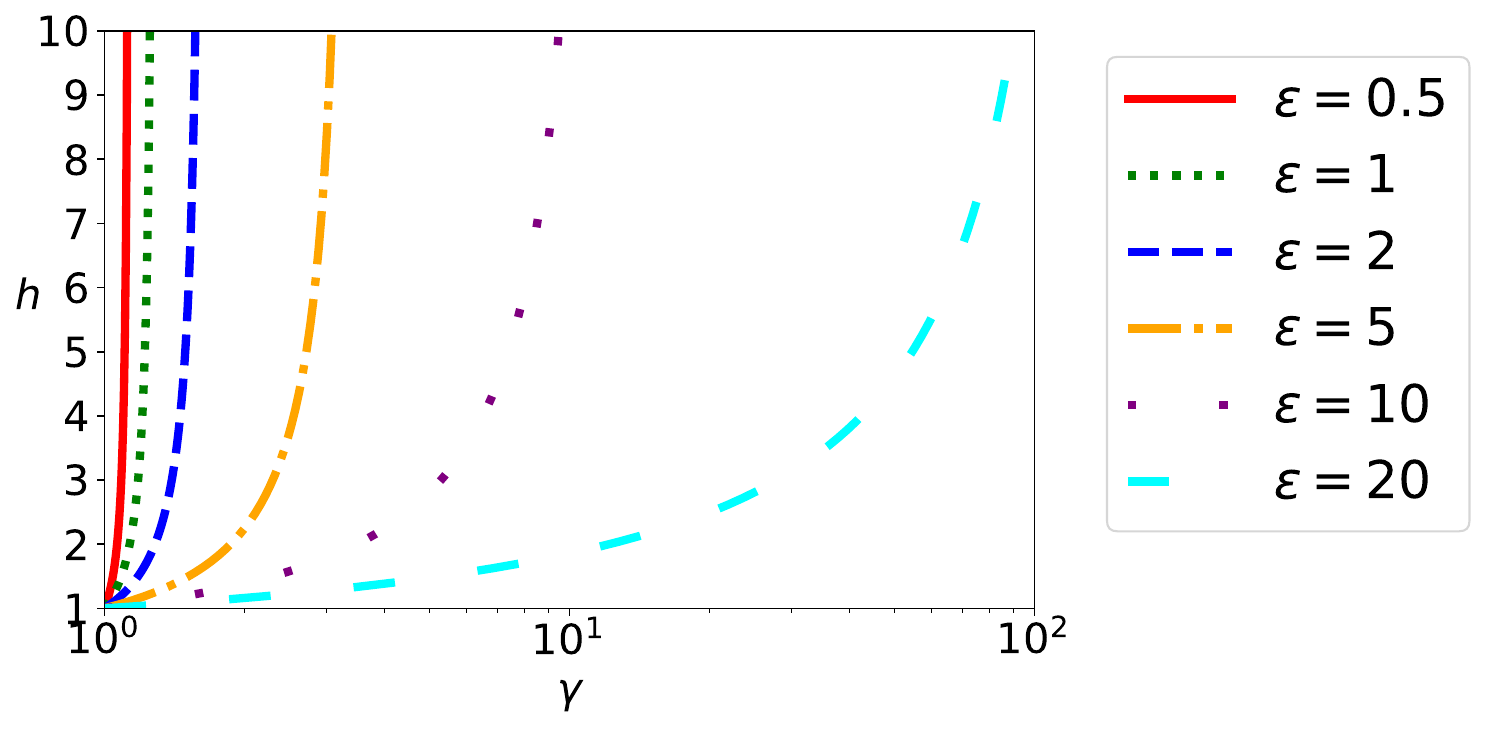}
    \caption[Relative Accuracy $h$ of an $\varepsilon$-BDP to an $\varepsilon$-DP mechanism for Markov Chain Data]{ Relative accuracy $h$ of an $\varepsilon$-BDP to an $\varepsilon$-DP mechanism for Markov chain data respect to $\gamma$. }
    %PREVIOUS CAPTION:
    %the maximum transition probability ratio $\gamma$ of a Markov chain so that the error of an $\varepsilon$-BDP mechanism $\mathcal{B}$ is at most $h$ times greater than the error of an $\varepsilon$-DP mechanism $\mathcal{M}_{\varepsilon, f}$.} 
    \label{fig:markov_utility2}
\end{figure}

\noindent
The Markov chain bound enables us to derive improved utility guarantees for the Laplace mechanism when $\gamma$ is sufficiently small.
\begin{corollary}\label{cor:markov_utility}
Let $\mathcal{M}_{\varepsilon, f}$ be the $\varepsilon$-Laplace mechanism that approximates the query $f : \S^n \to \mathbb{R}$ and inputs a database drawn from a Markov chain satisfying (H1) and (H2). 
If $\M_{\varepsilon,f}$ is $(\alpha,\beta)$-accurate w.r.t. $f$  and $\varepsilon\geq 4\ln(\gamma)$ then, there exists an $\varepsilon$-BDP mechanism $\mathcal{B}$ that is $(h \alpha, \beta)$-accurate w.r.t. $f$ with
    \begin{equation*}
       h=\frac{\varepsilon}{\varepsilon-4\ln(\gamma)}.
    \end{equation*}
\end{corollary}
\begin{proof}
The idea of this proof is to construct mechanism $\mathcal{B}$ with the Laplace mechanism as well, but to choose a carefully selected privacy leakage $\varepsilon' < \varepsilon$ so that mechanism $\mathcal{B}$ is (1) $\varepsilon$-BDP and (2) $(h \alpha, \beta)$-accurate. 

    First, we determine the accuracy of mechanism $\mathcal{M}_{\varepsilon, f_{I}}$. With \Cref{thm:laplace_mechanism_accuracy}, we know that the $(\alpha, \beta)$-accuracy of the Laplace mechanism for a given probability $\beta \in (0, 1]$ and privacy parameter $\varepsilon$ is 
    \begin{equation*}
        \alpha = \ln \left(\frac{1}{\beta}\right) \, \frac{\Delta f}{\varepsilon}.
    \end{equation*} 
    So this is the $(\alpha, \beta)$-accuracy of $\mathcal{M}_{\varepsilon, f}$.

    We have to show that there exists an $\varepsilon$-BDP mechanism $\mathcal{B}$ which is $(h \alpha, \beta)$-accurate.
    We choose $\mathcal{B}$ as the Laplace mechanism applied to $f$ with an adjusted privacy parameter $\varepsilon' > 0$. Thus, $\mathcal{B}$ will be $\varepsilon'$-DP. Therefore, we can use \Cref{cor:markov_bdpl_bound} to show that $\mathcal{B}$ is BDP. We must choose $\varepsilon'$ in a way that ensures that the BDPL is limited to $\varepsilon$, so that we have $\varepsilon$-BDP.
    With \Cref{cor:markov_bdpl_bound}, $\mathcal{B}$ is
    \begin{equation*}
        (\varepsilon' + 4 \ln\gamma)\textrm{-BDP.}
    \end{equation*}
    Therefore, to achieve $\varepsilon$-BDP, we must have
    \begin{equation*}
         \varepsilon' + 4 \ln\gamma = \varepsilon \Leftrightarrow \varepsilon'=\varepsilon-4 \ln\gamma
    \end{equation*}
    Now, we can calculate the accuracy of $\mathcal{B}$ because it also uses the Laplace mechanism. Then, we find an upper bound for this accuracy.
    Mechanism $\mathcal{B}$ is $(\alpha', \beta)$-accurate, with
    \begin{align*}
        \alpha' = \ln (\frac{1}{\beta}) \, \frac{\Delta f}{\varepsilon'}=\ln (\frac{1}{\beta}) \, \frac{\Delta f}{\varepsilon-4 \ln\gamma}
        = \alpha h.
    \end{align*}
\end{proof}
The statement of \Cref{cor:markov_utility} is visualized in \Cref{fig:markov_utility2}. This figure shows that in order to provide similar utility guarantees to DP, either the BDPL bound $\varepsilon$ has to be larger than $5$, or the ratio $\gamma$ between different transition probabilities must be smaller than $3$.

\subsubsection{Comparison with the state of the art}
\begin{figure}[t] % Use figure* for two-column wide figure
    \centering
    \begin{subfigure}{0.45\linewidth}
        \centering
        \includegraphics[width=\linewidth]{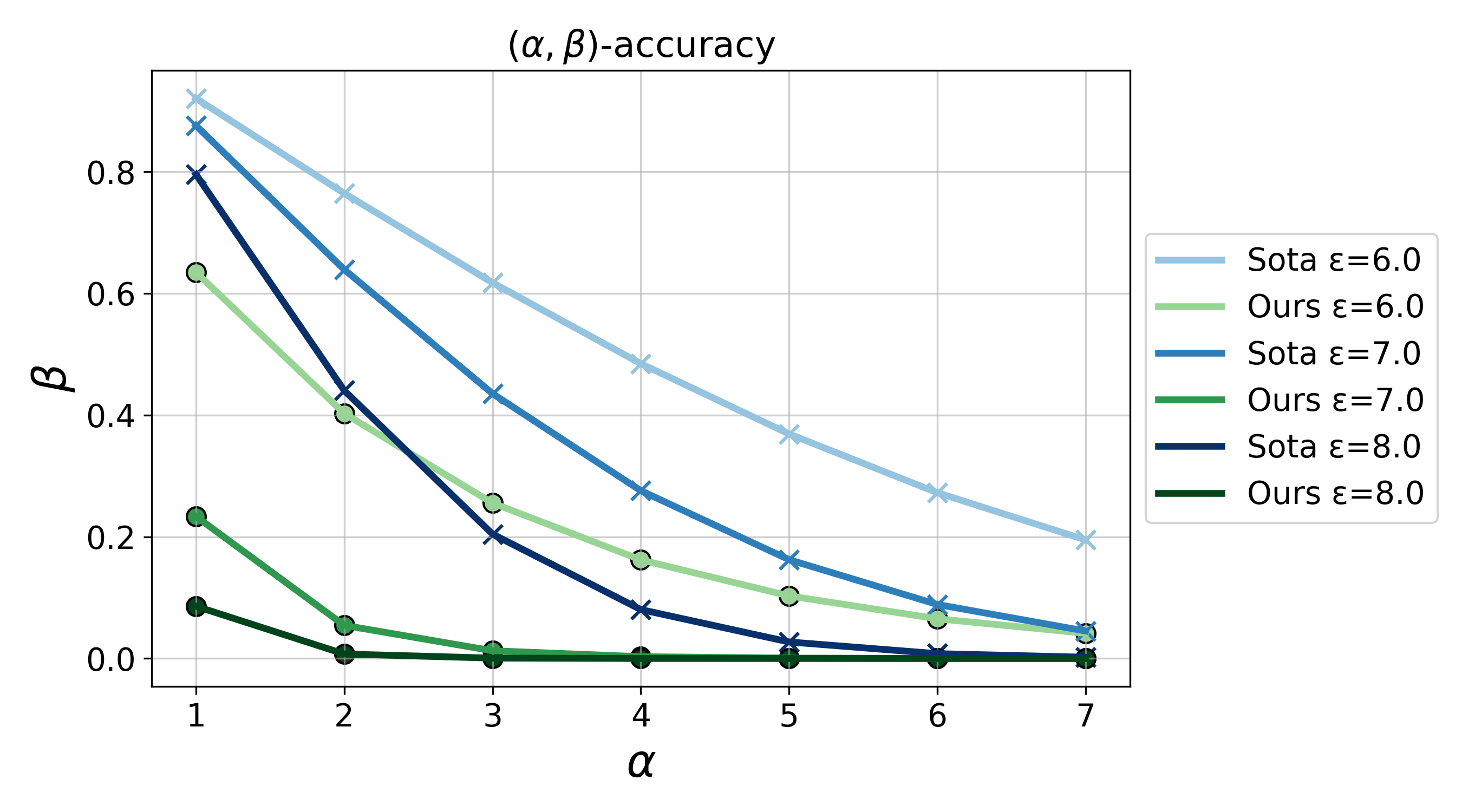}
        \caption{Self-transition probability $P_{ss}=0.8$.}
    \end{subfigure}
    \hfill
    \begin{subfigure}{0.45\linewidth}
        \centering
        \includegraphics[width=\linewidth]{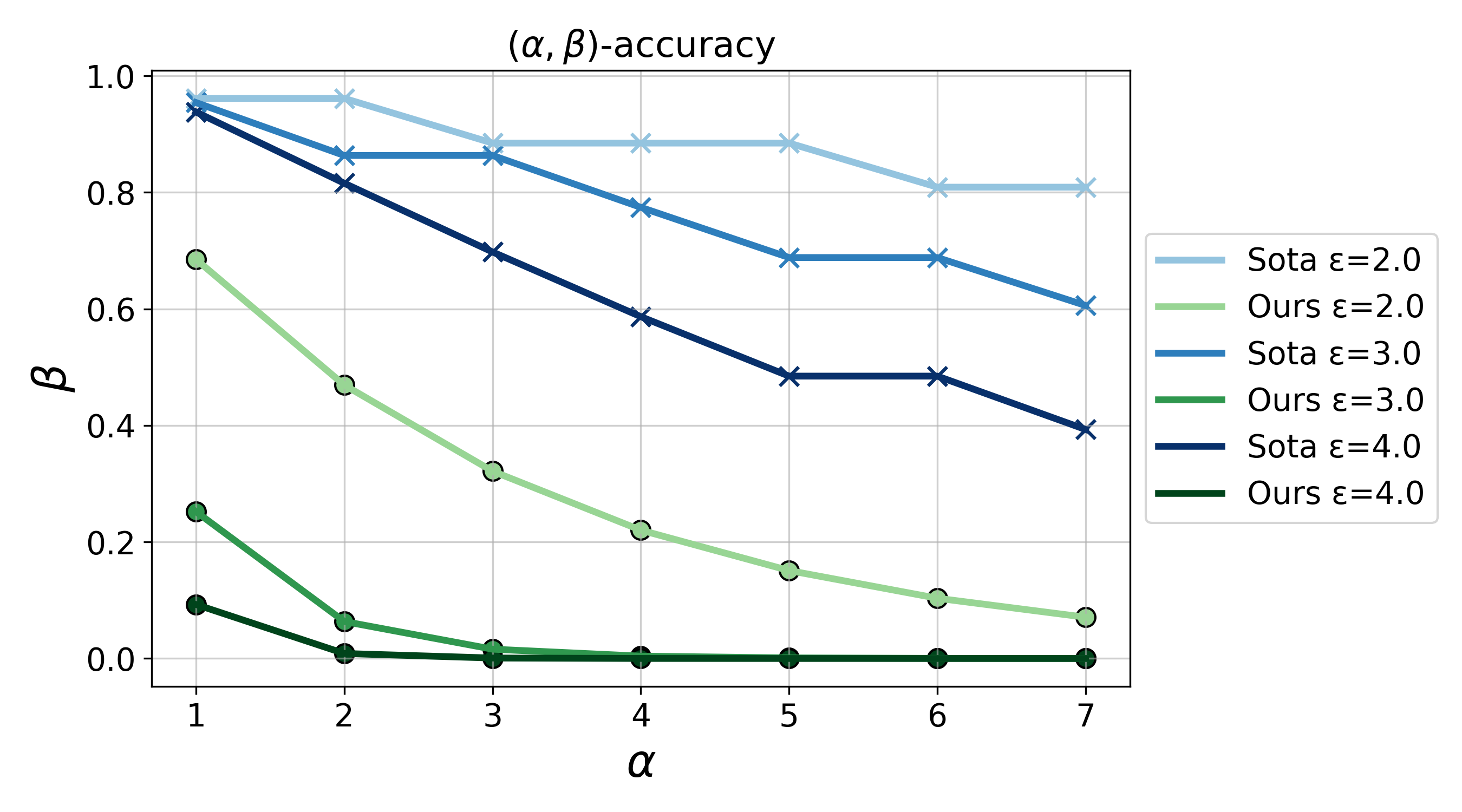}
        \caption{ Self-transition probability $P_{ss}=0.6$.}
    \end{subfigure}
    \caption[SOTA comparison]{$(\alpha,\beta)$-accuracy comparison of our mechanism vs. the s-o-t-a approach~\cite{chakrabarti2022optimal} for $n=500$.
    }
    \label{fig:markov_sota_long}
\end{figure}
\citeauthor{chakrabarti2022optimal}~\cite{chakrabarti2022optimal} propose a BDP adaptation of the randomized response mechanism for symmetric, lazy stationary Markov chains with binary states, i.e., a Markov chain with $s\in\{0,1\}$, $w=(0.5,0.5)$ and symmetric transition matrix
\[
P=
\begin{pmatrix}
1-r & r \\
r & 1-r
\end{pmatrix}
\]
with a constant self-transition probability $P_{ss}=r\in(0,0.5)$ for all $s\in\{0,1\}$, indicating the laziness, i.e., it is more likely to remain in the same state that a change.
They prove that the adapted randomized response such that
\[
\Pr_{\M}(Y_i\mid X_i)=\begin{cases}
    1-\rho & \text{ if }Y_i=X_i\\
    \rho & \text{ otherwise }
\end{cases}
\]
where \[
\rho \geq \frac{4+r(r \e^{\varepsilon}-2)-\sqrt{r^{2
} \e^{\varepsilon}(4+r(r \e^{\varepsilon}-4))}}{8+2r(r \e^{\varepsilon}+r-4)},\]
fulfills $\varepsilon$-BDP.
While they do not give any utility estimate or experiment, we compute the $(\alpha,\beta)$-accuracy of such mechanism to show that it provides worse results that our Laplace based-mechanism.

Given $y_i$ the noisy answer of $Y_i$, the unbiased estimator of the true number of $s=1$, denoted by $n_1$ for the randomized response with parameter $p=(1-\rho$) is~\cite{warner1965randomized}:
\[
\hat{n}_1 = \frac{\sum_{i=1}^n  y_i - n(1 - p) }{2p - 1}.
\]
Additionally, considering \( Z = \sum_{i=1}^{n} Y_i \), the random variable \( Z \) can be expressed as the convolution of two binomial distributions: \( Z = Z_0 + Z_1 \). Here, \( Z_0 \) represents the number of reported 1s originating from individuals with \( X_i = 0 \) who lied, and \( Z_1 \) represents the number of correctly reported 1s where \( X_i = 1 \) was preserved. Formally,
\[
Z_0\sim \mathrm{Bin}(N-n_1,\rho),\,Z_1\sim \mathrm{Bin}(n_1,1-\rho).
\]
Hence, by definition of $(\alpha,\beta)$-accuracy we obtain:
\begin{align*}
   \Pr\left[ |\hat{n}_1 - n_1| \geq \alpha \right]
   =&\Pr\left[|n_1-\frac{(Z - n(1 - p))}{(2p - 1)} |\geq \alpha\right]\\
   =& \Pr\left[ |n_1(2p -1)+n(1-p)-Z| \geq \alpha (2p - 1) \right]\\
\end{align*}
 Therefore, the probability of interest can be decomposed as:
\[
\Pr\left[\,|Z -(n_1(2p -1)+n(1-p))| \geq t\,\right] = \Pr\left[Z \leq \mu - t\right] + \Pr\left[Z \geq \mu + t\right],
\]
where $t=\alpha (2p-1)$ and $\mu=n_1(2p -1)+n(1-p)$. Since \(Z\) is a discrete random variable, for all $t\neq0$, this can be written as:
\[
    \sum_{k=0}^{\lfloor \mu - t \rfloor} \Pr[Z = k] + \sum_{k=\lceil \mu + t \rceil}^{n} \Pr[Z = k].
\]
Since, $Z$ is the convolution of two binomial random variables, its probability mass function is:
\[
\Pr(Z = k) = \sum_{i=0}^{k} \Pr(Z_1 = i) \cdot \Pr(Z_0 = k - i),
\]
therefore, the full expression becomes:
\begin{align}\label{eq:accu_bdp_rr}
\beta=&\Pr\left[\,|Z - \mu| \geq t\,\right]\notag\\
= &\sum_{k=0}^{\lfloor \mu - t \rfloor} \sum_{i=0}^{k}\binom{n_1}{i} (1-\rho)^k \rho^{\mu - k}\binom{n-\mu}{k-i} \rho^{k-i} (1 - \rho)^{n-\mu - k+i}\notag\\
+&
\sum_{k=\lceil \mu + t \rceil}^{n}\sum_{i=0}^{k}\binom{n_1}{i} (1-\rho)^k \rho^{n_1 - k}\binom{n-n_1}{k-i} \rho^{k-i} (1 - \rho)^{n-n_1 - k+i}\notag\\
=&
\sum_{k=0}^{\lfloor \mu - t \rfloor} \sum_{i=0}^{k}\binom{n_1}{i}\binom{n-n_1}{k-i} (1-\rho)^{n-n_1+i} \rho^{n_1 - i}
+
\sum_{k=\lceil \mu + t \rceil}^{n} \sum_{i=0}^{k}\binom{n_1}{i}\binom{n-n_1}{k-i} (1-\rho)^{n-n_1+i} \rho^{n_1 - i} 
\end{align}
where  $t=\alpha (2p-1)$ and $\mu=n_1(2p-1)+n(1-p)$.

Add the same time, since $\Delta f=1$ for binary counting queries, we have that the $(\alpha,\beta)$-accuracy of our Laplace-based mechanism is:
\[
\beta=\e^{-\alpha (\varepsilon-4\ln(\gamma))},
\]
where in this case $\gamma=\frac{1-P_{ss}}{P_{ss}}$.

\begin{figure}
    \centering
    \includegraphics[width=0.7\linewidth]{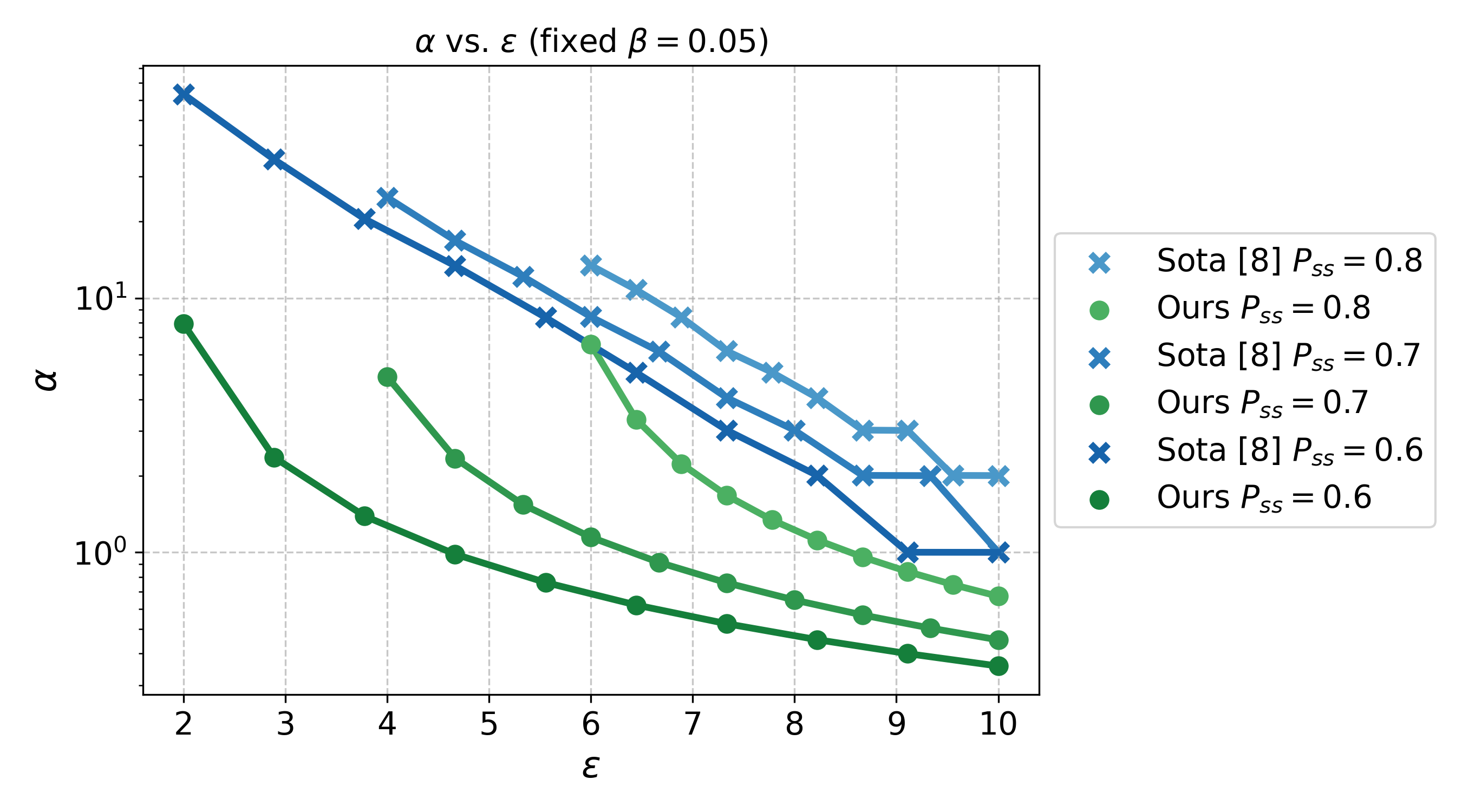}
    \caption{ Accuracy of our mechanism vs. the one proposed in~\cite{chakrabarti2022optimal} for $n=700$ and various self-transition probabilities~$P_{ss}$.}
    \label{fig:markov_sota}
\end{figure}

We compare both accuracies in~\Cref{fig:markov_sota_long}, showing that for all values, our mechanism has a better accuracy, i.e., lower $\beta$ for the same $\gamma$. Additionally, we provide the variation of $\alpha$ respect to different $\varepsilon$ values for a fix $\beta=0.05$, i.e. $95\%$ confidence in~\Cref{fig:markov_sota}. Note that, since Eq.~\ref{eq:accu_bdp_rr} is not invertible, to obtain ~\Cref{fig:markov_sota}, we numerically approximate $\alpha$ using the bisection method.

It is important to note that while their mechanism supports arbitrary BDPL, ours applies only for \( \varepsilon \geq 4 \ln(\gamma) \). However, our approach generalizes to arbitrary Markov chains, whereas theirs is limited to lazy, symmetric binary models. In the intersection of both applicability domains, our use of Laplace-based recalibration yields improved utility.

In conclusion, the Markov-specific bound improves upon the general bound under certain conditions and enables improved utility (\Cref{fig:markov_sota}) compared to prior work~\cite{chakrabarti2022optimal}. Its advantage is most notable when the number of correlated records is large, as it remains independent of dataset size--unlike the general bound, which grows linearly. However, this comes at the cost of a minimum privacy level determined by the data distribution, a limitation absent in the general bound and~\cite{chakrabarti2022optimal}.

\section{Utility Experiments}\label{sec:experiments}
\noindent
Theoretical bounds on privacy and utility do not always translate directly to practical implementations. For instance, while it may be theoretically feasible to achieve a given $(\alpha,\beta)$-accuracy, designing or implementing a mechanism that attains this in practice can be challenging. In this section, we use our theoretical results to construct a BDP mechanism and empirically evaluate its utility on real-world databases % \ts{hm. again "modeled by" seems strange, I guess it would be "on real-world data that follows" or "exhibits" "correlations described by multivariate Gaussian models or Markov chains". Sorry, I hadn't noticed this earlier - I guess I was not paying sufficient attention :-(}
that follow either multivariate Gaussian correlations or Markov chains. Our objective is to demonstrate that the utility gains predicted under specific correlation structures, rather than arbitrary ones, are indeed achievable in practice as well as measure the improvement over previous approaches. %through our approach.

We calibrated the Laplace mechanism using \Cref{th:gaussian_dp_bdpl_bound} and \Cref{cor:markov_bdpl_bound} to derive BDP mechanisms. 
We then ran these BDP mechanisms on the selected databases and compared the utility results with those of BDP mechanisms designed to protect against arbitrary correlation, in order to assess the improvements offered by the correlation-specific approach. 
Moreover, we also plot, when applicable, the accuracy results of the state-of-the-art solutions for Gaussian BDP ~\cite{Yang_2015_BDP}. 
Unfortunately, none of the evaluated datasets meet the strict assumptions needed to apply the only prior mechanism for Markov models~\cite{chakrabarti2022optimal}. Finally, we plot the utility of the classical DP Laplace mechanism as a baseline, representing the best-case utility achievable ignoring correlation.

\subsection{Databases}\label{sub:databases}\label{sec:utility_experiment_data_and_queries}
%why is this a good representative to give generalizable results for multivariate Gaussian correlation? How are the results limited? Which effect has this choice of assumptions on the results and on what we expect for the general case?
\noindent
We use four real-world databases, two for each correlation model. 
Additionally, we use a synthetic dataset to test scalability for Gaussian correlations. 
The selection criteria are public availability, quality of the databases, and the fulfillment of the theoretical assumptions, namely, following the correlation model and fulfilling the extra hypotheses of the corresponding theorem in each case, regarding the Pearson correlation coefficient and the transition matrix.

\subsubsection{Multivariate Gaussian: }
We use two datasets that align well with our modeling framework: the Galton Height Data~\cite{Galton_2017_Height}, a historical dataset originally compiled to study the correlation between parents' and children's heights, and the FamilyIQ dataset~\cite{FamilyIQ}, which includes IQ scores of gifted children and their parents.

The Galton Height Data--considered a classical example of linear correlation modeling, where regression and correlation are interpreted within the framework of a multivariate Gaussian distribution~\cite{Luo_1998_Height}--is especially well known in statistical analysis for introducing the very concept of regression~\cite{Brainard_1992_Height}. 
In contrast, several studies provide evidence that IQ scores in the general population are standardized to follow a multivariate Gaussian distribution, where non-zero correlations are observed only among close relatives~\cite{plomin_behavioral_2013}. 
These properties make both datasets well-suited for evaluating the practical transferability of our Gaussian-based bounds.  
Additionally, we generate the dataset SyntheticIQ to test the scalability of our approach. 
Following the findings among several populations summarized in~\cite{plomin_behavioral_2013}, we generate data following a Gaussian distribution with $\mu=100$, $\sigma^2=15$ and $\rho=0.45$ for parent-child.

%Technical data description.
To ensure bounded sensitivities, all records are clipped to the range of $1$cm to $254$cm ($0$ to $100$ inches) for Galton, and from $40$ to $160$ for IQ datasets as summarized in~\Cref{tab:data}.
\begin{table}
    \centering
    \begin{tabular}{l|l|l|l|l}
    \small
    \centering
     Database & n & m & Parameters & Sensitivity\\ \hline\hline
        Galton  & $ 897$       & $3$               & $\rho=0.275$ &$\Delta q = 254 cm$ \\ \hline
 FamilyIQ  & $ 868$       & $2$               & $\rho=0.4483$&$\Delta q = 120 $ \\ \hline
 SyntheticIQ  & $ 20000$       & $2$               & $\rho=0.45$ &$\Delta q =120$\\ \hline
 Activity  & $ 17568$       & $n$               & $\gamma=7.54 $& $\Delta q = 1$\\ \hline
  Activity Single Day  & $288$        & $n$               & $\gamma=7.54 $& $\Delta q = 1$\\ \hline
Electricity  & $731$       & $n$               &  \makecell{ 70 kWh, $\gamma=3.29$\\80 kWh, $\gamma=4.49$\\90 kWh, $\gamma=8.43$ }  &$\Delta q = 1$ \\ \hline
    \normalsize
    \end{tabular}
    \caption[Datasets parameters description]{Data description. $m$ is the max number of correlated records and $n$ the total amount.}
    \label{tab:data}
\end{table}
%The datasets consist of $n=897$ and $n=868$ records for the Galton and FamilyIQ datasets, respectively. Each record includes either the height (Galton) or IQ score (FamilyIQ) of an individual, along with identifiers that link them to their family members. The Galton dataset contains data from $205$ distinct families of parents and child, while FamilyIQ includes $434$ pairs of mother and child corresponding to a maximum of $m=3$ correlated individuals in Galton and $m=2$ in FamilyIQ.

%In the Galton dataset, the empirical Pearson correlation coefficient is $0.275$ between child and father, $0.202$ between child and mother, and $0.074$ between father and mother. In the FamilyIQ dataset, the corresponding values are \pc{add child-father correlation}, \pc{add child-mother correlation}, and \pc{add father-mother correlation}.

All explored datasets fulfill the conditions of our \Cref{th:gaussian_dp_bdpl_bound}: Galton Pearson correlation coefficient of $\rho=0.275$,  satisfies the condition $\rho=0.275<1=\frac{1}{m-2}$, hence our bounded-correlation assumptions hold. For $m=2$, the condition trivially holds for all $\rho$ values, so in particular for FamilyIQ and SyntheticIQ.

% https://www.kaggle.com/datasets/shambhavimalik/activity-data/data
\subsubsection{Markov Model: } 
We study two use cases--human activity and electricity consume--well-suited for Markov modeling. Human activity representations such as ``inactive'' versus ``active'' are modeled by Markov chains~\cite{huang2018hidden}. Similarly, electricity usage patterns, particularly transitions between high and low consume periods, have been effectively modeled using Markov processes~\cite{ardakanian2011markovian,munkhammar2021very,dalkani2021modelling}. 
We select a representative database for each domain to evaluate our framework. For human activity, we use Activity Data~\cite{Malik_2020_Activity}, which contains the time series of step counts recorded every 5 minutes from a personal activity monitoring device worn by a single individual during October and November 2012. This allows us to extract the ``active'' state if any steps are recorded and the ``inactive'' when the user does not move. 
Besides, to assess the data size impact, we split Activity data into $61$ unique subdatabases, each corresponding to the activity states of a single day.  For electricity usage, we use the Electricity Dataset~\cite{makonin2016electricity}, which captures a single residence electricity usage in Canada from 2012 to 2014. 
We classify each hour as low or high consume depending on whether the usage falls below or exceeds a fixed threshold of $80$ kWh--the central value of the range. Additionally, we study different threshold values, $70$ and $90$ kWh, to assess their impact on utility. 
In all cases, we evaluate event-level local privacy guarantees, assuming no trusted curator and focusing on user-side privacy protection~\cite{dwork2014algorithmic}. The technical details of the three datasets are summarized in~\Cref{tab:data}.

In order to fulfill the conditions of \Cref{cor:markov_bdpl_bound} we require the transition probabilities of the Markov chain to be positive. We calculate them empirically and receive the following transition matrices for Activity and Electricity $70,80,90$ kWh in this order:
{\small
\[
\begin{pmatrix}
0.882 & 0.117 \\
0.305 & 0.695\\
\end{pmatrix},\,  
\begin{pmatrix}
0.445& 0.555\\
0.149& 0.850
\end{pmatrix},
\,
\begin{pmatrix}
0.818 & 0.182\\
0.371 & 0.629
\end{pmatrix}\,
\begin{pmatrix}
0.894&  0.106\\
0.478& 0.522\\
\end{pmatrix},
\]
}
representing $P_{00},P_{01},P_{10},P_{11}$ with $s=0$ inactive/low consume  and $s=1$ active/high consume. Our theoretical results also require $w$ to be a stationary distribution. While $w$ can not be empirically computed since we only have one initial state, both Markov chains are irreducible, since both states are reachable from each other, aperiodic, since $P_{ss}\neq 0$ for both $s\in\{0,1\}$, and $P_{st}>0$ hence there exists a stationary initial distribution~\cite{ching2006markov}. 
Therefore, we conclude that the databases fulfill the conditions for testing our results. 

\subsection{Target Queries and Utility metrics}\label{sub:metrics}
%THORSTEN: what is the operator, to start with? Which query? Why?
\noindent
We focus our utility study on two concrete although commonly used queries: sum and counting queries. Formally, given a database  $D=(x_1,x_2,\dots,x_n)$, where each $x_i$ represents a numerical value, a sum query is defined as: 
$q_S(D)=\sum_{i=1}^n x_i.$
In the case of the Gaussian data, each $x_i$ corresponds to an individual's height or IQ. If each record is binary, i.e., $x_i\in\{0,1\}$, as is the case for the activity and electricity datasets, $q_S(D)$ is called a counting query since it outputs the count of states with the attribute~$1$.

Our theoretical results are expressed in terms of $(\alpha,\beta)$-accuracy. 
To evaluate empirical utility, we use the upper bound of a $(1-\beta)$ confidence interval for the absolute query error, which serves as a practical counterpart. 
Specifically, we report the upper limit of a $95\%$ confidence interval (i.e., $\beta = 0.05$), a standard choice in practice~\cite{Lee_2016_confidence}. A smaller upper bound indicates higher utility. 
When this bound is close to the theoretical error $\alpha$, it demonstrates a strong alignment between empirical and theoretical results, highlighting their practical applicability.
To facilitate comparison with our theoretical results, we plot the theoretical error tolerance $\alpha$ for each mechanism, derived from \Cref{thm:laplace_mechanism_accuracy} for the baseline DP mechanism and \Cref{cor:arbitrary_accuracy_laplace}, \Cref{cor:gaussian_accuracy_laplace}, and \Cref{cor:markov_utility} for the general bound, the Gaussian bound and the Markov chain bound respectively.
Additionally, to give an idea of the impact on utility in practice, we report the mean absolute percentage error (MAPE)  to estimate the expected relative error for a single execution. 
\begin{figure*}[t] % Use figure* for two-column wide figure
    \centering
    \subfloat[Galton, $n=897$ $m=3$]{\includegraphics[width=0.3\linewidth]{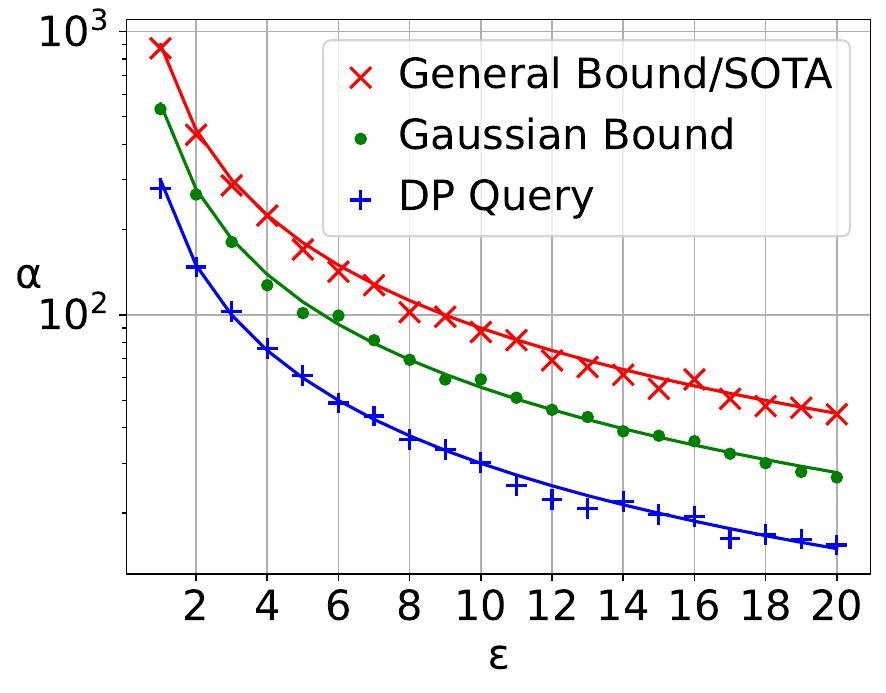}}
    \subfloat[FamilyIQ, $n = 868$, $m=2$.]{\includegraphics[width=0.3\linewidth]{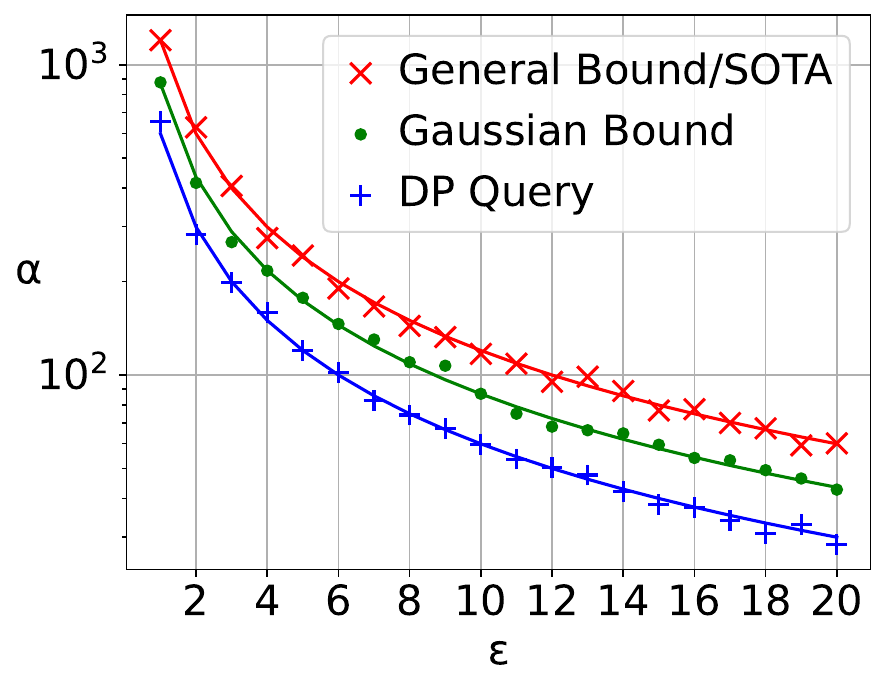}}
    \subfloat[SyntheticIQ, $n = 20000$, $m=2$.]{\includegraphics[width=0.3\linewidth]{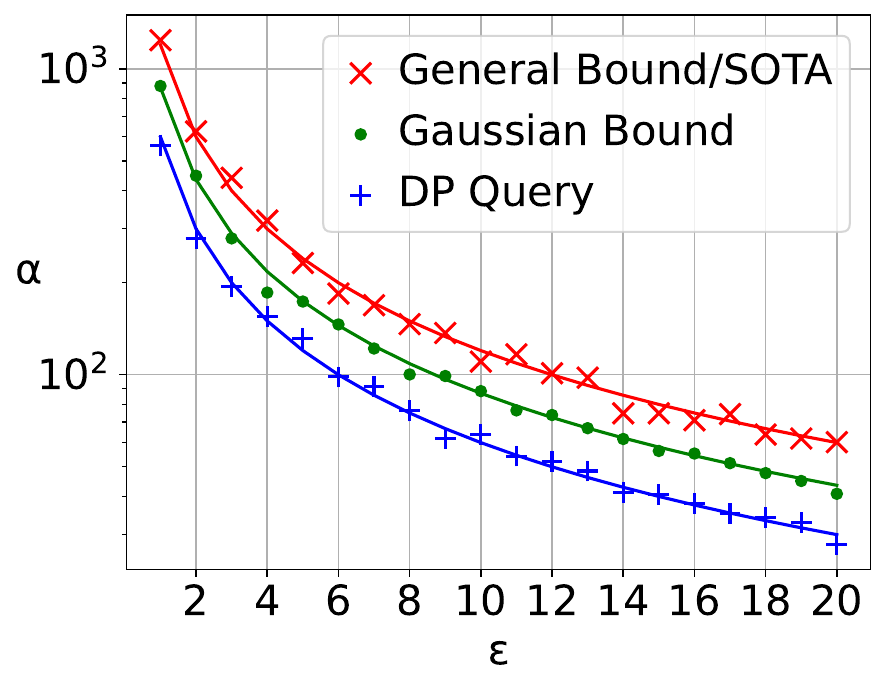}}
    \caption[Error of BDP]{Gaussian data results.  Lines show theoretical error at $\beta = 5\%$ and markers indicate empirical $95\%$ upper bounds.
    }
    \label{fig:results_gaussian}
\end{figure*}
\subsection{Mechanism and Experiment Design}\label{sec:experiment_mechanism_design}
\noindent
In order to provide BDP mechanisms that approximate the target queries presented in~\Cref{sub:metrics}, we use the Laplace mechanism with the noise calibrated through \Cref{thm:arbitrary_general_bound} for the  DP baseline, \Cref{th:gaussian_dp_bdpl_bound} for Gaussian data and \Cref{cor:markov_bdpl_bound} for Markov data. 
In this section, we refer to the DP privacy leakage by $\tau$, to avoid confusion with the actual maximum BDPL denoted by $\varepsilon$.

\subsubsection{Gaussian Data}
As explained in \Cref{sec:utility_experiment_data_and_queries}, we assume that the data is drawn from a multivariate Gaussian distribution with maximum number of correlated variables $m$ respectively.  
Both the general bound and state of the art~\cite{Yang_2015_BDP} indicate that for the Laplace mechanism $\mathcal{M}_{\tau, f}$, we have $\varepsilon = m \tau$, i.e., $\varepsilon= 3 \tau$ for Galton and $\varepsilon=2\tau$ for IQ datasets. Alternatively, according to the Pearson coefficients described in~\Cref{tab:data}, \Cref{th:gaussian_dp_bdpl_bound} tells us that $\mathcal{M}_{\tau, f}$ is $\varepsilon$-BDP, with $\varepsilon \approx 1.853 \tau,1.45\tau$  for Galton and IQ datasets respectively. Consequently, we fix BDPL values $\varepsilon\in (0, 20]$ and compute the corresponding $\tau$ using Eq.~\ref{eq:gaussian_tao}
 for the Gaussian-specific correlation approach and $\tau=\frac{\varepsilon}{3}$ for the general correlation and state of the art. For $\varepsilon \in (0, 5)$, we ensure strong theoretical privacy guarantees, while also considering the higher range $\varepsilon \in [5, 20]$, which has shown empirical resilience to certain privacy attacks~\cite{Near_2022_epsilon,Carlini_2022_MIA_Epsilon}.

\subsubsection{Markov Data}
As discussed in~\Cref{sub:databases} we assume that the data follows a Markov chain. According to the $\gamma$ values summarized in~\Cref{tab:data}, \Cref{cor:markov_bdpl_bound} tells us that the Laplace mechanism $\mathcal{M}_{\tau, f}$ applied to a counting query $f$ is $\varepsilon$-BDP, with
\begin{equation}\label{eq:markov_tao}
    \varepsilon_{A} = \tau + 8.05,\, 
    \varepsilon_{E,70} \approx \tau+ 4.7,\,\varepsilon_{E,80} \approx \tau+ 6.03,\,
    \varepsilon_{E,90} \approx \tau+ 8.54,
\end{equation}
In comparison, with the general bound we have $\varepsilon = n \tau $ for mechanism $\mathcal{M}_{\tau, f}$. 
Similar to Gaussian data, we apply the Laplace mechanism to compute the sum query of each subgroup with BDPL values $\varepsilon\in (0, 20]$ and compute the corresponding $\tau$ using Eq.~\ref{eq:markov_tao} for the Markov-specific mechanism and taking $\tau=\frac{\varepsilon}{n}$ for the general correlation approach. 
However, none of the datasets provide a symmetric transition matrix, which means that the proposal in ~\cite{chakrabarti2022optimal} is not applicable, making an empirical comparison impossible.

Note that, while $\varepsilon$-BDP can be provided for all values using the general bound and state of the art~\cite{chakrabarti2022optimal}, Eq.\ref{eq:markov_tao} only allows for $\varepsilon\geq 8.05,6.9,4.7$ and $8.45$ for Activity and Electricity data respectively, since $\tau$ must be positive (See~\Cref{sec:markov}).

In both Markov and Gaussian experiments, 
to calculate empirical confidence intervals, we execute the mechanism for each dataset $1000$ times. Since Activity Single Day provides $56$ unique datasets, we average the result over them.

\subsection{Results and Discussion}
%REPORT
\begin{figure}[tp] % Use figure* for two-column wide figure
    \centering
    \subfloat[Electricity $70$ kWh, $n =731$.]{ \includegraphics[width=0.3\linewidth]{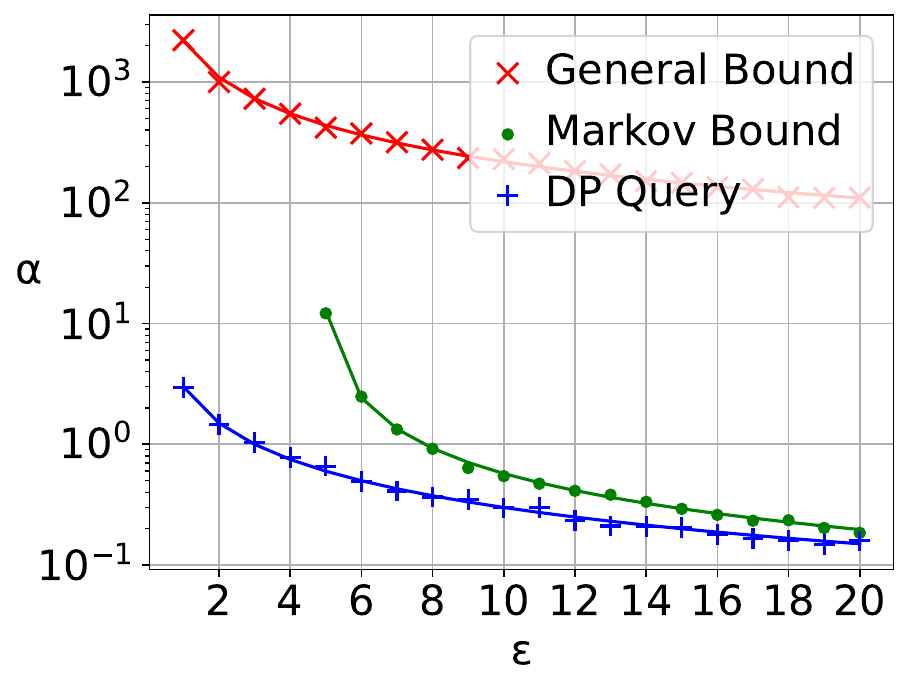}}
    \hfill
    \subfloat[Electricity $80$ kWh, $n =731$.]{ \includegraphics[width=0.3\linewidth]{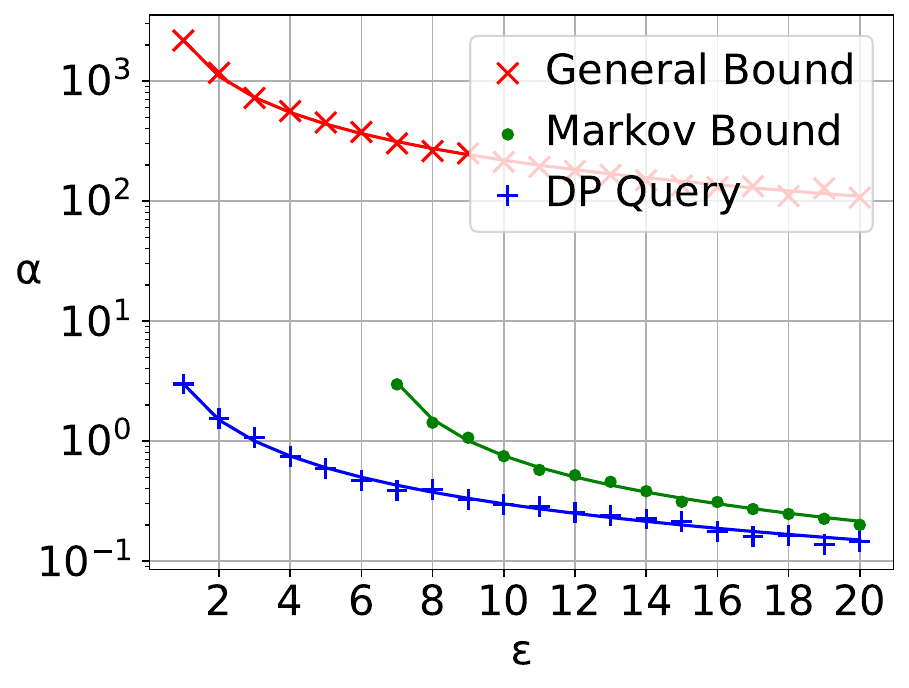}} 
    \hfill
     \subfloat[Electricity $90$ kWh, $n =731$.]{ \includegraphics[width=0.3\linewidth]{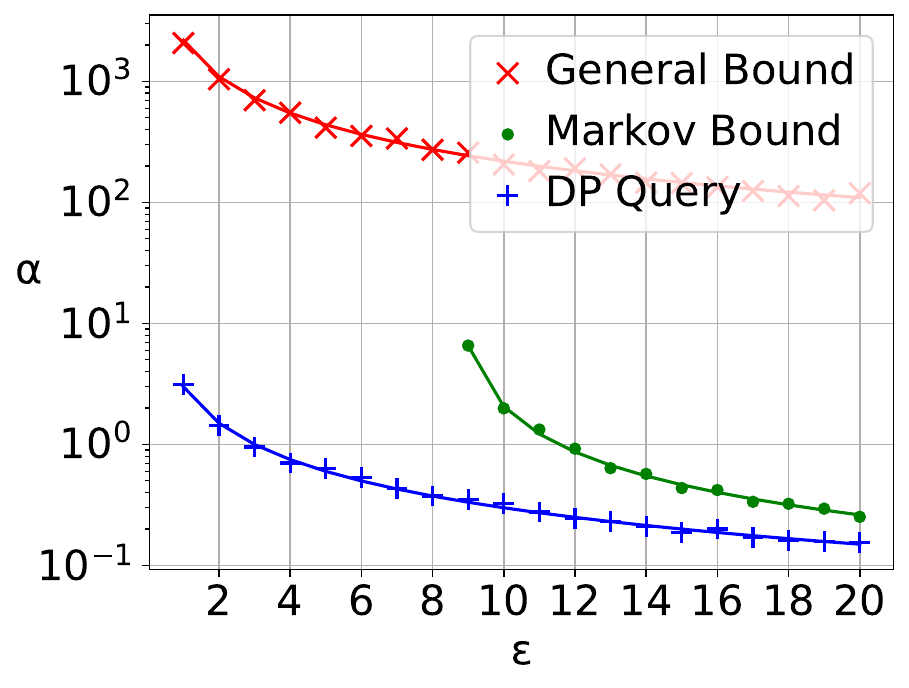}} 
     \hfill
     \subfloat[Activity Single Day, $n =288 $.]{ \includegraphics[width=0.3\linewidth]{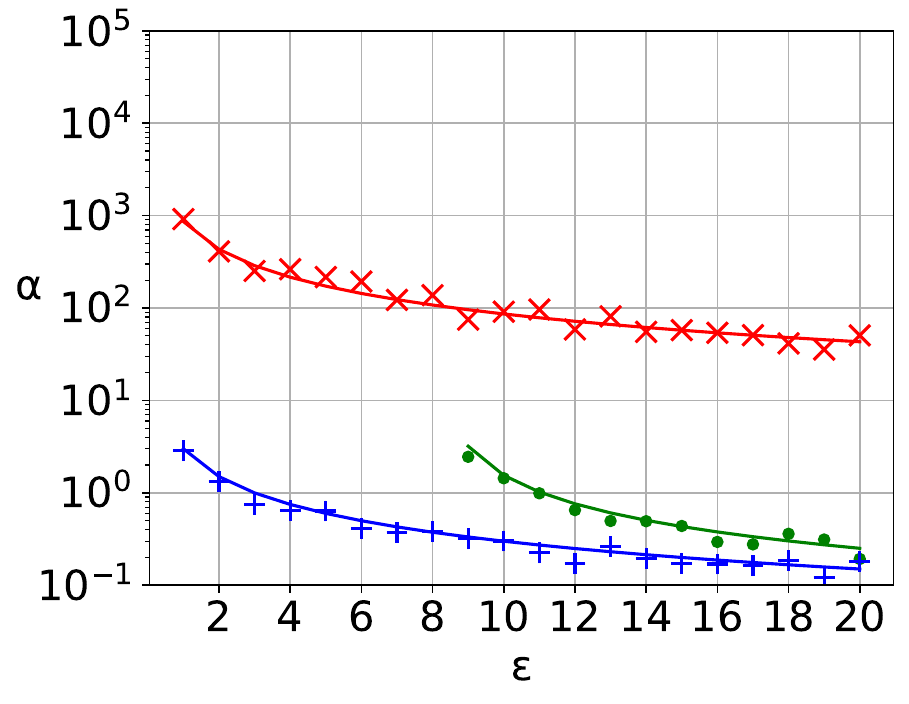}}
    \subfloat[Activity, $n = 17 \, 568$.]{ \includegraphics[width=0.3\linewidth]{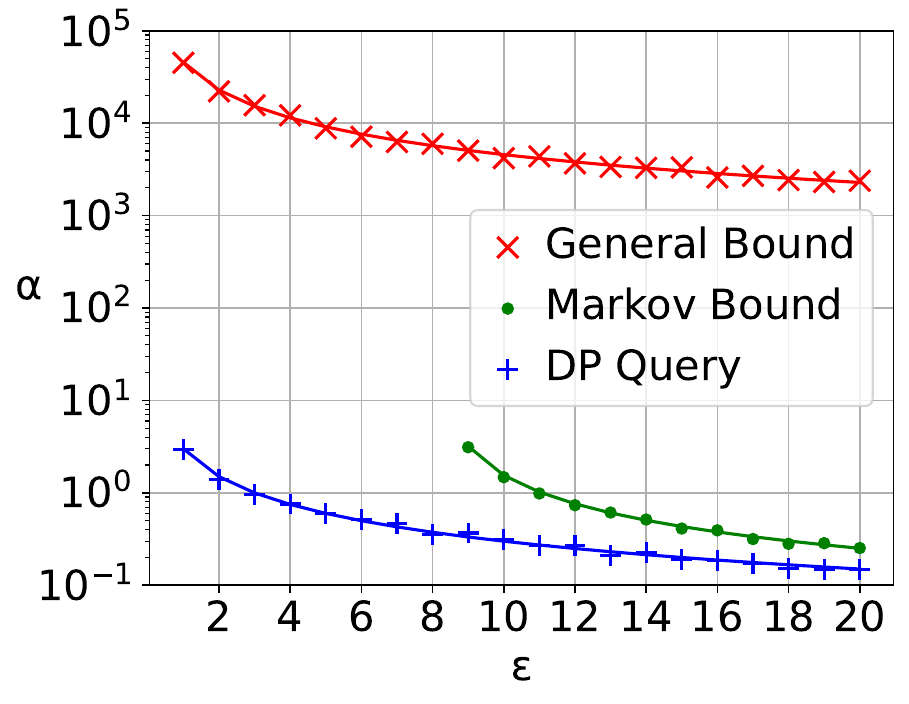}} 
    \caption[Error of BDP]{Markov Data Results. 
    Lines show theoretical error at $\beta = 5\%$ and markers indicate empirical $95\%$ upper bounds.
    }\label{fig:results_Electricity}
\end{figure}
\noindent
\Cref{fig:results_gaussian} presents the results for the Gaussian models, including our Gaussian-specific bound, the state-of-the-art bound from~\cite{Yang_2015_BDP} (which coincides with the general bound), and the DP Laplace mechanism for sum queries. We plot the DP mechanism as the baseline for the best possible utility; however, it is important to note that DP does not offer meaningful protection in this experiment, given correlation. Among the correlation-protecting mechanisms, those that use the Gaussian bound consistently outperform the s-o-t-a mechanism~\cite{Yang_2015_BDP} for all $\varepsilon$ in all datasets. Note that we plot all results on a logarithmic scale. This makes it harder to visually see the substantial reduction of error achieved by our mechanisms--particularly for small values of $\varepsilon$. For instance, for $\varepsilon=1$ the error is reduced by more than 400 units for both IQ datasets and 200 inches for the Galton. Note that the Galton height data uses imperial units (inches), thus the errors are also interpreted in inches.

The results for Markov chains are shown in \Cref{fig:results_Electricity}. Again, we use the DP mechanism as the baseline for the best possible utility, not as a comparable protective mechanism. For BDP mechanisms, we observe that the different Markov models tested lead to varying minimum achievable BDPL levels, as determined by our Markov-based bound: Electricity $70$ kWh yields the most favorable case with a minimum $\varepsilon = 4.9$, while $90$ kWh imposes the weakest bound with a minimum $\varepsilon = 8.45$. In contrast, the general bound supports all $\varepsilon > 0$.
In all cases where the Markov chain bound is applicable, mechanisms using it significantly outperform those relying on the general bound. While the error of mechanisms based on the general bound increases sharply, the error of both the Markov chain–based mechanism and the standard DP mechanism remains stable. The larger $n$, the larger the improvement of our approach respect to the general bound. For the largest dataset--Activity--the general bound results in a  $10^5$ times larger error than that of our proposed Markov chain bound. This is because the general bound scales with the size of the database $n$, while the Markov bound is independent of $n$, highlighting the huge benefit of using our novel bound for large datasets.

%CONCLUSIONS
The results demonstrate that BDP mechanisms calibrated with our newly proven Gaussian and Markov chain bounds outperform prior BDP mechanisms and mechanisms calibrated with the general bound in terms of utility on real-world data. 
Moreover, the empirical errors from our experiments closely align with our theoretical utility results, validating the practical applicability of our theorems.

\begin{figure*}[t] % Use figure* for two-column wide figure
     \centering
    \begin{subfigure}{0.3\linewidth}
        \centering
        \includegraphics[width=\linewidth]{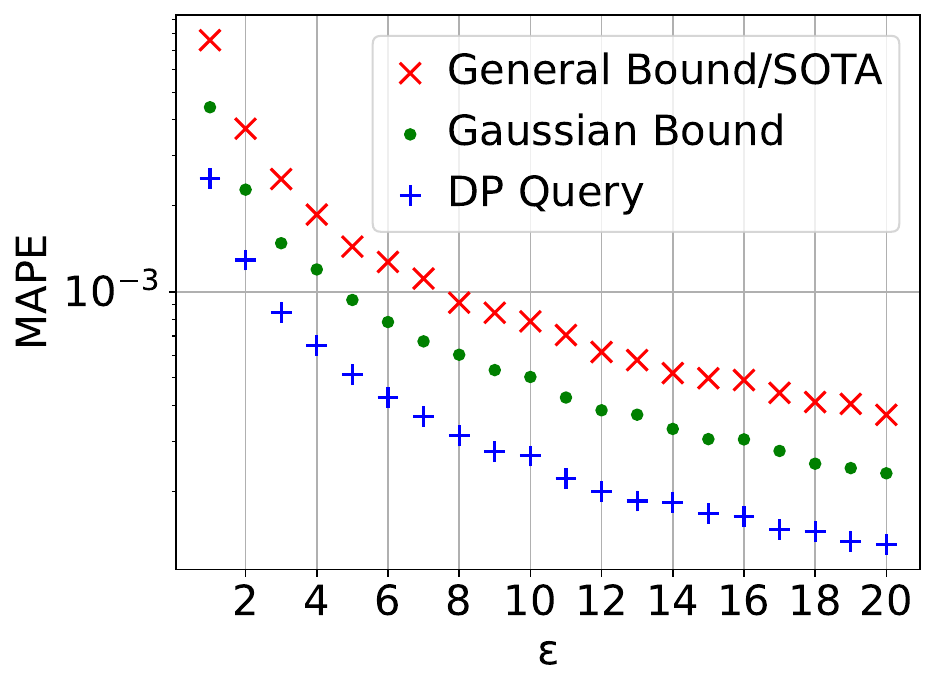}
        \caption{Galton height data, $n=897$}
    \end{subfigure}
    \hfill
    \begin{subfigure}{0.3\linewidth}
        \centering
        \includegraphics[width=\linewidth]{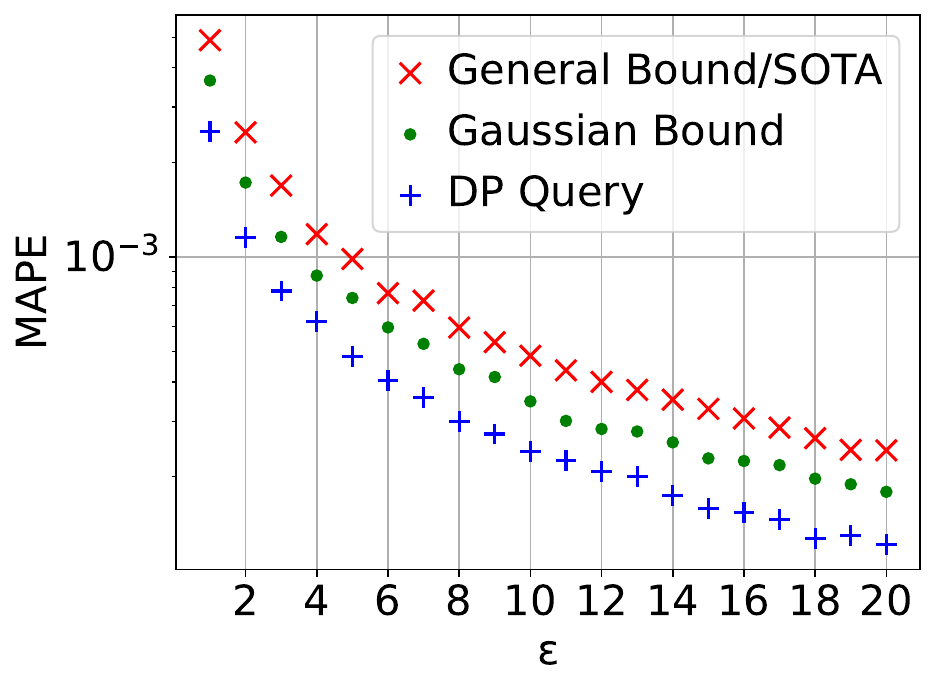}
        \caption{\centering FamiliyIQ, $n = 868 $}
    \end{subfigure}
    \hfill
    \begin{subfigure}{0.3\linewidth}
        \centering
        \includegraphics[width=\linewidth]{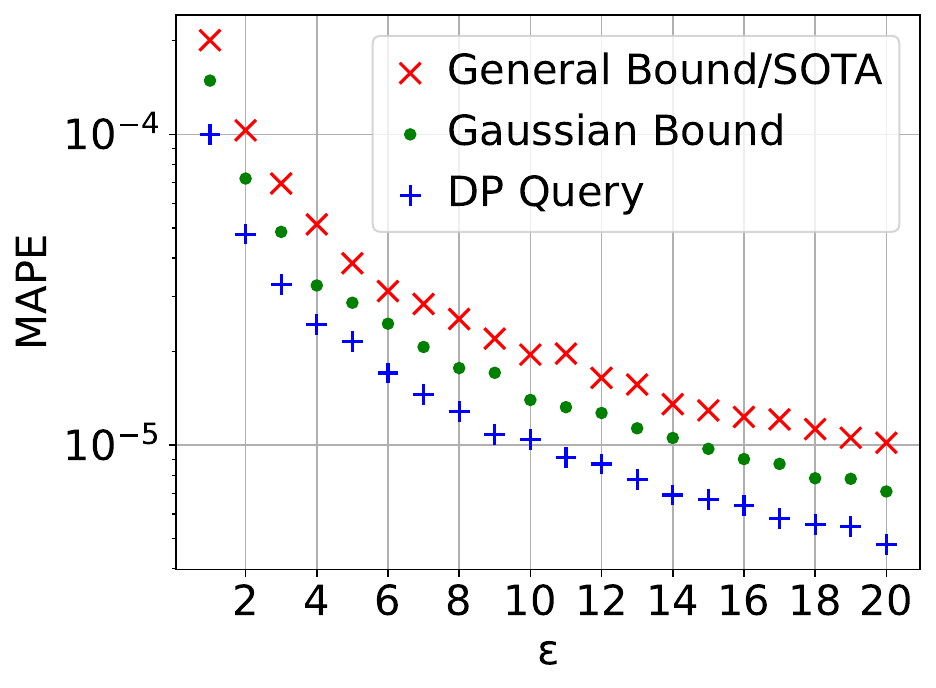}
        \caption{\centering SyntheticIQ,  $n = 20 \, 000$}
    \end{subfigure}
    \caption[[MAPE of BDP queries]{MAPE of private sum queries on data correlated according to a Gaussian mutivariate distribution.
    }
    \label{fig:mape_gaussian}
\end{figure*}

\begin{figure*}[t] % Use figure* for two-column wide figure
     \centering
    \begin{subfigure}{0.3\linewidth}
        \centering
        \includegraphics[width=\linewidth]{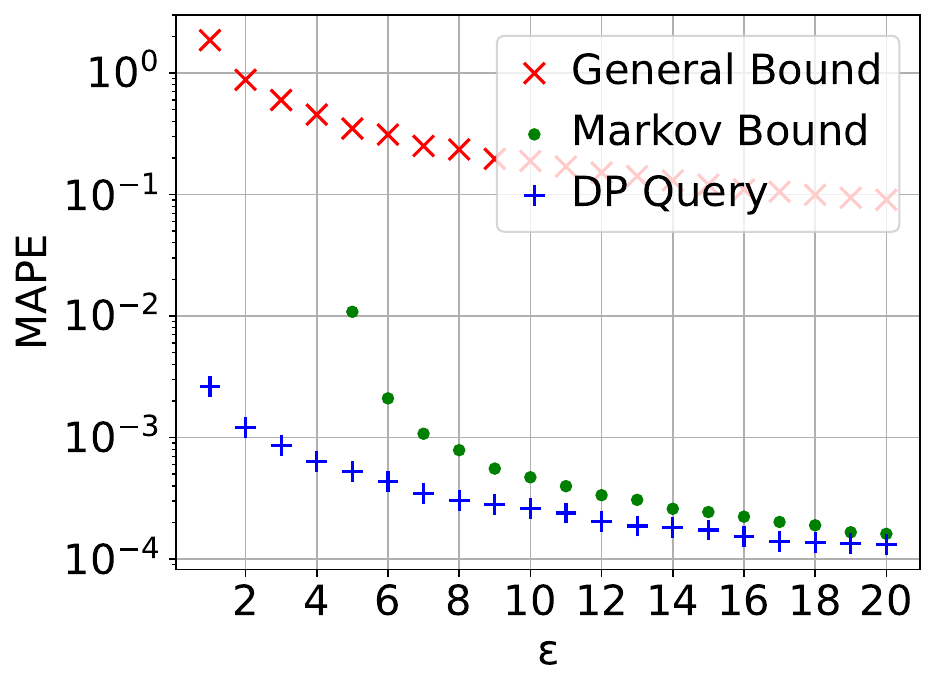}
        \caption{Electricity $70$ kWh, $n=731$}
    \end{subfigure}
    \hfill
    \begin{subfigure}{0.3\linewidth}
        \centering
        \includegraphics[width=\linewidth]{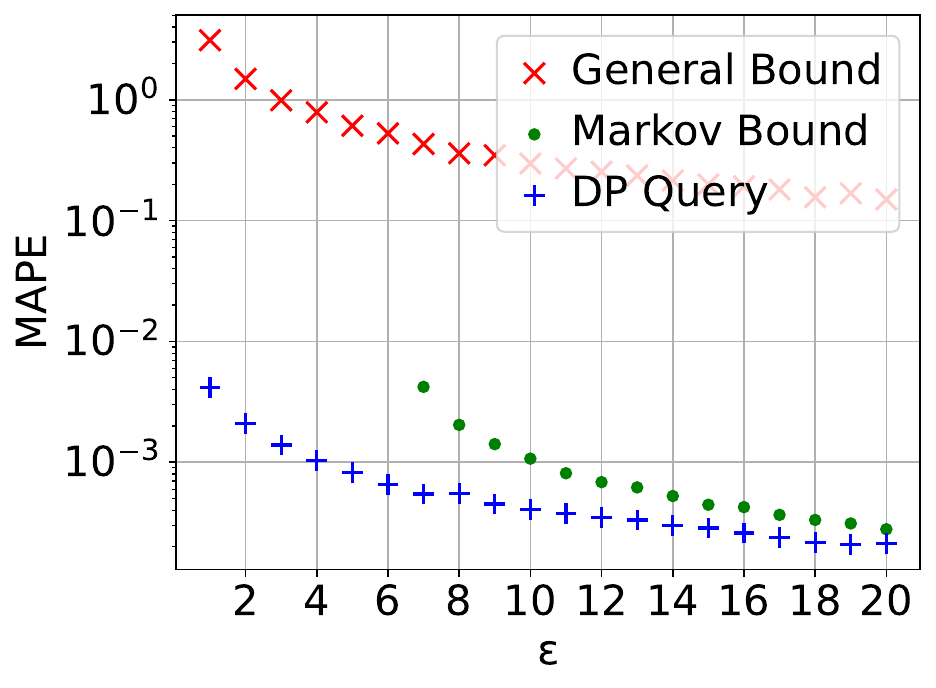}
        \caption{Electricity $80$ kWh, $n=731$}
    \end{subfigure}
    \hfill
     \begin{subfigure}{0.3\linewidth}
        \centering
        \includegraphics[width=\linewidth]{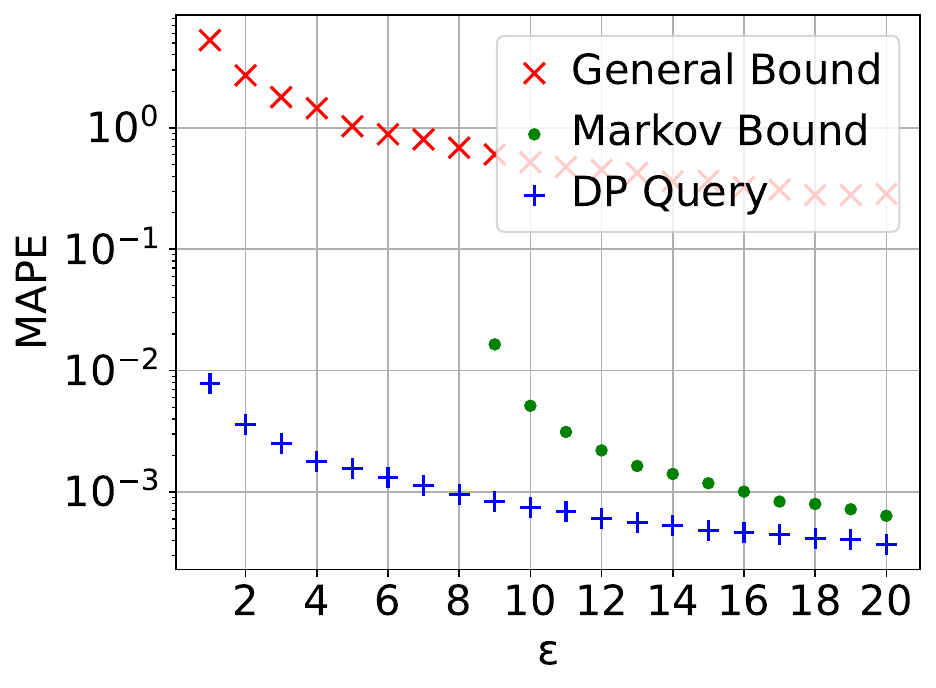}
        \caption{Electricity $90$ kWh, $n=731$}
    \end{subfigure}
    \hfill
    \begin{subfigure}{0.3\linewidth}
        \centering
        \includegraphics[width=\linewidth]{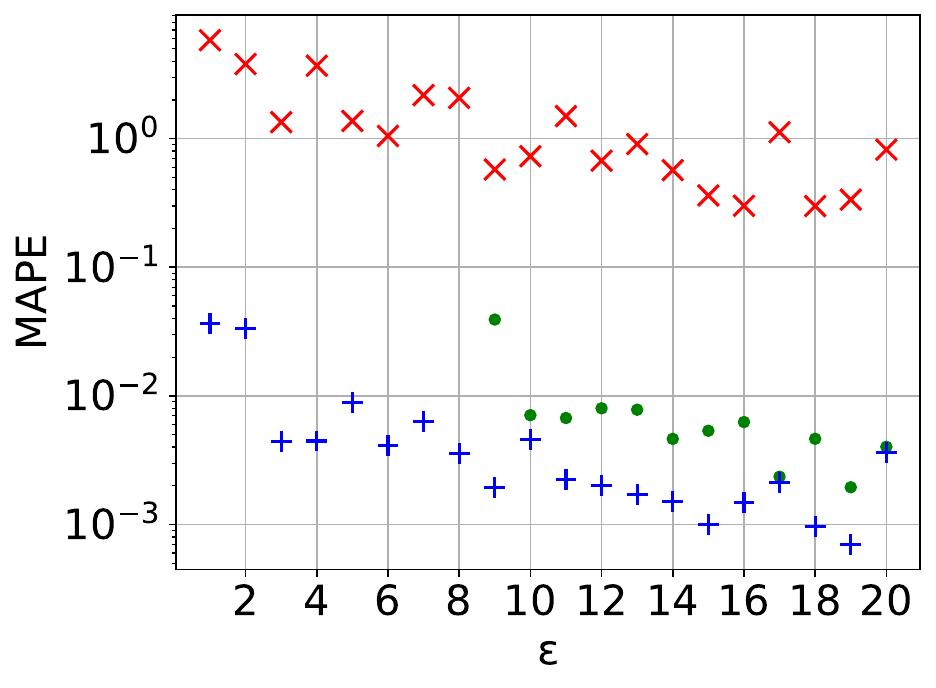}
        \caption{\centering Activity single day, $n = 288$}
    \end{subfigure}
    \begin{subfigure}{0.3\linewidth}
        \centering
        \includegraphics[width=\linewidth]{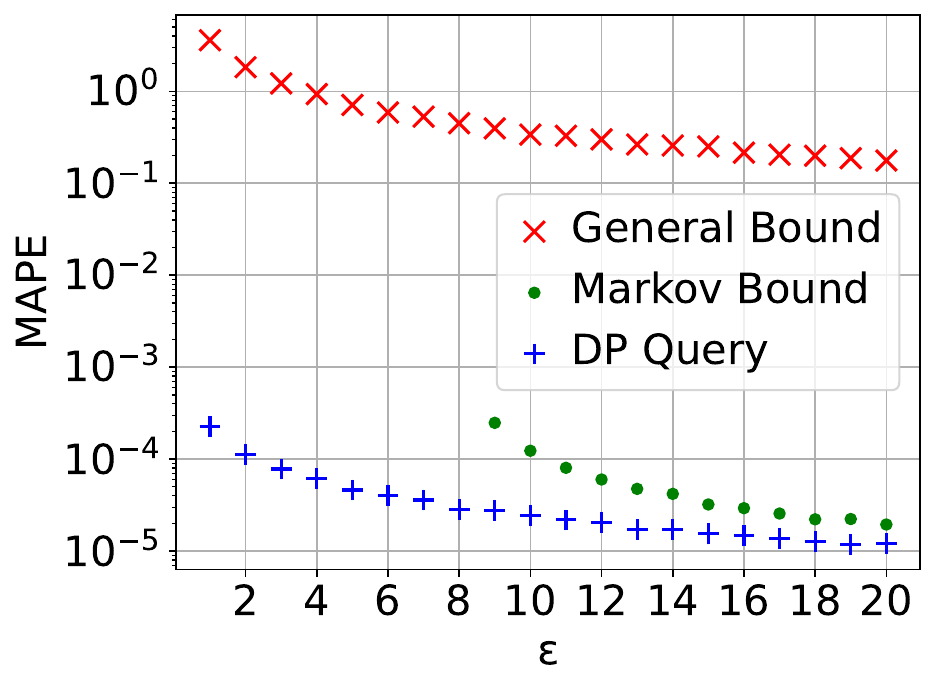}
        \caption{\centering Activity,  $n = 17 \, 568$}
    \end{subfigure}
    \caption[[MAPE of BDP queries]{MAPE results for databases following a Markov distribution.
    }
    \label{fig:mape_markov}
\end{figure*}

We extend this study with the analysis of the relative error. \Cref{fig:mape_gaussian} show the MAPE for Galton and IQ datasets (Gaussian correlation model) and \Cref{fig:mape_markov} Activity and Electricity data (Markov chain model). As expected, the DP query has the lowest MAPE however it does not offer protection against correlation. When we offer BDP protections we see that the correlation-specific BDP mechanisms (i.e., using the Gaussian bound or the Markov chain bound) outperform the BDP mechanism protecting against arbitrary correlation with the general bound following the same trend as for the $(\alpha,\beta)$-accuracy. The benefit is particularly prominent for data following a Markov chain where the MAPE of the general bound reaches values above $100\%$, resulting in an error as large as the ground truth itself. In comparison, the Markov chain bound achieves errors below $10\%$ for single day activity data, and below $0.1\%$ for Activity and Electricity datasets.

We acknowledge certain limitations when extrapolating our results. The validity of our experimental findings is constrained by the specific databases used. While the Galton height data serves as a well-known example of record correlation, it reflects only one of many possible correlation patterns. Similarly, most practical applications of a Markov chain would involve more than two states, introducing complexity beyond the binary-state model used in our study. Nevertheless, our results provide valuable insight into the practical applicability of our theorems and indicate their potential for real-world scenarios. Furthermore, these experiments demonstrate that achieving meaningful utility while protecting against correlation is feasible in practice.

\section{Conclusion}\label{sec:conclusions}
\noindent
In this paper, we explored the utility of BDP mechanisms for correlated data. We addressed prior limitations by analyzing broader correlation models and providing a detailed study of privacy-utility trade-offs, supported by theoretical results and empirical evidence. Specifically, we established new connections between DP and BDP mechanisms and demonstrated how they can be leveraged for privacy protection under correlation.

We proved that any $\varepsilon$-DP mechanism satisfies $m\varepsilon$-BDP, where $m$ is the size of the correlated group, and showed this bound is tight. We then improved upon it by considering multivariate Gaussian and Markov models, deriving novel bounds on BDP leakage that provide stronger utility guarantees than the s-o-t-a approaches under the same privacy constraints. The advantage of our correlation-specific bounds is particularly evident under Markov-modeled correlations. While mechanisms based on the general bound exhibit high sensitivity to the number of correlated records, our Markov-based bound remains robust and stable regardless of the dataset size.

While it remains a futile attempt to apply BDP without assuming a specific correlation model, both our theoretical and experimental results demonstrate that it is possible to achieve better utility without weakening the adversary model in practical scenarios: (a) when the number of correlated records is small, (b) when the data follows a weakly correlated Gaussian model, or (c) when the data is a time series following a Markov chain with
sufficiently similar transition probabilities.

Overall, our results~\Cref{thm:arbitrary_general_bound,th:gaussian_dp_bdpl_bound,cor:markov_bdpl_bound} advance the theoretical and practical understanding of BDP, enabling the reuse of DP mechanisms in correlated settings. This opens future directions for deriving correlation-specific bounds to design more accurate BDP mechanisms protecting against real-world correlation-based attacks.

\printbibliography
\end{document}